\documentclass[aps,prd,preprint,nofootinbib,superscriptaddress,tightenlines]{revtex4-1}
\usepackage{amsfonts}
\usepackage{mathrsfs}
\usepackage{graphicx}
\usepackage{amsmath}
\usepackage{amssymb}
\usepackage{multirow}
\usepackage{subfigure}
\usepackage{epsfig}
\usepackage{graphicx}
\usepackage{booktabs}
\usepackage{array}
\usepackage{tabularx}
\usepackage{slashed}
\usepackage{ulem}
\usepackage{bm}
\usepackage{threeparttable}
\usepackage{float}
\usepackage{comment}
\usepackage{placeins}
\usepackage{hyperref}
\usepackage{appendix}

\newcommand{\bqa}{\begin{eqnarray}}
\newcommand{\eqa}{\end{eqnarray}}
\newcommand{\beq}{\begin{equation}}
\newcommand{\eeq}{\end{equation}}
\graphicspath{{fig/}{dia/}} \DeclareGraphicsExtensions{.eps}
\begin{document}
\baselineskip 20pt
\title{Planar Property and Long-range Azimuthal Correlation in $e^+e^-$ Annihilation}

\author{Xuan Chen}
\email{xuan.chen@sdu.edu.cn}
\affiliation{School of Physics, Shandong University, Jinan, Shandong 250100, China}

\author{Yuesheng Dai}
\email{yueshengdai@mail.sdu.edu.cn}
\affiliation{School of Physics, Shandong University, Jinan, Shandong 250100, China}

\author{Shi-Yuan Li}
\email{lishy@sdu.edu.cn}
\affiliation{School of Physics, Shandong University, Jinan, Shandong 250100, China}

\author{Zong-Guo Si}
\email{zgsi@sdu.edu.cn}
\affiliation{School of Physics, Shandong University, Jinan, Shandong 250100, China}

\author{Huiting Sun}
\email{huiting.sun@mail.sdu.edu.cn}
\affiliation{School of Physics, Shandong University, Jinan, Shandong 250100, China}

\begin{abstract}

The $e^+e^-$ annihilation of unpolarized beams is free from initial hadron states or initial anisotropy around the azimuthal angle, hence ideal for studying the correlations of dynamical origin via final state jets. We investigate the planar properties of the multi-jet events employing the relevant event-shape observables at next-to-next-to-leading order ($\mathcal{O}$($\alpha_{s}^{3}$)) in perturbative QCD; particularly, the azimuthal angle correlations on the long pseudo-rapidity (polar angle) range  (Ridge correlation) between the inclusive jet momenta are calculated. We illustrate that the significant planar properties and the strong correlations as the consequence  are natural results of the energy-momentum conservation of the perturbative QCD radiation dynamics. Our study provides benchmarks of hard strong interaction background for the investigations on  the collective and/or thermal effects via the Ridge-like correlation observables for various scattering processes.

\vspace {2mm} 
\end{abstract}

\maketitle

\section{Introduction}

The Ridge  phenomena, i.e., the azimuthal angle ($\phi$) correlations on the long pseudo-rapidity (or polar angle $\theta$) range between the triggered hadrons and associated hadrons, were first observed in relativistic heavy ion collisions at the RHIC \cite{Adams:2005ph,Alver:2008gk,Abelev:2009af,Alver:2009id} and then at the LHC \cite{Chatrchyan:2011eka,Chatrchyan:2012wg,Aamodt:2010pa,ATLAS:2012at,CMS:2013bza}. They are intensively studied and  taken as important  signals of the thermal collective effects of the expansion of the hot dense matter such as Quark-Gluon Plasma (QGP). The  hydrodynamic flows   
   have been  correspondingly measured for the nontrivial azimuthal distributions  in most of the above mentioned experiments \cite{Adams:2005ph,Alver:2008gk,Abelev:2009af,Alver:2009id,Chatrchyan:2012wg,Aamodt:2010pa,ATLAS:2012at,CMS:2013bza}.
Similar phenomena are also observed from events  with large hadronic multiplicities in various small system collision processes  such as proton-proton  collisions \cite{Khachatryan:2010gv,Aad:2015gqa,Khachatryan:2015lva,Khachatryan:2016txc},
proton-nucleus  collisions \cite{CMS:2012qk,alice:2012qe,Aad:2012gla,Aaij:2015qcq,ABELEV:2013wsa,Khachatryan:2014jra,Khachatryan:2015waa,ATLAS:2017hap,ATLAS:2017rtr,PHENIX:2018lia},
as well as  lighter nucleus-nucleus collisions \cite{Adamczyk:2015xjc,Adare:2015ctn,Aidala:2017ajz,PHENIX:2018lia}. In other scattering processes, such as  $e^+e^-$ annihilation  at low energy \cite{Badea:2019vey,Belle:2022fvl}, electron-proton collisions \cite{ZEUS:2019jya,ZEUS:2021qzg}, photon-proton collision \cite{CMS:2022doq} and photon-nuclear reactions  \cite{ATLAS:2021jhn} systems, this correlation is also studied.
Recently,  interesting implications regarding the  Ridge  phenomena have been found in the studies of single hard jet by CMS collaboration \cite{Baty:2021ugw,CMS:2023iam} and in
$e^+e^-$ annihilation  at LEPII energies \cite{Chen:2023njr}.
 In these  small system  scatterings, the collective behaviors and the production of QGP droplet  which had
been assumed to be absent traditionally, were introduced  to explain the observed Ridge correlations
 \cite{Vertesi:2024fwl,Zhao:2024wqs,Zheng:2024xyv,Nagle:2013lja,Schenke:2014zha,Shen:2016zpp,Weller:2017tsr,Mantysaari:2017cni,
Zhao:2017rgg,Bierlich:2019wld,Katz:2019qwv,Zhao:2022ayk,Zhao:2022ugy,Bzdak:2014dia,He:2015hfa,Lin:2015ucn,Nagle:2017sjv,
Kurkela:2018qeb,Kurkela:2019kip,Zhao:2021bef,Schenke:2019pmk,Giacalone:2020byk,Schenke:2021mxx,Grosse-Oetringhaus:2024bwr}.
However,  the  `precursors' \cite{Grosse-Oetringhaus:2024bwr} or the dynamics of the anisotropy which lead to  the    Ridge correlations in the small system
scatterings remain unclear.
In this paper, we study  the azimuthal anisotropy of  multiple  hard jets system  in $e^+e^-$ annihilation  and the corresponding   Ridge correlations,  within the framework of   perturbative Quantum Chromodynamics (pQCD).

\begin{figure}[hbt!]
  \centering
\includegraphics[width=0.8\textwidth]{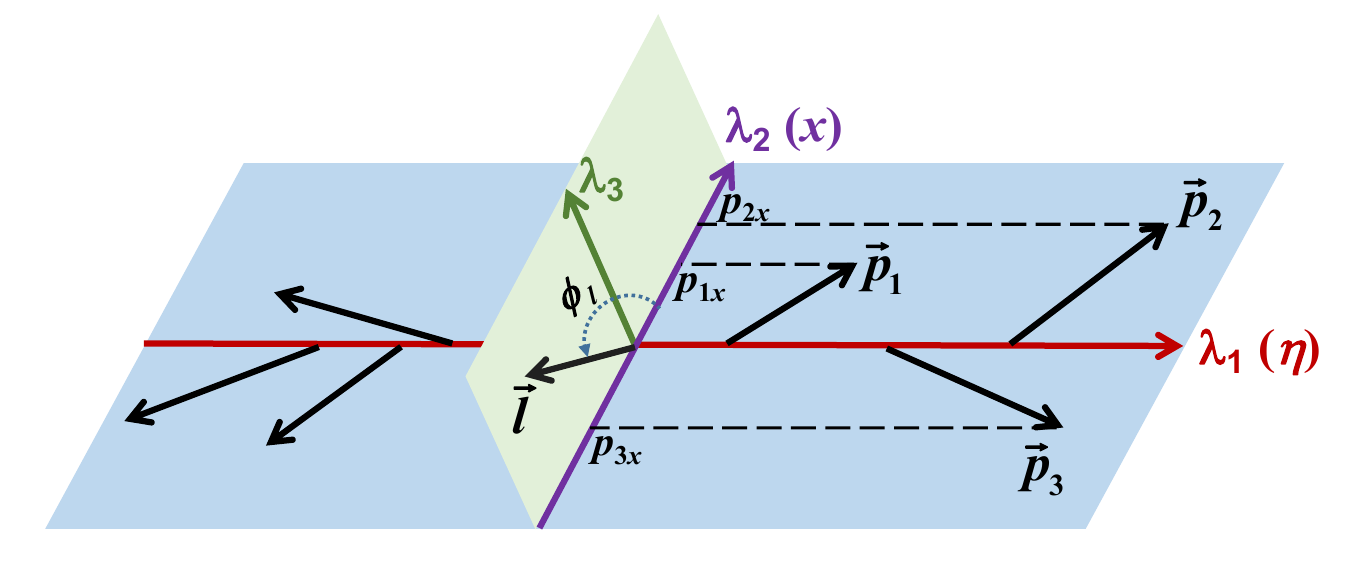}
\caption{An extreme planar event sample with all hadrons in the same plane. The three Sphericity axis (see Section \ref{sec:event shape}) are marked by the eigenvalues  $\mathbf{\lambda_1}$, $\mathbf{\lambda_2}$ and $\mathbf{\lambda_3}$ (=0).}
\label{plane}
\end{figure}

The Ridge correlation is  the  global property of  the event and   manifests
that  the global  phase space is planar.
Two key ingredients lead to the Ridge phenomena in relativistic heavy ion collisions: one is   the initial geometric anisotropy of non-central collisions as input;  the other is  the intermediate  multi-interaction and collective behaviors of the bulk system which  can transfer this anisotropy to the phase space of the final state hadron system.
As a result, the plane is stretched by the pseudo-rapidity direction (to define the polar angle) and a special transverse direction (to define the azimuthal angle) for each specific event. The hadronic momentum distribution is favored around this plane.
We illustrate in Figure~\ref{plane} the most extreme case where all hadron momenta are on the same plane.
One can observe a joint distribution on $\Delta \phi$ and  $\Delta \eta$ as
  \begin{align}
       f(\Delta \phi, \Delta \eta)= (c_1 \delta(\Delta \phi)+ c_2 \delta(\Delta\phi-\pi)) \times \theta(\eta_0-|\Delta \eta|) \times g(\Delta \eta),
       \label{eq of fig1}
   \end{align}
where $\theta (x)$ is the Heaviside step function with $\eta_0$ $(>0)$ a parameter for the range  of the `Ridge' in the pseudo-rapidity direction determined  by the available hadron pseudo-rapidities, and $ g(\Delta \eta)$ is a slow-varying function. The real data are obtained by  smearing and fluctuating this extreme distribution and hence look like the `Ridge'
\footnote{The equation \eqref{eq of fig1} shows the double Ridge correlation which is observed in most  heavy ion collisions, only that for the most central scatterings, some measurements indicate the away-side Ridge is almost flatted.}.

The above analysis  emphasizes  that  {\it  the initial geometric  anisotropy} is transferred by the  collective behaviors to the final state hadrons as the Ridge correlation in heavy ion collisions.
If  any other kind  of  anisotropy exists and could  be translated to the final state, a
similar  behavior could be observed.
We have suggested \cite{ppt} that the
 hard  scatterings and radiations can  lead to significantly  planar phase space. This dynamical anisotropy can be one of the precursors of  the final state Ridge phenomena  in the small system collisions.
In heavy ion collisions, dynamical anisotropy could  be smeared in the thermalization
or by multi-parton interactions, but still provides important  information  for probing the properties of the QGP.

We take the $e^+e^-$ annihilation process to study the anisotropy from the hard interactions and radiations since it is free from initial state hadrons or initial anisotropy for unpolarized beams.
For the 3-jet events, the jet momenta construct  a triangle in the center-of-mass frame because of momentum conservation, and must be in a plane since our space is Euclidean.
This is not only a theoretical expectation, but plays the key role in the discovery of the gluon jet. The first 3-jet event caught by the TASSO collaboration critically employed the planar property of the corresponding final state hadrons \cite{Ellis:1976uc,Wu:2015yoa}.
For an event with higher multiplicity of jets, their momenta are not necessarily restricted to a plane. Our previous study \cite{ppt}, by employing the event generator PYTHIA~\cite{SJOSTRAND2015159}, shows that most of
the events are not isotropic but planar for the reason of energy momentum conservation.
The planar events are not strongly correlated with hadron multiplicity.
To quantify the precise impact of hard scatterings among multiple jets,
we study the above mentioned planar properties of the inclusive 3-jet production at the full NNLO pQCD \footnote{In this paper, `inclusive n-jet' means the final state with n or more jets.}.
The results obtained with the  pQCD  well-defined jet momenta to make the investigations show significant  planar property and the Ridge-like correlations \footnote{We use the terminology  'ridge-like' to distinguish with those obtained with the hadron momenta}.
So consequently
our pQCD calculation   provides benchmarks for further  study of the effects from pQCD resummation,  parton shower,
multi-parton interaction,
preconfinement,  as well as hadronization and (possible) thermal behaviors in both partonic and hadronic phases \cite{Grosse-Oetringhaus:2024bwr,ppt}. As to estimate the affections of the planar property from the hard interactions to the final states,  we will also review
part of the results of \cite{ppt}.
In this paper,  we concentrate on the    $e^+e^-$ annihilation events  at the Z-pole. From the investigations  in \cite{ppt}, such events
share the similar planar property as events in LEPII \cite{Chen:2023njr}. The methods and conclusions of this paper can be applied to LEPII as well as
CEPC energies.
Here we also point out  that the event shape had  also been  measured in proton-proton scatterings  at the LHC \cite{ATLAS:2012tch}. An extension of current study could be performed at the NNLO accuracy based on the inclusive hadronic 3-jet production at the LHC \cite{Czakon:2021mjy,Chen:2022ktf,Gehrmann:2023dxm}.

The calculation of jet observables from $e^+e^-$ annihilation with higher order QCD corrections requires consistent cancellation of the infrared singularities.
Three independent studies have achieved the NNLO accuracy for the inclusive 3-jet production~\cite{Gehrmann-DeRidder:2007nzq,Gehrmann-DeRidder:2007vsv,Gehrmann-DeRidder:2008qsl,Weinzierl_2008,Gehrmann-DeRidder:2009fgd,Weinzierl:2009ms,Gehrmann-DeRidder:2014hxk,DelDuca2016ThreeJetPI,Gehrmann:2017xfb}.
Our fixed order calculation is based on the publicly available code EERAD3~\cite{Gehrmann-DeRidder:2007vsv,Gehrmann-DeRidder:2008qsl,Gehrmann:2017xfb}, which is later implemented in the NNLOJET event generator~\cite{Gehrmann:2017xfb,Huss:2025iov}.
The antenna subtraction method~\cite{GehrmannDeRidder:2005cm,Gehrmann-DeRidder:2005svg,Daleo:2006xa,Currie:2013vh,Chen:2022ktf,Gehrmann:2023dxm} is applied to remove the infrared divergences such that the multi-dimensional phase space integration could be performed with the aid of adaptive Monte Carlo integrator VEGAS~\cite{Lepage:1977sw}. We implement the event shape observables studied in this work in NNLOJET~\cite{Huss:2025iov}.
During this study, an efficient framework utilizing generalized antenna functions has been applied to the inclusive 3-jet production at NNLO~\cite{Fox:2024bfp}.

The paper is organized as follows. In Section~\ref{sec:tools}, we  give the introduction to the  event shape observables employed to investigate the planar property. In Section~\ref{sec:numerics}, we study the differential distribution of these observables and Ridge-like correlations with various values of jet resolution parameter and the results of the scale variation are provided. Finally, the conclusion and discussion are presented in Section~\ref{sec:conclusions}.

\section{Analysis framework and tools}
\label{sec:tools}

\subsection{Event shape observables}
\label{sec:event shape}

\indent We focus on the event shape observables such as the linearized sphericity tensor ($S^{jk}$) and its three eigenvalues ($\lambda_{1}, \lambda_{2}$ and $\lambda_{3}$)
, modified planarity ($\Tilde{P}$) and the Ridge correlation with respect to the jet momenta in the inclusive 3-jet events at NNLO and the inclusive 4-jet events at NLO.

These observables are all related to the planarity of the events. Specifically, \(\lambda_1\), \(\lambda_2\), \(\lambda_3\) and $\tilde{P}$ are infrared (IR)-safe at the parton level. In contrast, the Ridge correlation is not IR-safe at the parton level; thus, we calculate its distribution at the jet level with the Durham jet algorithm. Accordingly, the observables related to event planarity—including \(\lambda_1\), \(\lambda_2\), \(\lambda_3\) and $\tilde{P}$—are also computed at the jet level to corroborate the Ridge correlation. These jet-defined observables can be directly compared with experimental data.

The linearized sphericity tensor is ~\cite{SJOSTRAND2015159,Bjorken:1969wi,Andersson1983PartonFA}:
\begin{align}
    S^{jk}=\frac{\sum_i \frac{p_i^j p_i^k}{\left|\mathbf{p}_i\right|}}{\sum_i\left|\mathbf{p}_i\right|},
\end{align}
where $\mathbf{p}_{i}$ denotes the three-momentum of jet $i$ and the summation is performed over all final state jets in an event. Here, $j,k=1,2,3$ correspond to the three spatial components. The symmetric 3$\times$3 matrix $S$ has three real eigenvalues and three mutually orthogonal eigenvectors corresponding to these eigenvalues. The determination of these eigenvalues involves solving the characteristic equation of $S$, which is a cubic equation and the roots can be analytically solved using the Shengjin Formula~\cite{fan1989new} (see Appendix~\ref{app:Shengjin Formula}).
For Euclidean space, it is sufficient to consider only the cases with real solutions.
These three eigenvalues can be ordered as $\lambda_{1} \ge \lambda_{2} \ge \lambda_{3}\ge 0$, with $\lambda_{1} + \lambda_{2} + \lambda_{3}=1$, such that $\lambda_1$ is constrained within the range $\left[\frac{1}{3}, 1\right]$.
Events in which all hard partons are closely aligned in a collinear direction (the back-to-back limit) correspond to $\lambda_1 \approx 1$, while an approximately isotropic event corresponds to $\lambda_1 \approx \frac{1}{3}$. The $\lambda_{2}$ is restricted to the interval $\left[0, \frac{1}{2}\right]$. In a planar event, the plane is defined by the eigenvectors corresponding to $\lambda_{1}$ and $\lambda_{2}$.
The $\lambda_{3}$ is bounded between $\left[0, \frac{1}{3}\right]$. It measures the transverse momentum component out of the event plane: a planar event has $\lambda_{3}=0$ whereas an approximately isotropic event has $\lambda_{3} \approx \frac{1}{3}$.

We define the modified planarity $\Tilde{P}$ as:
\begin{align}
\label{tildeP}
    \Tilde{P}=\frac{\lambda_{2}-\lambda_{3}}{\lambda_{2}+\lambda_{3}},
\end{align}
which is constrained to the range $\left[0, 1\right]$.
It measures the planarity of an event: a planar event has $\Tilde{P}=1$ whereas an approximately isotropic one has $\Tilde{P} \approx 0$.
For $e^{+}e^{-}$ annihilation to the inclusive 3 jets, the $\lambda_{2}$ is rather small even though much larger than $\lambda_{3}$ for most of the events (which are planar).  Consequently, if we only
use $\lambda_{2}-\lambda_{3}$, it will always peak around small values and cannot show whether the event is planar or spherical.

\indent The Ridge correlation is the joint distribution of $\Delta \phi$ and  $\Delta \eta$ between final state jets momenta from the same event, defined as
\begin{align}
    R(\Delta \phi, \Delta \eta)=\frac{1}{\sigma}\frac{d^2\sigma}{d\Delta \phi d\Delta \eta}=\frac{1}{\sigma}\sum_{i<j}\int d\Delta\phi_{ij}d\Delta\eta_{ij}(\frac{d^2\sigma}{d\Delta \phi_{ij} d\Delta \eta_{ij}})\delta(\Delta\phi_{ij}-\Delta\phi)\delta(\Delta\eta_{ij}-\Delta\eta),
    \label{EEC-definition}
\end{align}
where $i,j=1,2,...,n$, and $n$ denotes the jet multiplicity of the event; $\Delta \phi$ and $\Delta \eta$ represent the differences in azimuthal angles and pseudo-rapidities of jets, respectively.
In our study, we use the direction of the eigenvector associated with the largest eigenvalue $\lambda_1$ to define the polar angle.
The eigenvector associated with $\lambda_1$ is obtained using the power method \cite{golub2013matrix}. Initially, an initial vector $\mathbf{x}_0$ is repeatedly multiplied by the linearized sphericity matrix $S$. This process ensures that the resulting vector converges towards the eigenvector associated with the largest eigenvalue $\lambda_1$.
However, if $\mathbf{x}_{1}$ is an eigenvector corresponding to $\lambda_1$, then $-\mathbf{x}_{1}$ is equally valid as an eigenvector for the same eigenvalue. Selecting the eigenvector with forward rapidity in the lab frame as our convention, we ensure that our final result is symmetric in $\Delta\eta$:
\begin{align}
    R(\Delta\phi,\Delta\eta)=R(\Delta\phi,-\Delta\eta),
\end{align}
which is a consequence of the parity conservation in QCD.
Subsequently, the angle between the jet momentum $\mathbf{p}$ and the axis $\mathbf{x}_{1}$ is given by:
\begin{align}
    \theta=\arccos (\mathbf{x}_{1} \cdot \mathbf{p}/|\mathbf{x}_{1}||\mathbf{p}|).
\end{align}
Then we can calculate the pseudo-rapidity
\begin{align}
    \eta=-\ln (\tan (\theta/2)).
\end{align}
The $\Delta\phi_{ij}$ between two jets with momenta $\mathbf{p}_{i}$ and $\mathbf{p}_{j}$ is:
\begin{align}
\Delta \phi_{ij}=\arccos \left(\mathbf{p}_{i\perp} \cdot \mathbf{p}_{j\perp} /\left|\mathbf{p}_{i\perp}\right|\left|\mathbf{p}_{j\perp}\right|\right)
\end{align}
where $\mathbf{p}_{\perp}$ is the momentum perpendicular to the axis $\mathbf{x}_{1}$.

When employing a small jet resolution parameter, one encounters contributions from relative soft jets.
To extract the properties of hard interactions, we introduce the energy-weighted Ridge correlation~\cite{Basham:1978bw,Basham:1978zq},
\begin{align}
    R_{EE}(\Delta \phi, \Delta \eta)=\frac{1}{\sigma}\sum_{i<j}\int d\Delta\phi_{ij}d\Delta\eta_{ij}\frac{4E_i E_j}{s}(\frac{d^2\sigma}{d\Delta \phi_{ij} d\Delta \eta_{ij}})\delta(\Delta\phi_{ij}-\Delta\phi)\delta(\Delta\eta_{ij}-\Delta\eta),
\end{align}
where $s$ is the squared centre-of-mass energy; $E_i$ and $E_j$ denote the energies of each of two jets. We note that $R_{EE}(\Delta \phi, \Delta \eta)$ is also infrared safe at parton level.

\subsection{Setup of the simulations and validation}
\label{sec:setup}
We choose the centre-of-mass energy at the Z-pole with $\sqrt{s}=m_Z=91.19\ \mathrm{GeV}$.
The dominant contribution to jets production is $e^{+}e^{-}\rightarrow
\gamma^{\ast} / Z\rightarrow$ jets.
And we set the strong coupling constant at $\alpha_s(m_{Z})=0.118$.
For the electroweak gauge coupling constant, we use the $G_{\mu}$-scheme \cite{Buccioni:2019sur} with $G_{\mu}=1.1663787\cdot10^{-5}\thinspace\mathrm{GeV}^{-2}$ and $m_{W}=80.379 \thinspace \mathrm{GeV}$.
The central values of the renormalization and factorization scale are set to the centre-of-mass energy,
$\mu_{R}=\mu_{F}=m_{Z}$.
To estimate the systematic uncertainty in our calculations, we vary $\mu_{R}$ and $\mu_{F}$ following (the 7-point scale variations):
\begin{align}
\label{scale}
   \begin{aligned}
& \mu_{R}\to k_{\mu_{R}}\mu_{R},\\
& \mu_{F}\to k_{\mu_{F}}\mu_{F},
    \end{aligned}
\end{align}
with
$
    (k_{\mu_{R}},k_{\mu_{F}})\in\{(1,1),(1,\frac{1}{2}),(1,2),(\frac{1}{2},1),(2,1),(\frac{1}{2},\frac{1}{2}),(2,2)\}.
$

Throughout the computation, we employ the exclusive Durham jet algorithm~\cite{WJStirling_1991} with the E-scheme to cluster partons into jets with the resolution parameter $y_{\text{cut}} \in [10^{-4},10^{-2}]$~\cite{ALEPH:2003obs}. As $y_{\text{cut}}$ approaches very small values (i.e., $y_{\text{cut}} \sim 10^{-4}$), events approaching the back-to-back limit are allowed, revealing the large logarithmic corrections of pQCD.

To validate the NNLO calculation of $e^+e^-\rightarrow 3$ jets implemented in NNLOJET, we compare the perturbative coefficients \cite{Weinzierl:2010cw} $C_3$, $C_4$ of jet rates between NNLOJET and CoLoRFulNNLO~\cite{Verbytskyi:2019zhh}, and $C_{5}$ between NNLOJET, CoLoRFulNNLO and Sherpa~\cite{Sherpa:2019gpd} with different $y_\text{{cut}}$ values. The results of $C_3$, $C_4$, and $C_{5}$  between NNLOJET and CoLoRFulNNLO are consistent within the margin of error, except for $y_{\text{cut}}=2.49\cdot 10^{-4}$. The results of $C_{5}$ between NNLOJET and Sherpa are consistent within the margin of error. Further detailed analysis can be found in Appendix~\ref{app:C3,C4,C5}.

\section{Numerical results}
\label{sec:numerics}
In this section, we present the numerical results of the event shape observables relevant to planarity as discussed in Section~\ref{sec:event shape}. In Section~\ref{sec:cross section}, we compute the cross sections for the production of the inclusive n-jet events (for n=3, 4, 5) at fixed order, with various values of $y_{\text{cut}}$. In Section~\ref{sec:eaxis}, we examine the differential cross sections of $\lambda_1$, $\lambda_2$, $\lambda_3$ and $\tilde{P}$. Finally, the joint distribution observables $R(\Delta \phi, \Delta \eta)$ and $R_{EE}(\Delta \phi, \Delta \eta)$ are explored in Section~\ref{sec:ridge}.
\subsection{Fiducial cross sections}
\label{sec:cross section}

The cross sections for the production of the inclusive n-jet (where n=3, 4, 5) at fixed order for various values of $y_{\text{cut}}$ demonstrate how the fraction of the exclusive n-jet events shifts with changes in $y_{\text{cut}}$. Table~\ref{table:inclusive-cross-section} presents the cross sections (in fb) for the production of the inclusive 3-jet at NNLO, the inclusive 4-jet at NLO, and the inclusive 5-jet at LO for $y_{\text{cut}} = 10^{-2},10^{-2.5},10^{-3},10^{-3.5},10^{-4}$. Each row corresponds to a distinct $y_{\text{cut}}$ value, while each column corresponds to a different jet multiplicity. The numbers as superscripts and subscripts indicate the systematic errors arising from the scale variations according to the equation \eqref{scale}, and those within parentheses denote the statistical errors.

By examining the difference between the cross sections of the inclusive 3-jet at NNLO and the inclusive 4-jet at NLO, we observe that the majority of events are the exclusive 3-jet events (87\%) at $y_{\text{cut}}=10^{-2}$. As $y_{\text{cut}}$ decreases, the cross section of the inclusive 3-jet at NNLO increases. However, there is a notable decrease in the proportion of the exclusive 3-jet events, accompanied by a corresponding increase in the proportion of the inclusive 4-jet events. 
Notably, the exclusive 3-jet cross section at NNLO turns negative at $y_{\text{cut}}=10^{-4}$ due to the large logarithmic corrections associated with the back-to-back events. The scale uncertainty for the inclusive 3-jet cross section at NNLO is: $_{-1.9\%}^{+1.1\%}$ at $y_{\text{cut}}=10^{-2}$, $_{-1.3\%}^{+0.3\%}$ at $y_{\text{cut}}=10^{-2.5}$ and $_{-1.5\%}^{+0.2\%}$ at $y_{\text{cut}}=10^{-3}$. 
For $y_{\text{cut}}$ $\ge 10^{-2.5}$, the large exclusive 3-jet rates indicate that
the majority of the events are planar.
On the other hand, the inclusive 4-jet events lead to deviations of event shapes from a plane.
The details are illustrated through the distributions of the eigenvalues from $S^{jk}$ in the next subsection.

\begin{table}[hbt!]
	\centering
	\begin{threeparttable}
        \renewcommand{\arraystretch}{1.4}
		\begin{tabular}{|c|c|c|c|}\hline
			\makebox[2.5cm][c]{$y_{\text{cut}}$} & \makebox[4cm][c]{3-jet (NNLO)} & \makebox[4cm][c]{4-jet (NLO)} & \makebox[4cm][c]{5-jet (LO)}\\ \hline
			$1\cdot10^{-2.0}$& \makebox[4cm][c]{$1.594(3)^{+0.018}_{-0.030}\cdot 10^{7}$}& \makebox[4cm][c]{$2.0016(7)_{-0.245}^{+0.280}\cdot 10^{6}$} &\makebox[4cm][c]{$4.775(4)_{-1.129}^{+1.833}\cdot 10^{4}$} \\ \hline
			$1\cdot10^{-2.5}$& \makebox[4cm][c]{$2.59(1)^{+0.007}_{-0.033}\cdot10^{7}$}& \makebox[4cm][c]{$7.343(4)^{+0.959}_{-0.862}\cdot10^{6}$} &\makebox[4cm][c]{$5.523(4)_{-1.129}^{+2.476}\cdot 10^{5}$}\\ \hline
			$1\cdot10^{-3.0}$& \makebox[4cm][c]{$3.55(1)^{+0.006}_{-0.053}\cdot 10^{7}$}& \makebox[4cm][c]{$1.8516(7)_{-0.198}^{+0.205}\cdot10^{7}$} &\makebox[4cm][c]{$3.017(2)_{-0.777}^{+1.192}\cdot 10^{6}$}\\ \hline
			$1\cdot10^{-3.5}$& \makebox[4cm][c]{$4.27(2)^{+0.095}_{-0.163}\cdot10^{7}$}& \makebox[4cm][c]{$3.656(2)^{+0.276}_{-0.322}\cdot10^{7}$} &\makebox[4cm][c]{$1.1181(5)_{-0.288}^{+0.442}\cdot 10^{7}$}\\ \hline
			$1\cdot10^{-4.0}$& \makebox[4cm][c]{$4.47(3)^{+0.257}_{-0.343}\cdot10^{7}$}& \makebox[4cm][c]{$5.954(4)_{-0.329}^{+0.080}\cdot10^{7}$} &\makebox[4cm][c]{$3.261(5)_{-0.841}^{+1.288}\cdot 10^{7}$}\\ \hline
		\end{tabular}
            \caption{ \label{table:inclusive-cross-section}The inclusive n-jet cross sections (in fb) at fixed order with different $y_{\text{cut}}$ values. Each row corresponds to a distinct $y_{\text{cut}}$ value, while each column corresponds to a different jet multiplicity. The numbers as superscripts and subscripts indicate the systematic errors arising from the scale variations according to the equation \eqref{scale}, and those within parentheses denote the statistical errors.}
	\end{threeparttable}
\end{table}

Table~\ref{table:exclusive-3-jet-cross-section} presents the cross sections for the exclusive 3-jet production at LO, NLO and NNLO with $y_{\text{cut}}=10^{-2}$, $10^{-3}$ and $10^{-4}$.
Each row corresponds to a distinct $y_{\text{cut}}$ value, while each column corresponds to a different fixed order.
For $y_{\text{cut}}=10^{-2}$, the reduction of the relative scale uncertainty is $67\%$ between LO and NLO and $89\%$ between NLO and NNLO. The relative corrections are $18\%$ between LO and NLO and $-0.9\%$ between NLO and NNLO. 
We observe that the effects of large-logarithm corrections are remarkably prominent around \(y_{\text{cut}} = 10^{-4}\). Consequently, in the subsequent analysis, we restrict our attention to the results corresponding to \(y_{\text{cut}}\) in the range \(10^{-2}\) to \(10^{-3}\).

\begin{table}[hbt!]
	\centering
	\begin{threeparttable}
       \renewcommand{\arraystretch}{1.4}
		\begin{tabular}{|c|c|c|c|}\hline
			\makebox[2.5cm][c]{$y_{\text{cut}}$} & \makebox[4cm][c]{$\sigma_{LO}$} & \makebox[4cm][c]{$\sigma_{NLO}$} & \makebox[4cm][c]{$\sigma_{NNLO}$}\\ \hline
			$1\cdot10^{-2.0}$& \makebox[4cm][c]{$1.1878(3)^{+0.140}_{-0.112}\cdot 10^{7}$}& \makebox[4cm][c]{$1.4054(4)_{-0.054}^{+0.046}\cdot 10^{7}$} &\makebox[4cm][c]{$1.393(2)_{-0.011}^{+0.000}\cdot 10^{7}$} \\ \hline
			$1\cdot10^{-3.0}$& \makebox[4cm][c]{$3.2104(8)^{+0.377}_{-0.304}\cdot 10^{7}$}& \makebox[4cm][c]{$2.409(3)_{-0.222}^{+0.104}\cdot10^{7}$} &\makebox[4cm][c]{$1.70(1)_{-0.257}^{+0.204}\cdot 10^{7}$}\\ \hline
			$1\cdot10^{-4.0}$& \makebox[4cm][c]{$6.275(2)^{+0.736}_{-0.594}\cdot10^{7}$}& \makebox[4cm][c]{$5.3(6)_{-159.2}^{+104.1}\cdot10^{5}$} &\makebox[4cm][c]{$-1.48(2)_{-0.583}^{+0.419}\cdot 10^{7}$}\\ \hline
		\end{tabular}
  	\caption{ \label{table:exclusive-3-jet-cross-section}The exclusive 3-jet cross sections (in fb) at fixed order with different  $y_{\text{cut}}$ values. Each row corresponds to a distinct $y_{\text{cut}}$ value, while each column corresponds to a different fixed order. The numbers as superscripts and subscripts indicate the systematic errors, which are obtained from the equation \eqref{scale}, and those within parentheses denote the statistical errors.}
	\end{threeparttable}
\end{table}

\FloatBarrier

\subsection{Differential cross sections}
\label{sec:eaxis}

\begin{figure}[hbt!]
\includegraphics[width=0.49\textwidth]{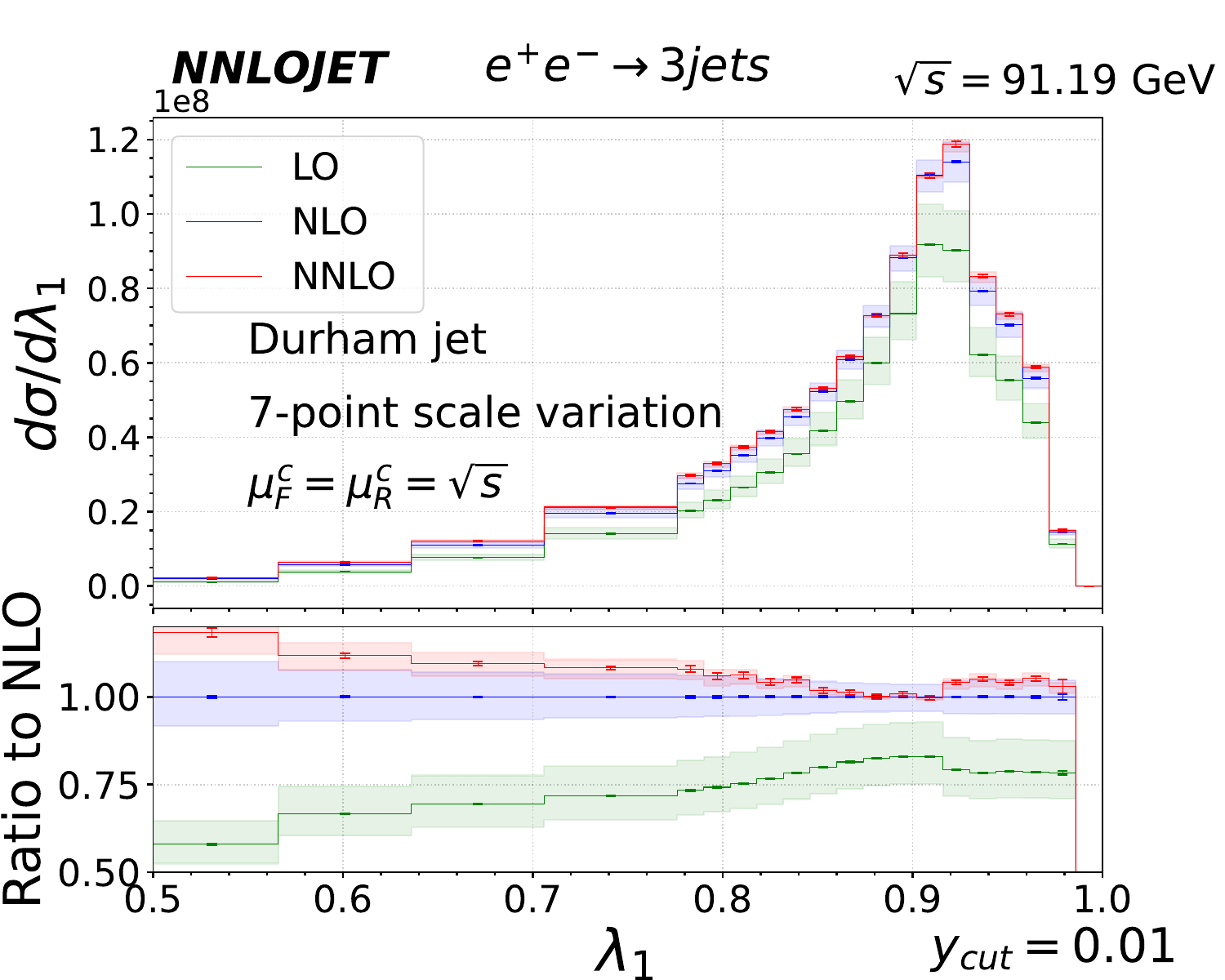}
\includegraphics[width=0.49\textwidth]{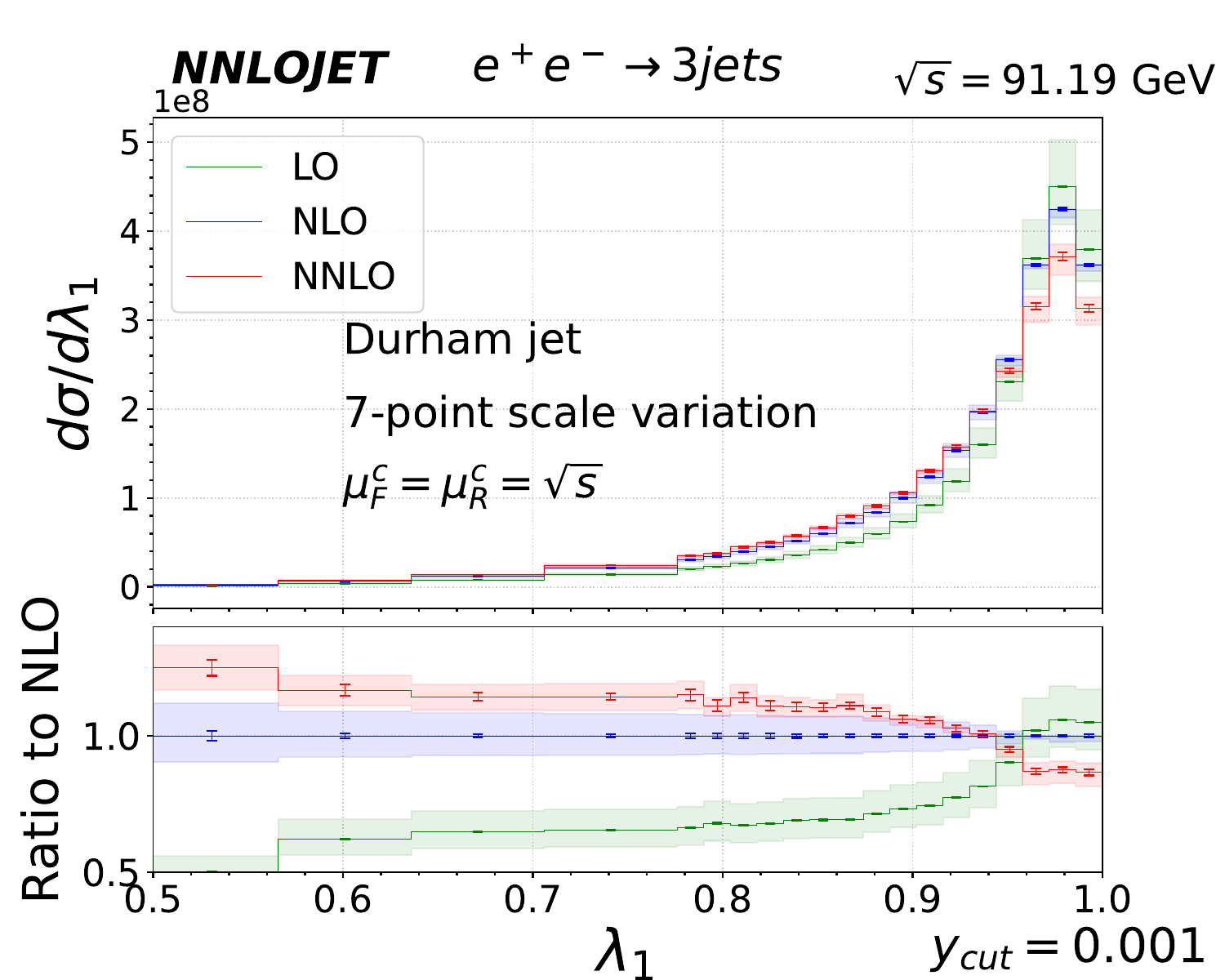}
\caption{The distribution of the event shape observable \(\lambda_1\) in the inclusive 3-jet events at the center-of-mass energy $\sqrt{s}=m_Z$ for \(y_{\text{cut}} = 10^{-2}\) (left panel) and \(10^{-3}\) (right panel). The upper panels present the predictions at LO (in green), NLO (in blue), and NNLO (in red). The lower panels display the ratios of the NLO predictions and the LO predictions over the NLO predictions. The statistical errors are represented by the error bars, while the light-shaded bands indicate the systematic uncertainties arising from the scale variations according to the equation \eqref{scale}.}
\label{fig:lambda1-3jets-lin}
\end{figure}

\begin{figure}[hbt!]
\includegraphics[width=0.49\textwidth]{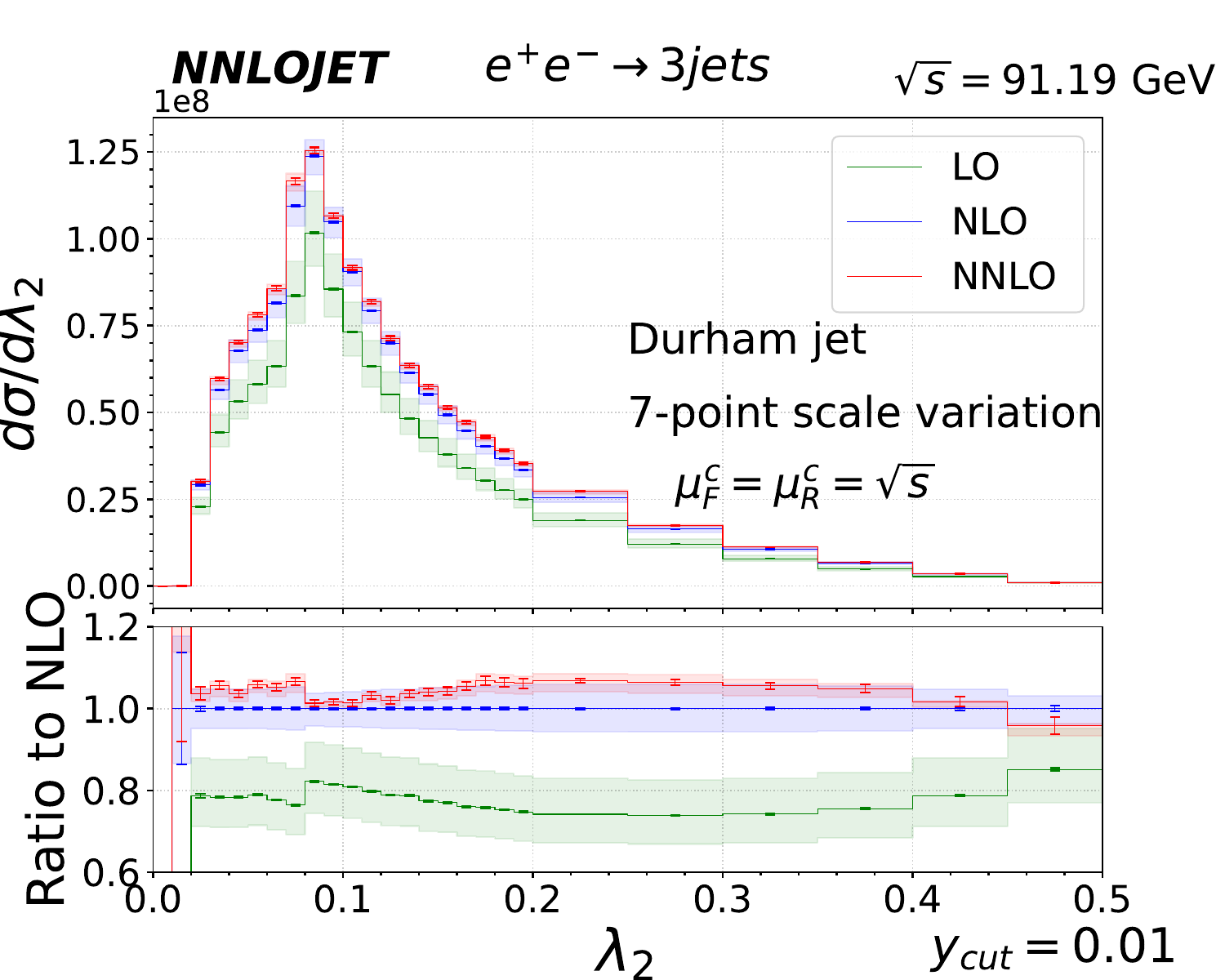}
\includegraphics[width=0.49\textwidth]{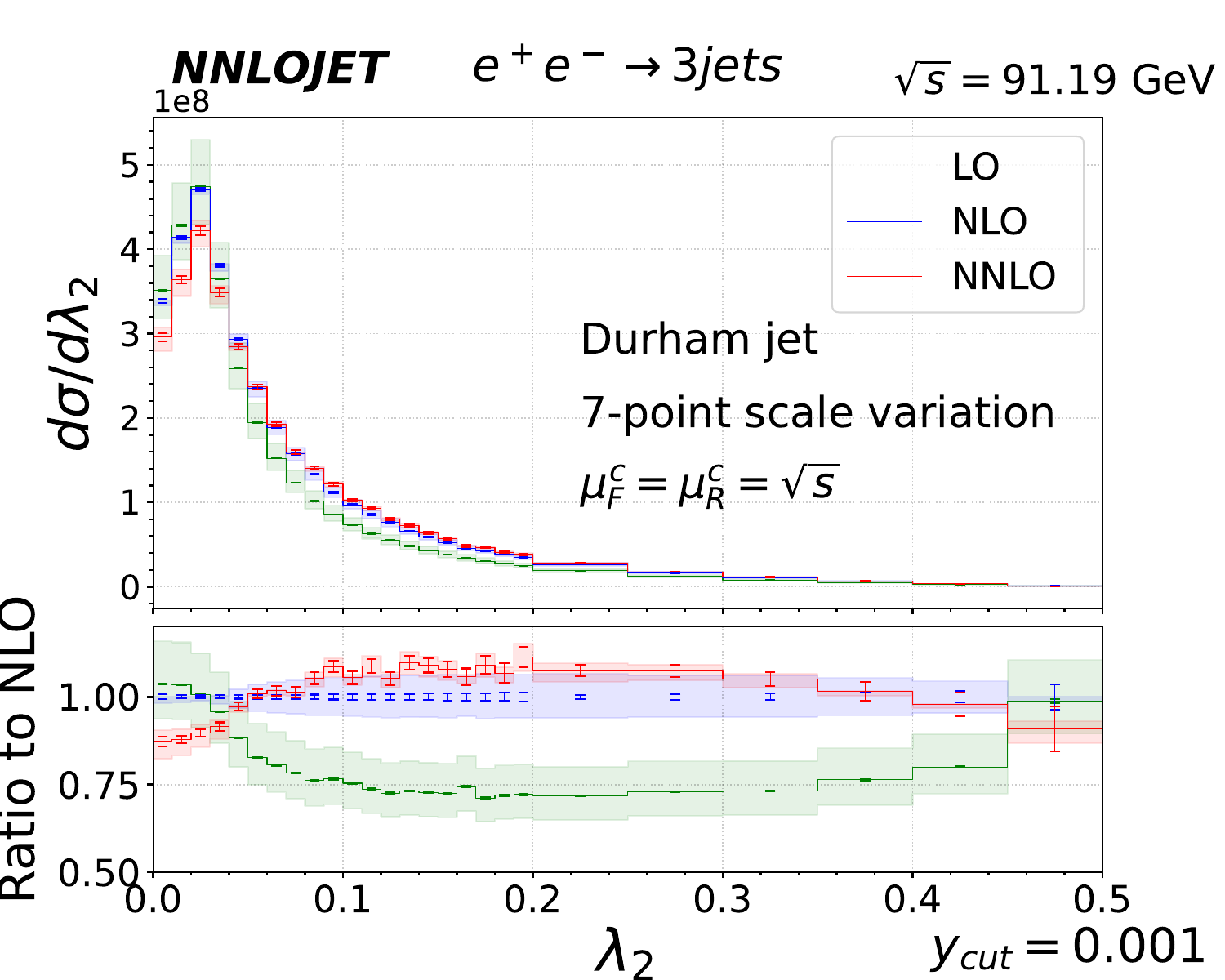}
\caption{The distribution of the event shape observable \(\lambda_2\) in the inclusive 3-jet events at the center-of-mass energy $\sqrt{s}=m_Z$ for \(y_{\text{cut}} = 10^{-2}\) (left panel) and \(10^{-3}\) (right panel). The upper panels present the predictions at LO (in green), NLO (in blue), and NNLO (in red). The lower panels display the ratios of the NLO predictions and the LO predictions over the NLO predictions. The statistical errors are represented by the error bars, while the light-shaded bands indicate the systematic uncertainties arising from the scale variations according to the equation \eqref{scale}.}
\label{fig:lambda2-3jets-lin}
\end{figure}

We study the differential cross sections for $\lambda_1$, $\lambda_2$, $\lambda_3$ and $\tilde{P}$ with different values of $y_{\text{cut}}$ to analyze the planarity of multi-jet production in the inclusive 3-jet events up to NNLO and in the inclusive 4-jet events up to NLO.
Large logarithmic divergences in the back-to-back fiducial regions limit the prediction power of fixed order QCD calculations. We use PYTHIA 8.3~\cite{Bierlich:2022pfr} to resum the leading log divergences for inclusive 3, 4-jet events to extend and exam our study.
Figure~\ref{fig:lambda1-3jets-lin} and Figure~\ref{fig:lambda2-3jets-lin} show the differential distribution of $\lambda_1$ and $\lambda_2$ in the inclusive 3-jet events for \(y_{\text{cut}} = 10^{-2}\) (left panel) and \(10^{-3}\) (right panel). The upper panels present the predictions at LO (in green), NLO (in blue), and NNLO (in red). The lower panels display the relative ratios over the NLO predictions. The statistical errors are represented by the error bars, while the light-shaded bands indicate the systematic uncertainties arising from the 7-point scale variations. 

Both figures show no shift of the peak position with higher order pQCD corrections, which however change the shape and theory uncertainty of the distributions.
The peak of the $\lambda_1$ ($\lambda_2$) distribution is near 0.92 (0.08) at $y_{\text{cut}} = 10^{-2}$.
In the peak region, the relative scale uncertainties are reduced by approximately $63\%$ ($62\%$) between LO and NLO, and by $100\%$ ($90\%$) between NLO and NNLO. The relative corrections are about $21\%$ ($22\%$) between LO and NLO and $0\%$ ($0.8\%$) between NLO and NNLO.
In the tail region, the relative scale uncertainties are reduced by about $44\%\sim65\%$ ($51\%\sim60\%$) between LO and NLO, and by $58\%\sim100\%$ ($71\%\sim88\%$) between NLO and NNLO. The relative corrections are approximately $23\%\sim41\%$ ($27\%\sim33\%$) between LO and NLO and $8.1\%\sim13\%$ ($2.1\%\sim6.3\%$) between NLO and NNLO. These results suggest that the pQCD calculations converge for the plotted regions at $y_{\text{cut}}=10^{-2}$.

As $y_{\text{cut}}$ decreases, the peak of $\lambda_{1}$ ($\lambda_{2}$) shifts towards 1 (0) indicating that more events approaching the back-to-back limit ($\lambda_{1}=1$ and $\lambda_{2}=0$) are adopted. For $y_{\text{cut}}=10^{-3}$, the relative scale uncertainty in the peak region increases about $301\%$ ($479\%$) between NLO and NNLO. And the relative corrections come to about $-5.5\%$ ($-0.4\%$) between LO and NLO and $-12.7\%$ ($-10.4\%$) between NLO and NNLO. These facts indicate effects of the large logarithmic corrections in the peak region at $y_{\text{cut}}=10^{-3}$.
In contrast, the fixed order results converge well in the tail region. The reductions of the relative scale uncertainty are about $24\%$ ($41\%$) between LO and NLO and $41\%$ ($61\%$) between NLO and NNLO. The relative corrections are about $51\%$ ($40\%$) between LO and NLO and $15\%$ ($7.6\%$) between NLO and NNLO.

\begin{figure}[hbt!]
\includegraphics[width=0.49\textwidth]{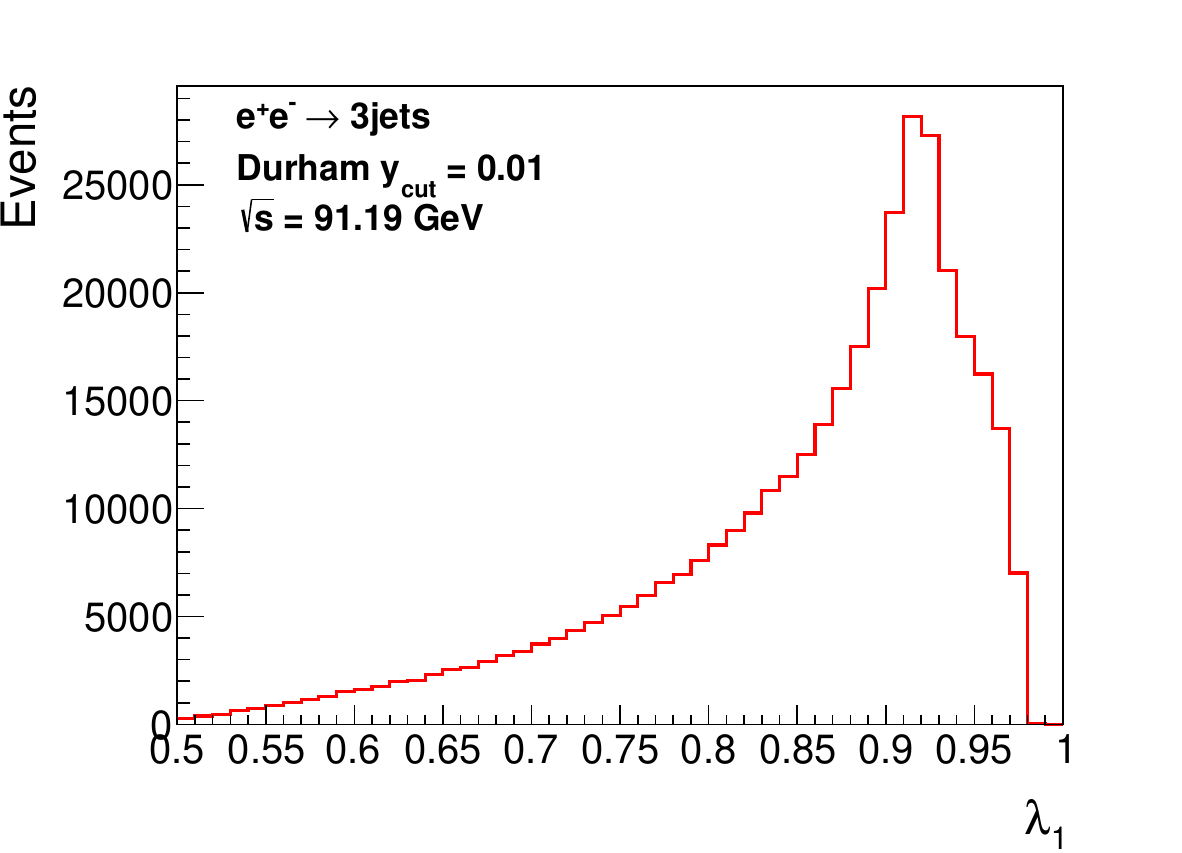}
\includegraphics[width=0.49\textwidth]{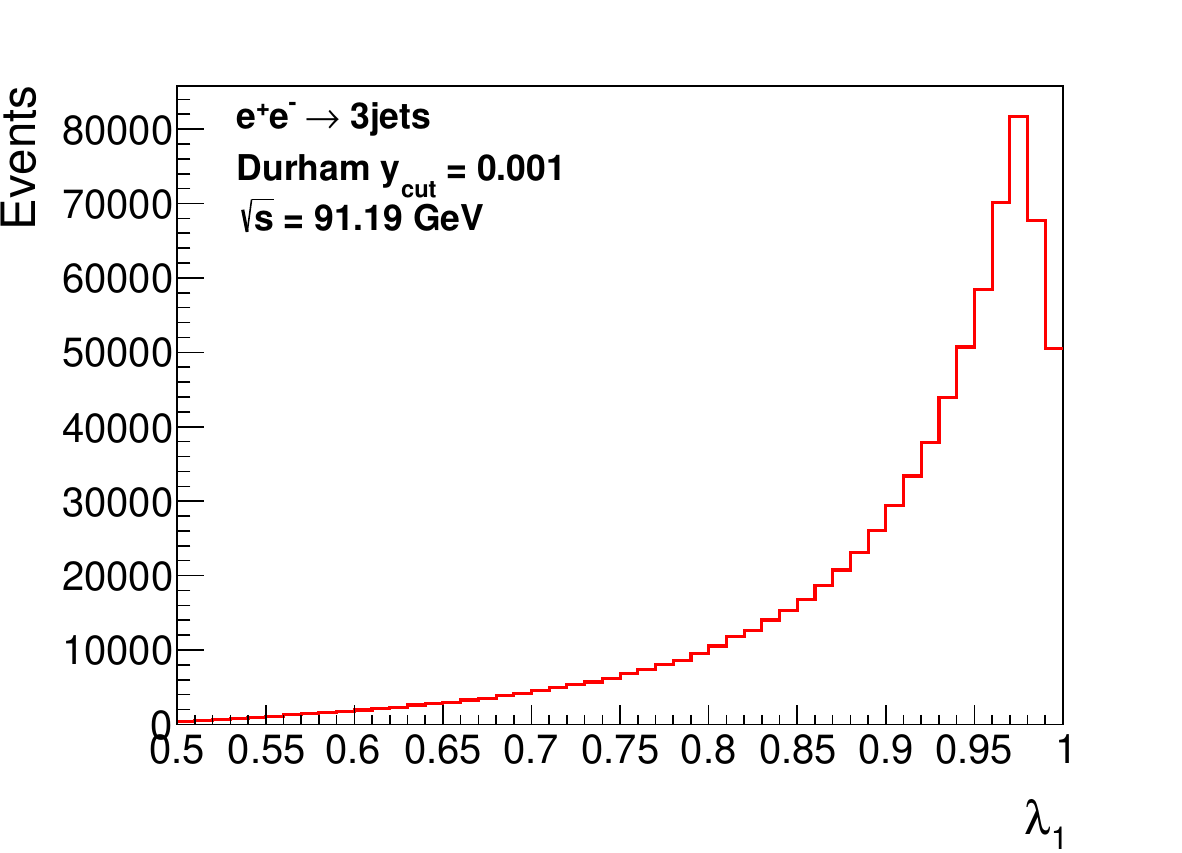}
\caption{The distribution of the event shape observable \(\lambda_1\) in the inclusive 3-jet events at the center-of-mass energy $\sqrt{s}=m_Z$ for \(y_{\text{cut}} = 10^{-2}\) (left panel) and \(10^{-3}\) (right panel), based on PYTHIA 8.3~\cite{Bierlich:2022pfr}.}
\label{fig:lambda1-3jets-lin-pythia}
\end{figure}

\begin{figure}[hbt!]
\includegraphics[width=0.49\textwidth]{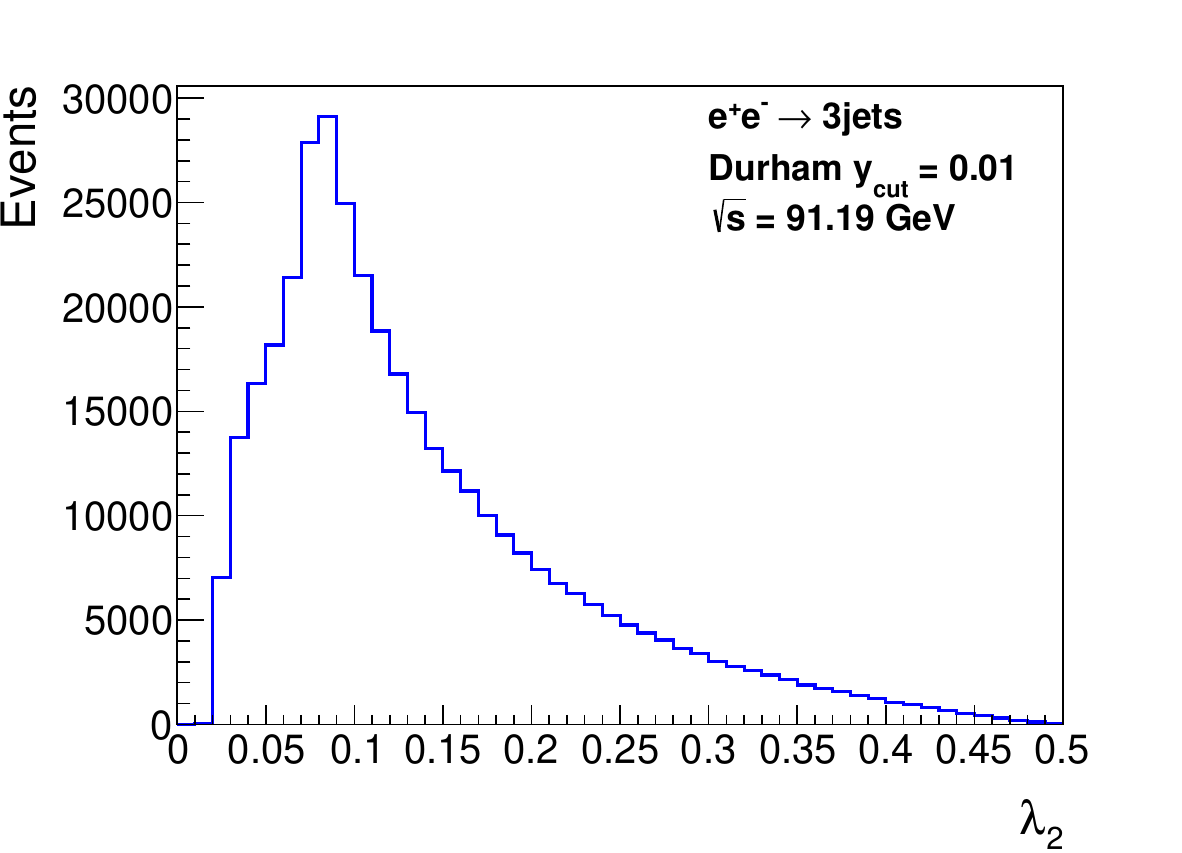}
\includegraphics[width=0.49\textwidth]{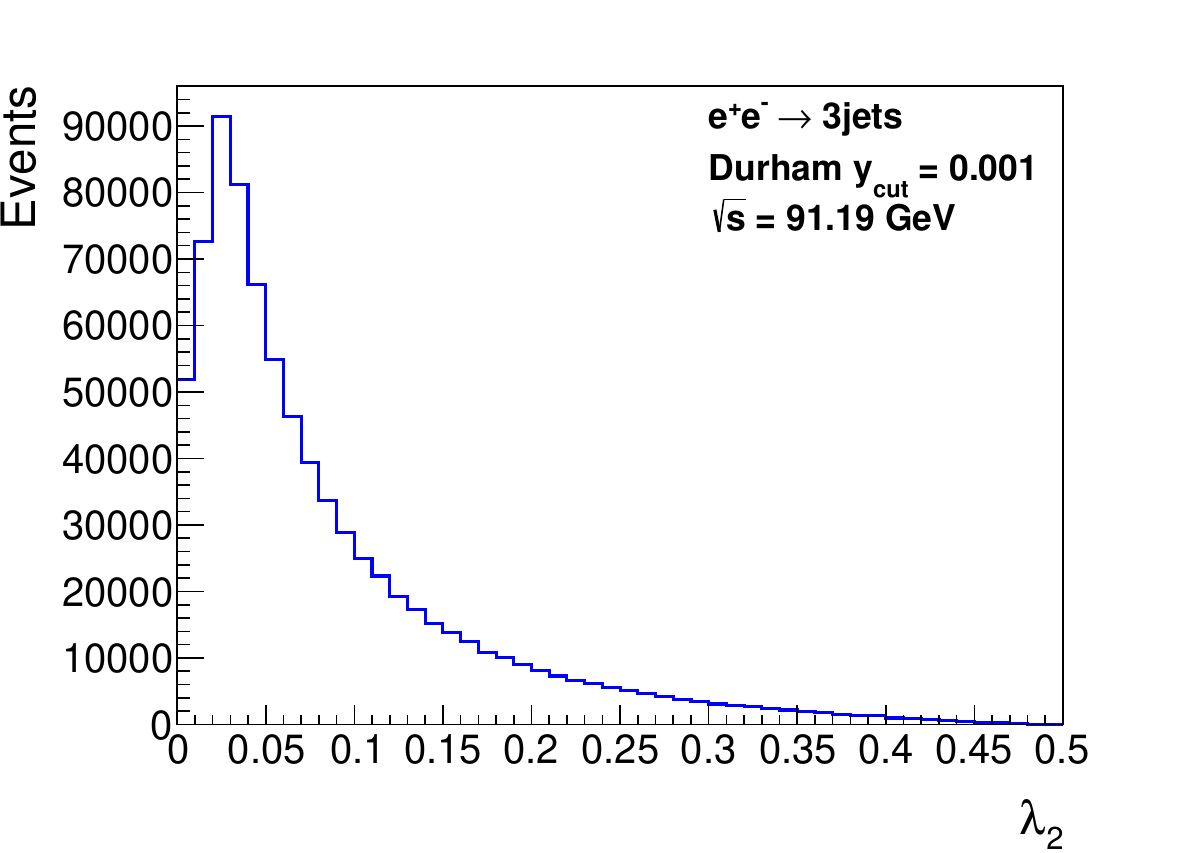}
\caption{The distribution of the event shape observable \(\lambda_2\) in the inclusive 3-jet events at the center-of-mass energy $\sqrt{s}=m_Z$ for \(y_{\text{cut}} = 10^{-2}\) (left panel) and \(10^{-3}\) (right panel), based on PYTHIA 8.3~\cite{Bierlich:2022pfr}.}
\label{fig:lambda2-3jets-lin-pythia}
\end{figure}

Large logarithmic divergences in the back-to-back fiducial regions limit the prediction power of fixed order QCD calculations. We use PYTHIA 8.3~\cite{Bierlich:2022pfr} to resum the leading log divergences for inclusive 3, 4-jet events to extend and exam our study. Figure~\ref{fig:lambda1-3jets-lin-pythia} and Figure~\ref{fig:lambda2-3jets-lin-pythia} show the distributions of \(\lambda_1\) and \(\lambda_2\) in inclusive 3-jet events for \(y_{\text{cut}} = 10^{-2}\) (left panel) and \(10^{-3}\) (right panel). The peak positions and full width at half maximum remain similar to those from the fixed-order calculation in Figure~\ref{fig:lambda1-3jets-lin} and Figure~\ref{fig:lambda2-3jets-lin}.
For $y_{\text{cut}}=10^{-2}$, the peak of the $\lambda_1$ ($\lambda_2$) distribution is near 0.92 (0.08).
As $y_{\text{cut}}$ decreases, the peak of $\lambda_{1}$ ($\lambda_{2}$) shifts towards 1 (0), indicating that the resummation effects do not change the overall picture: more events approach the back-to-back limit ($\lambda_{1}=1$ and $\lambda_{2}=0$).

\begin{figure}[hbt!]
\includegraphics[width=0.49\textwidth]{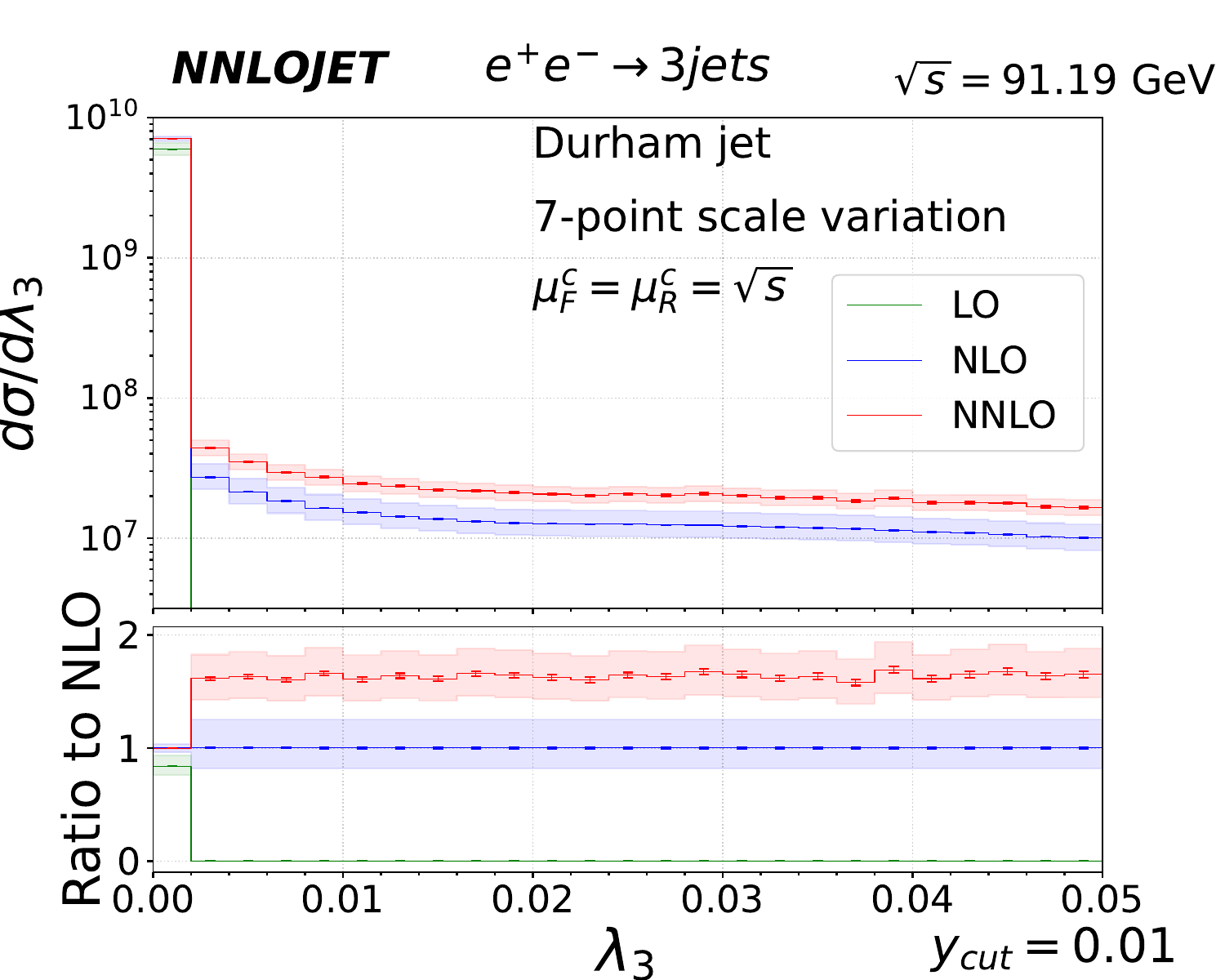}
\includegraphics[width=0.49\textwidth]{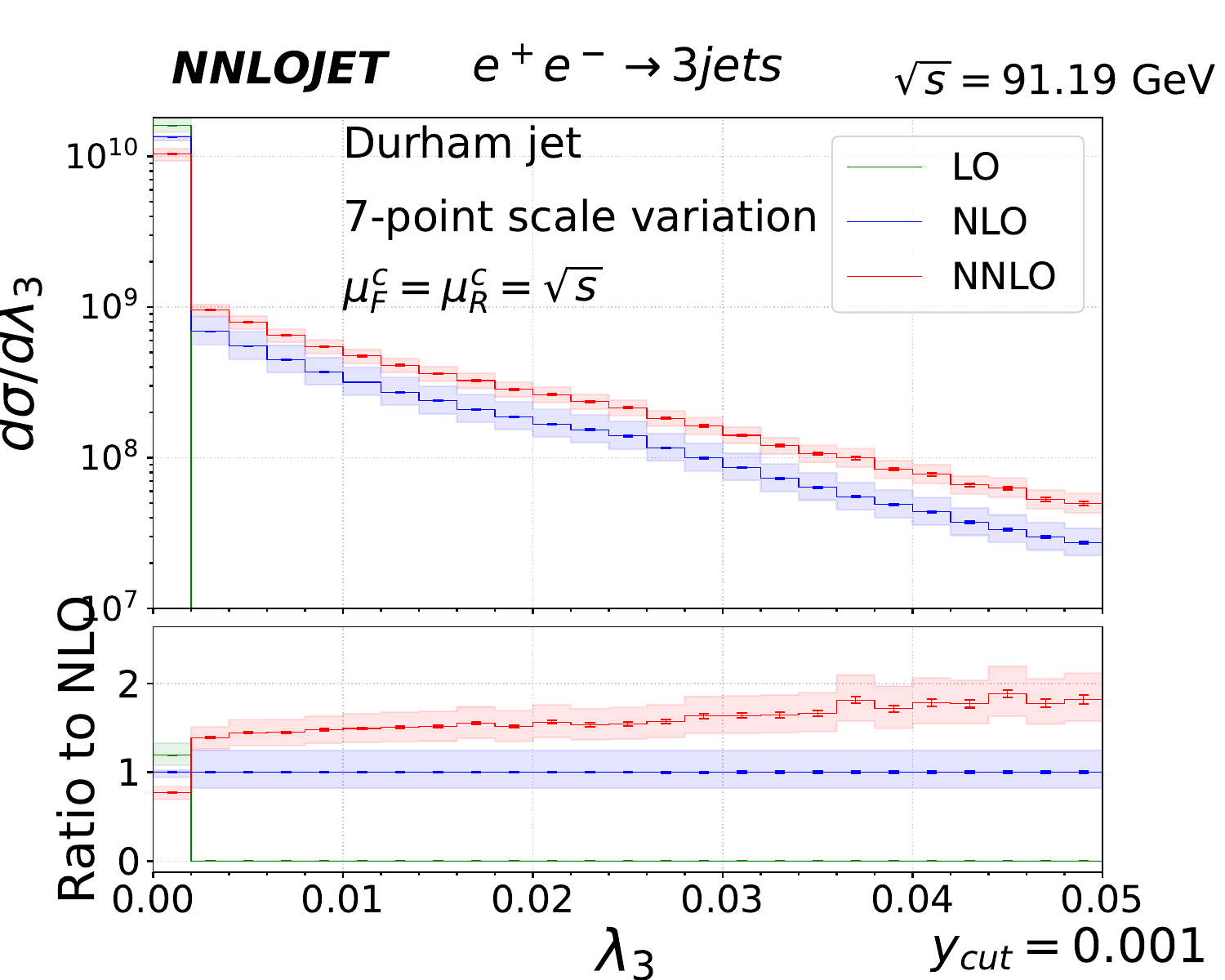}
\caption{The distribution of the event shape observable \(\lambda_3\) in inclusive 3-jet events at center-of-mass energy $\sqrt{s}=m_Z$ for \(y_{\text{cut}} = 10^{-2}\) (left panel) and \(10^{-3}\) (right panel). The upper panels present predictions at LO (in green), NLO (in blue), and NNLO (in red). The lower panels display the ratios of NNLO, NLO and LO over NLO predictions. Statistical errors are represented by error bars, while light-shaded bands indicate systematic uncertainties arising from scale variations according to equation \eqref{scale}.}
\label{fig:lambda3-3jets-lin}
\end{figure}

Figure~\ref{fig:lambda3-3jets-lin} shows the differential distribution of $\lambda_3$ in inclusive 3-jet events for \(y_{\text{cut}} = 10^{-2}\) (left panel) and \(10^{-3}\) (right panel).
We observe that the distribution of $\lambda_3$, which indicates the planarity of an event, is concentrated around 0 for these two \(y_{\text{cut}}\) values. The closer $\lambda_3$ is to 0, the higher the planarity. Consequently, the peaks in the distribution suggest that the events exhibit high planarity.
The distribution at LO is confined to the first bin as there are only exclusive 3-jet events ($\lambda_{3}=0$) at this order. And we can observe that as $y_{\text{cut}}$ decreases, the degree of event clustering at NLO and NNLO decreases.
In Table~\ref{table:inclusive-cross-section}, we observe that a decrease in \(y_{\text{cut}}\) leads to a reduction in the proportion of 3-jet events, accompanied by an increase in the proportions of 4-jet and 5-jet events. As the jet multiplicity increases, the planarity of the events decreases accordingly. This trend is reflected in the enhanced distribution within the tail region, as illustrated in Figure~\ref{fig:lambda3-3jets-lin}.

\begin{figure}[hbt!]
\includegraphics[width=0.49\textwidth]{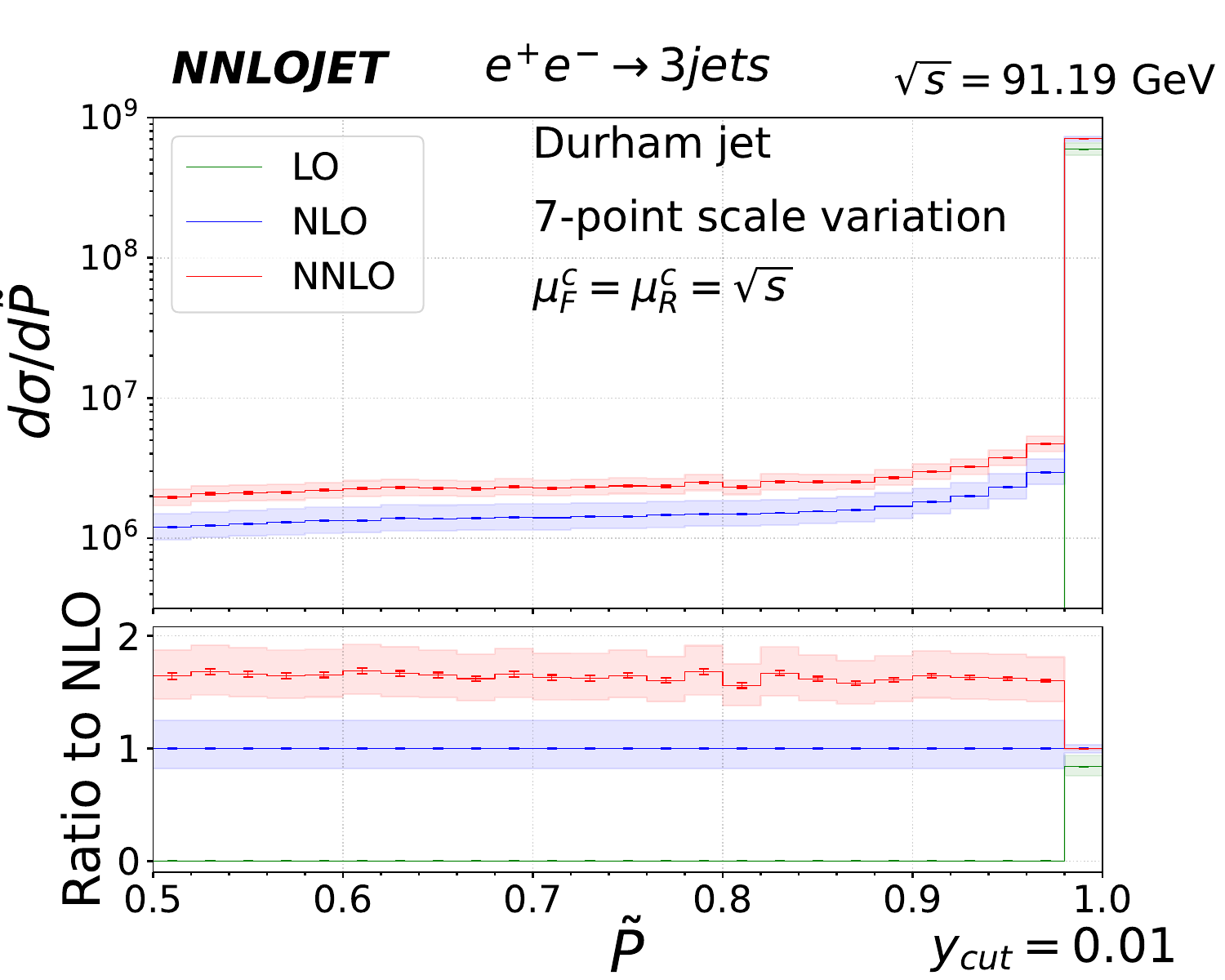}
\includegraphics[width=0.49\textwidth]{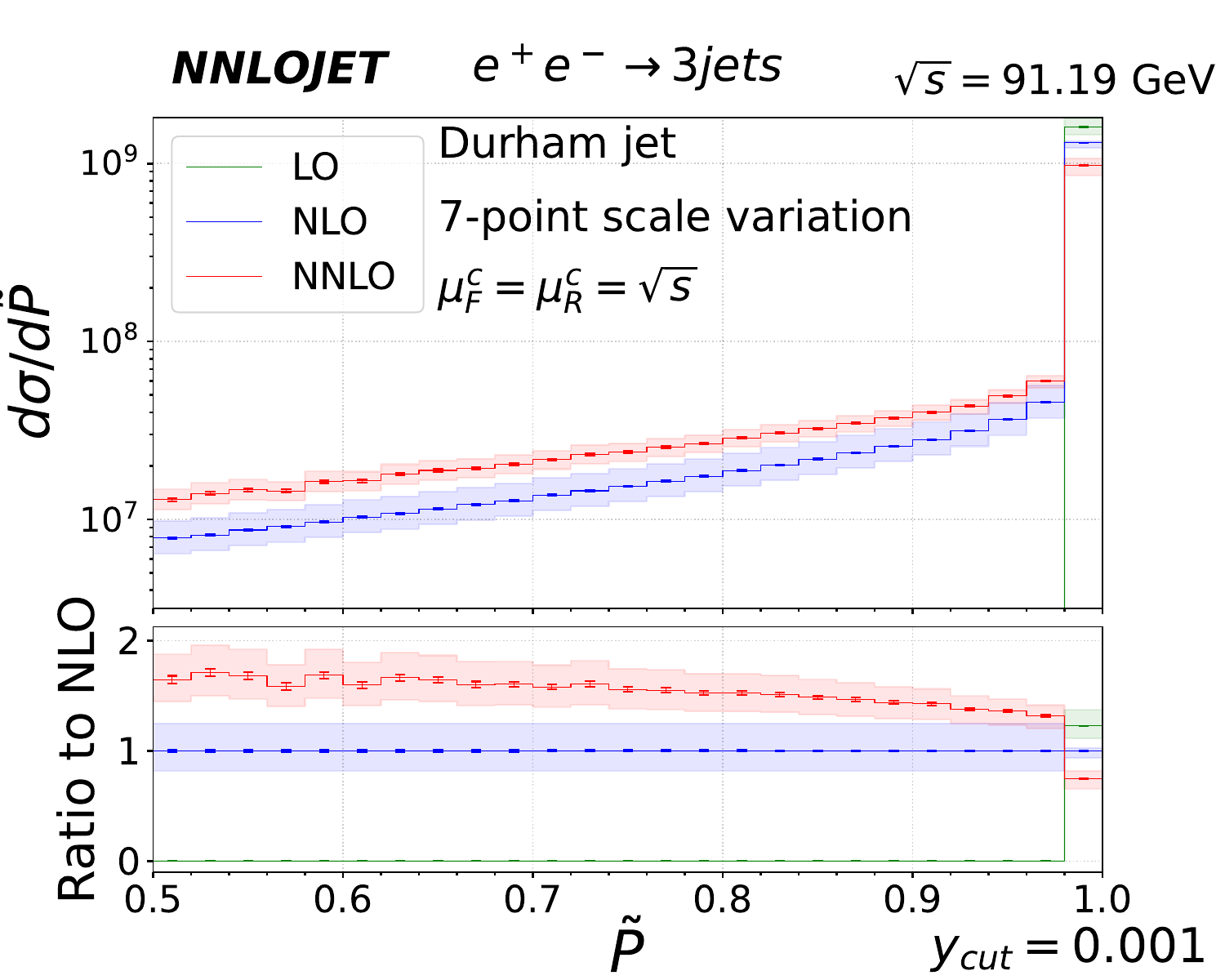}
\caption{The distribution of the event shape observable \(\tilde{P}\) in inclusive 3-jet events at center-of-mass energy $\sqrt{s}=m_Z$ for \(y_{\text{cut}} = 10^{-2}\) (left panel) and \(10^{-3}\) (right panel). The upper panels present predictions at LO (in green), NLO (in blue), and NNLO (in red). The lower panels display the ratios of NNLO, NLO and LO over NLO predictions. Statistical errors are represented by error bars, while light-shaded bands indicate systematic uncertainties arising from scale variations according to equation  \eqref{scale}.}
\label{fig:P-3jets-lin}
\end{figure}

The $\lambda_{1}$ distribution demonstrates that the majority of events exhibit a `back-to-back-like' pattern, which is a well known picture of the global event shape in high energy $e^+e^-$ annihilation processes.
The line shapes of the observables $\lambda_{2}$ and $\lambda_{3}$ are similar, i.e., both peak near the minimum values. The modified planarity $\Tilde{P}$ depends on the relative ratio of $\lambda_{2}$ and $\lambda_{3}$, which can reveal the difference of the concrete peak region of $\lambda_{2}$ and $\lambda_{3}$.
From the upper panel of Figure~\ref{fig:P-3jets-lin}, we observe that the distribution of $\tilde{P}$ is concentrated around 1 all the time at these two \(y_{\text{cut}}\) values similar to $\lambda_{3}$ around 0 in Figure~\ref{fig:lambda3-3jets-lin}.
And the relative scale uncertainty for $\tilde{P}$ between NLO and NNLO and the relative correction between NLO and NNLO around 1 are similar to these for $\lambda_{3}$ around 0. We also calculate $\lambda_{3}$ and $\tilde{P}$ using PYTHIA, and observe the same shape and peak region as in the above fixed order calculations (see Appendix~\ref{app:PYTHIAresults}). Therefore, the parton shower does not change the conclusion that most of events have high planarity.

\begin{figure}[hbt!]
\includegraphics[width=0.49\textwidth]{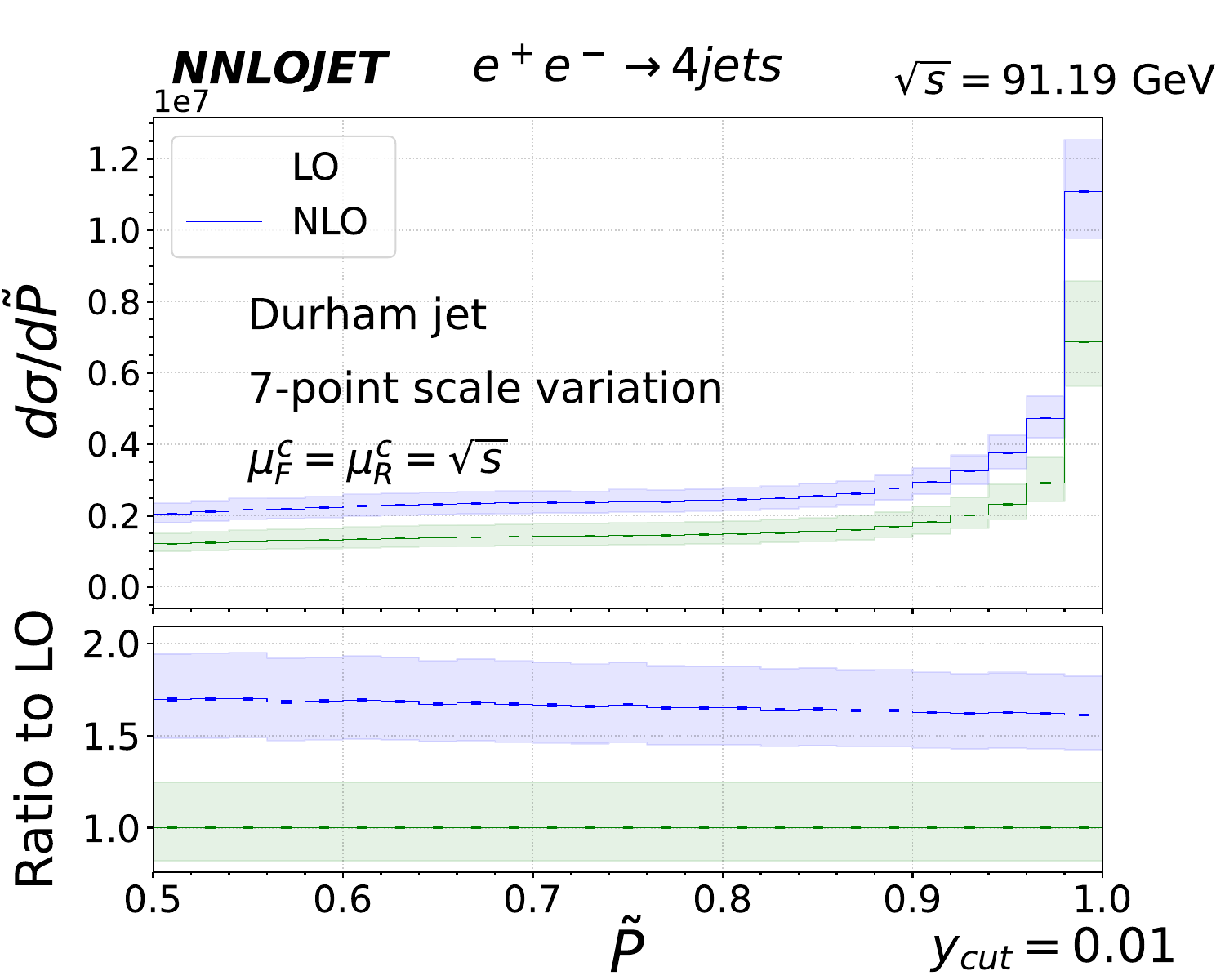}
\includegraphics[width=0.49\textwidth]{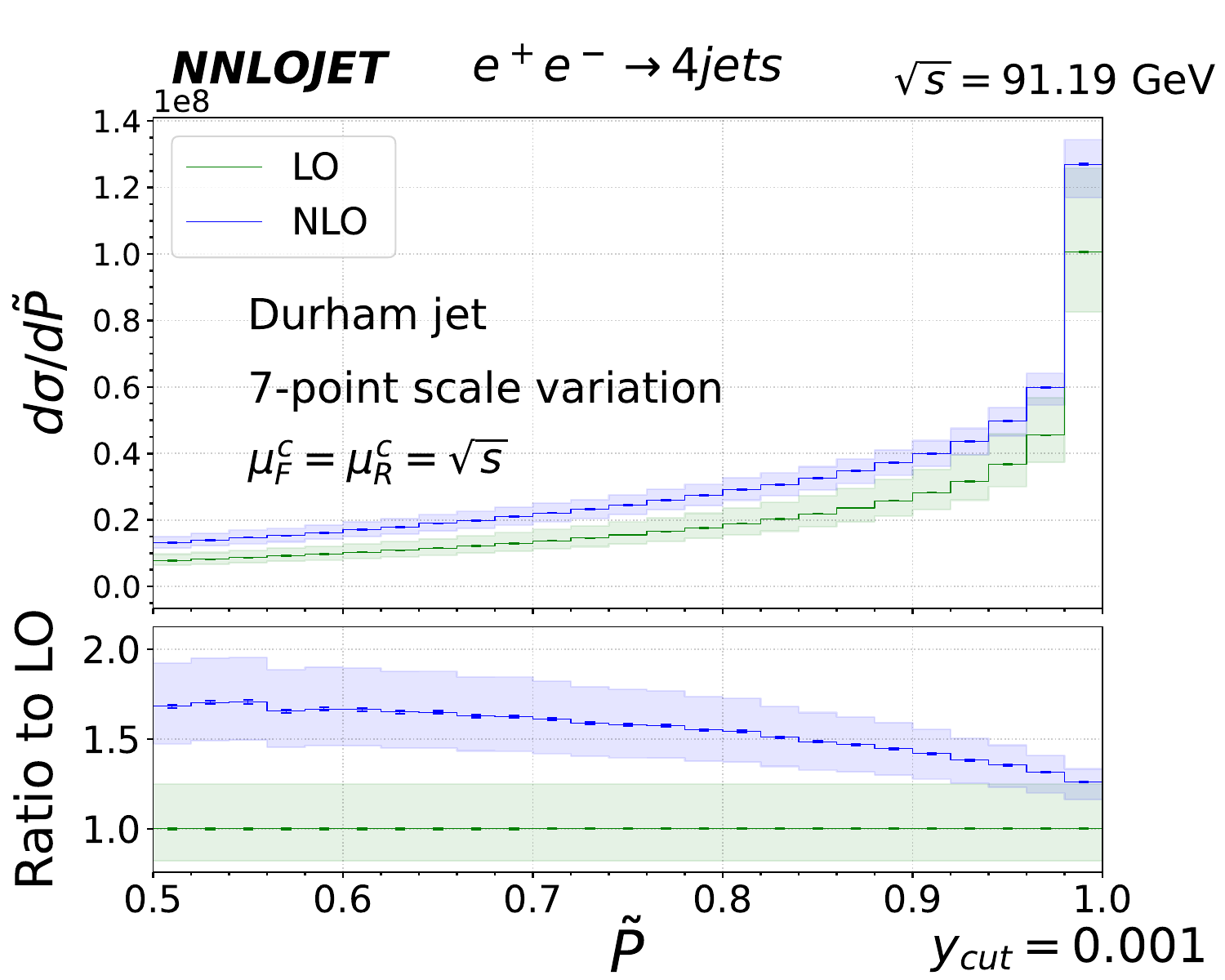}
\caption{The distribution of the event shape observable \(\tilde{P}\) in the inclusive 4-jet events at the center-of-mass energy $\sqrt{s}=m_Z$ for \(y_{\text{cut}} = 10^{-2}\) (left panel) and \(10^{-3}\) (right panel). The upper panels present the predictions at LO (in green) and NLO (in blue). The lower panels display the ratios of the NLO predictions and the LO predictions over the LO predictions. The statistical errors are represented by the error bars, while the light-shaded bands indicate the systematic uncertainties arising from the scale variations according to the equation \eqref{scale}.}
\label{fig:P-4jets-lin}
\end{figure}

In the above analysis of the inclusive 3-jet events, we noted that exclusive 3-jet events are fully planar. We therefore present the \(\tilde{P}\) distributions for the inclusive 4-jet events, as shown in Figure~\ref{fig:P-4jets-lin}, to analyze the planarity of events with higher jet multiplicities.
We observe that the distribution is concentrated around 0, which indicates that most of the inclusive 4-jet events have a planar event shape.
The reduction of the relative scale uncertainty and the relative correction between LO and NLO for $\tilde{P}$ around 1 are similar to those for $\lambda_{3}$ around 0.
The reduction of the relative scale uncertainty for $\tilde{P}$ between LO and NLO around the peak region is: $48\%$ at $y_{\text{cut}}=10^{-2}$ and $69\%$ at $y_{\text{cut}}=10^{-3}$ and the relative correction between LO and NLO is: $62\%$ at $y_{\text{cut}}=10^{-2}$ and $26\%$ at $y_{\text{cut}}=10^{-3}$.
At \(y_{\text{cut}} = 10^{-3}\), we observe reductions in the relative scale uncertainty between LO and NLO within the peak regions which differs from that observed for inclusive 3-jet events. This suggests that the back-to-back events are primarily derived from the 3-jet events.
We also calculate $\tilde{P}$ for inclusive 4-jet events using PYTHIA, and observe the same shape and peak region as in the above fixed order calculations (see Appendix~\ref{app:PYTHIAresults}).

The analysis of the above observables well demonstrates the planarity of the events driven by pQCD including those with a higher multiplicity of jets. When there are more than three jets, generally, the momenta are not restricted in a plane, but the whole phase space of an event is still significantly planar.
Considering the fact that the higher order corrections are relatively small and have less uncertainty of scale variation in the entire region for large $y_{\text{cut}}$ values (around $10^{-2}$), we conclude that hard scatterings which are dictated by pQCD lead to a high planarity.
This will be further illustrated by the study of the Ridge correlations in the following subsection.

On the other hand, for small values of $y_{\text{cut}}$ (around $10^{-3}$), the non-perturbative corrections could be large. The resummation is required in the peak region to obtain physically valid results. And we found that the parton shower does not change this picture that most of events have high planarity around this $y_{\text{cut}}$ value.

\FloatBarrier

\subsection{Ridge}
\label{sec:ridge}

\begin{figure}[hbt!]
\centering
\includegraphics[width=1.0\textwidth]{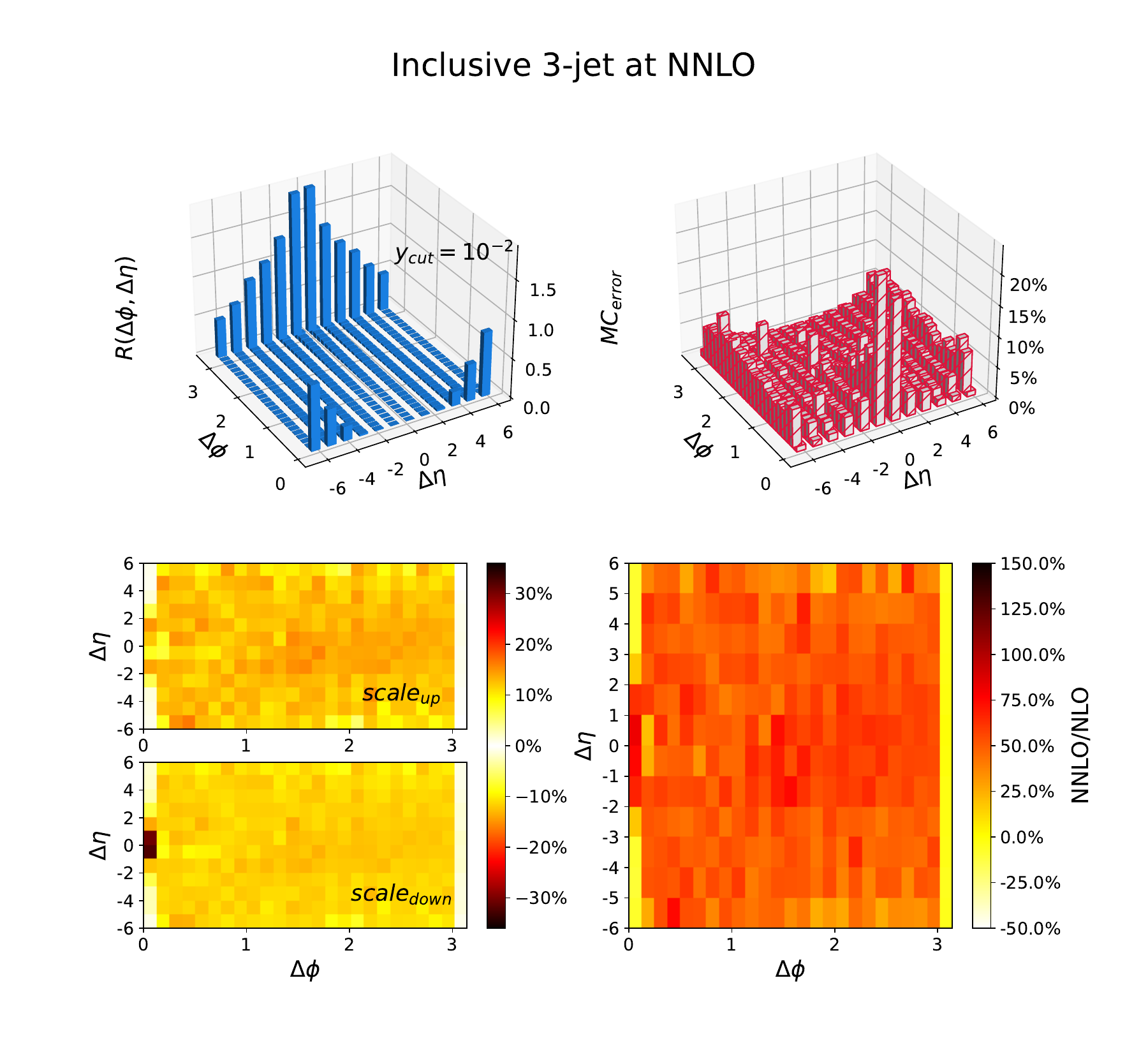}
\caption{The long pseudo-rapidity range $\phi$ correlation (Ridge) distribution in the inclusive 3-jet events at NNLO with \(y_{\text{cut}} = 10^{-2}\). The upper left panel shows the result at the central scale choice; the upper right panel shows the relative statistical error; the lower left panel illustrates the systematic error due to the scale variation, as given by the Eq. \eqref{scale}, where $\text{scale}_{\text{up}}$ ($\text{scale}_{\text{down}}$) means the largest positive (negative) correction relative to the central scale; and the lower right panel displays the relative correction of the NNLO results with respect to the NLO results at the central scale.}
\label{fig:ridge-3+4+5-d-2}
\end{figure}

In Section \ref{sec:eaxis}, our analysis demonstrate that the events driven by pQCD exhibit high planarity. This property leads to significant Ridge correlations. Figure~\ref{fig:ridge-3+4+5-d-2} shows such distributions in the inclusive 3-jet events at NNLO with $y_{\text{cut}} = 10^{-2}$. The upper left panel shows the result at the central scale choice; the upper right panel shows the relative statistical error;
the lower left panel illustrates the systematic error due to the scale variation, as given by the Eq. \eqref{scale}, where $\text{scale}_{\text{up}}$ ($\text{scale}_{\text{down}}$) means the largest positive (negative) correction relative to the central scale; and the lower right panel displays the relative correction of the NNLO results with respect to the NLO results at the central scale.

From the upper left panel of Figure~\ref{fig:ridge-3+4+5-d-2}, we observe that the events cluster mainly at $\Delta \phi = 0$ and $\pi$, which indicates that the inclusive 3-jet events have high planarity. The concentration at $\Delta \phi = \pi$ (the away-side) is notably higher than that at $\Delta \phi = 0$ (the near-side), and it shows a pronounced `Ridge' shape. This trend is consistent with the data presented in Table~\ref{table:inclusive-cross-section}, which highlights the dominance of the 3-jet events at $y_{\text{cut}} = 10^{-2}$. In any 3-jet event, the azimuthal angle $\Delta \phi$ between any two jets is restricted to these two values, and $\Delta \phi = \pi$ occurs twice as frequently as $\Delta \phi = 0$. This characteristic feature of the 3-jet events is responsible for shaping the observed distribution.
We also notice that only a minimal number of events are distributed near the $(\Delta \eta, \Delta \phi) = (0,0)$ region. As a result, there is a significantly larger relative statistical error in the upper right panel of Figure~\ref{fig:ridge-3+4+5-d-2}. The relative statistical error increases with larger $\Delta \eta$, indicating a scarcity of the events with jets at large pseudo-rapidities.
In the lower left panel of Figure~\ref{fig:ridge-3+4+5-d-2}, it is evident that the uncertainty of the scale variation is significantly lower in the $\Delta \phi = 0, \pi$ regions ( $\pm5\%$) compared to the intermediate $\Delta \phi$ region ($\pm15\%$). Given that the distribution in the intermediate $\Delta \phi$ region is exclusively associated with the inclusive 4-jet events, this discrepancy in the uncertainty is attributed to the different orders of the perturbative calculations: NNLO for the near-side region and NLO for the intermediate region.
From the lower right panel of Figure~\ref{fig:ridge-3+4+5-d-2}, we observe that the relative correction is small (approximately $\pm 10\%$) in the region near $\Delta\phi = 0, \pi$ but large (approximately $25\% \sim 75\%$) in the middle region. On both sides of the plot, the dominant contribution stems from the 3-jet events, for which the correction reaches the NNLO accuracy. In contrast, in the middle region, the contribution is derived from the inclusive 4-jet events, and the correction is of the NLO accuracy. The difference in the order of accuracy between these two regions is responsible for the varying magnitudes of the relative correction.

\begin{figure}[hbt!]
\centering
\includegraphics[width=1.0\textwidth]{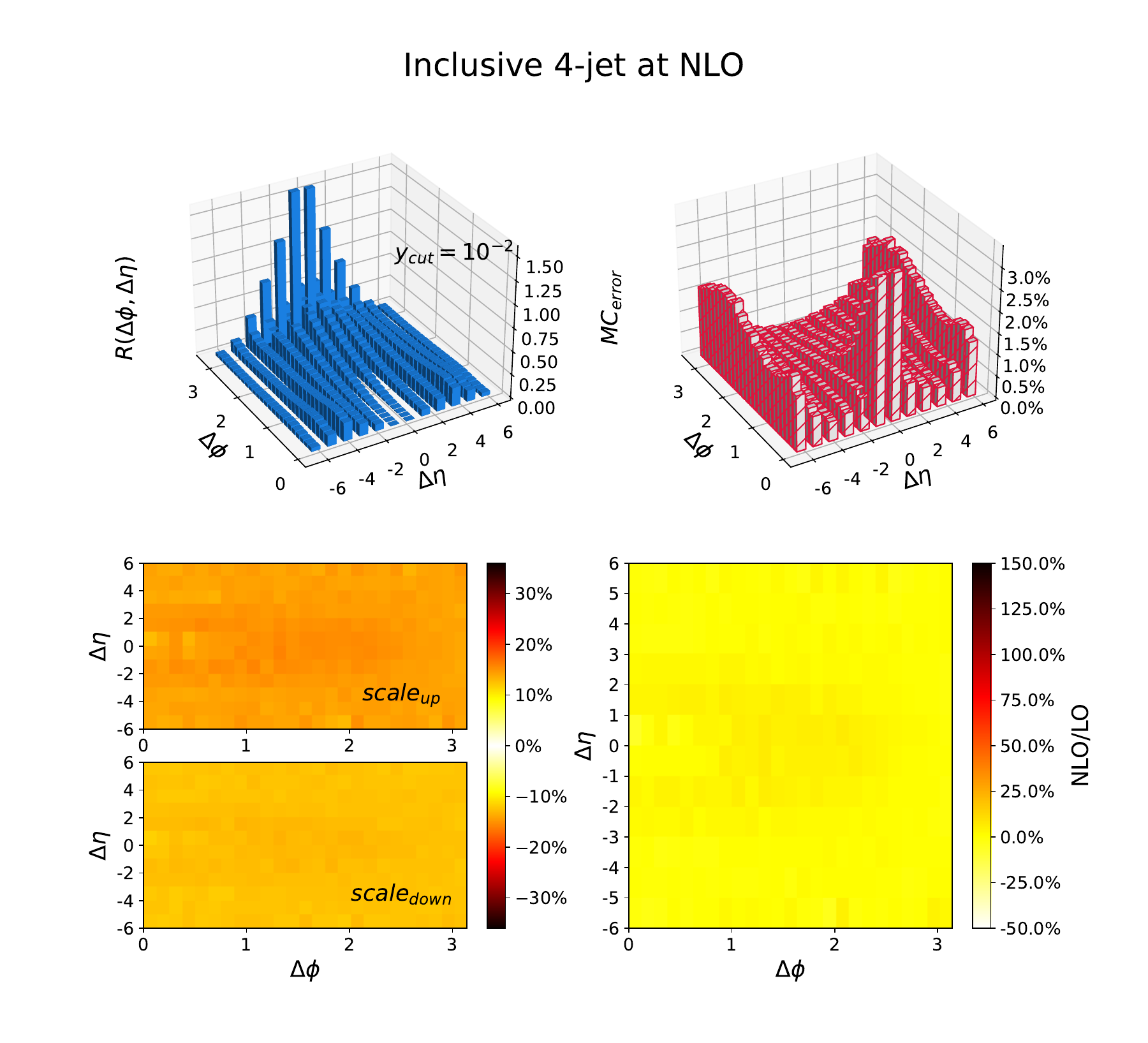}
\caption{The long pseudo-rapidity range $\phi$ correlation (Ridge) distribution in the inclusive 4-jet events at NLO with  \(y_{\text{cut}} = 10^{-2}\). The upper left panel shows the result at the central scale choice; the upper right panel shows the relative statistical error; the lower left panel illustrates the systematic error due to the scale variation, as given by the Eq. \eqref{scale}, where $\text{scale}_{\text{up}}$ ($\text{scale}_{\text{down}}$) means the largest positive (negative) correction relative to the central scale; and the lower right panel displays the relative correction of the NLO results with respect to the LO results at the central scale.}
\label{fig:ridge-4+5-d-2}
\end{figure}

As analyzed above, when $y_{\text{cut}} = 10^{-2}$, the 3-jet events predominate and are primarily concentrated in the regions around $\Delta\phi = 0$ and $\Delta\phi = \pi$, making the distribution in the intermediate areas less visible. Therefore, we present the distribution within the context of the inclusive 4-jet events to offer a clearer view of the intermediate $\Delta\phi$ region.
Figure~\ref{fig:ridge-4+5-d-2} shows the distribution of the Ridge correlation in the inclusive 4-jet events at NLO with $y_{\text{cut}} = 10^{-2}$.
In the upper left panel of Figure~\ref{fig:ridge-4+5-d-2}, we observe obvious distributions in the intermediate $\Delta \phi$ region, while the main distributions are still in the $\Delta \phi = \pi$ region, along with the `Ridge' shape. These observed features suggest that the 4-jet and 5-jet events also have high planarity with $y_{\text{cut}}=10^{-2}$, which is consistent with our finding in Section~\ref{sec:eaxis}.
The plot of the relative statistical error shows that the error is generally minimal, except in the $(\Delta\eta, \Delta\phi)=(0,0)$ region and at the positive and negative boundaries of $\Delta\eta$. The reason for these discrepancies is similar to that in the case of the inclusive 3-jet events.
In the lower left plot of Figure~\ref{fig:ridge-4+5-d-2}, the scale variation remains consistent at approximately $\pm14\%$ across the entire $(\Delta\eta, \Delta\phi)$ region. This uniformity can be ascribed to the NLO accuracy throughout this region.
In the lower right panel of Figure~\ref{fig:ridge-4+5-d-2}, the relative correction is observed to be approximately $\pm5\%$. This value is smaller than the relative correction observed in the inclusive 3-jet events shown in Figure~\ref{fig:ridge-3+4+5-d-2}. This reduction in the relative correction can be attributed to the significant differences in the total cross section between the NLO results and the LO results for the inclusive 4-jet events. Specifically, the total cross section at NLO is $2.0016(7)_{-0.245}^{+0.280}\cdot 10^{6}$ fb, while at LO it is $1.2010(2)_{-0.22}^{+0.30}\cdot 10^{6}$ fb.

\begin{figure}[hbt!]
\centering
\includegraphics[width=1.0\textwidth]{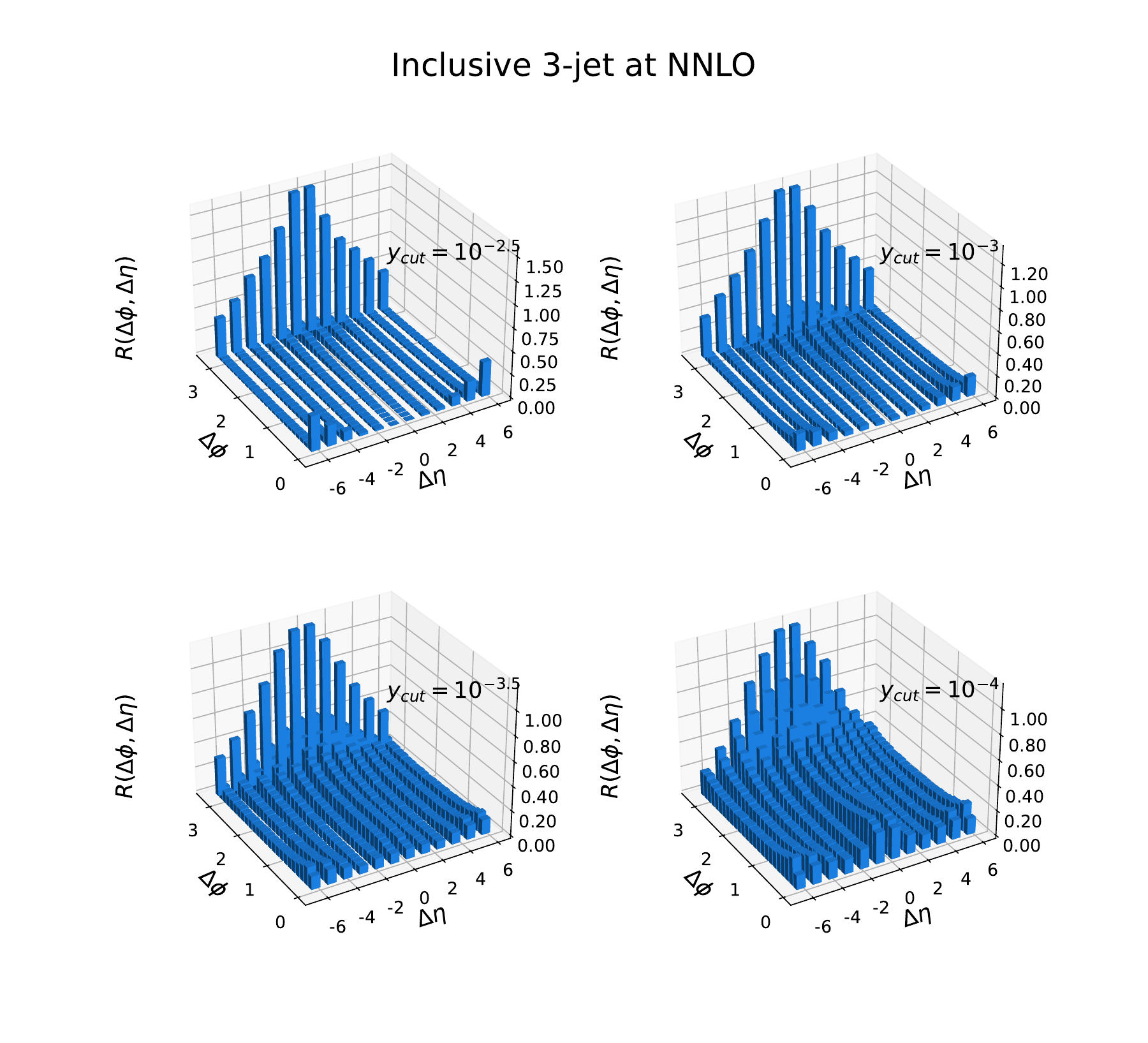}
\caption{The long pseudo-rapidity range $\phi$ correlation (Ridge) distribution in the inclusive 3-jet events at NNLO for $y_{\text{cut}}=10^{-2.5}$(upper left), $10^{-3}$(upper right), $10^{-3.5}$(lower left) and $10^{-4}$(lower right).}
\label{fig:ridge-3+4+5}
\end{figure}

\begin{figure}[hbt!]
\centering
\includegraphics[width=1.0\textwidth]{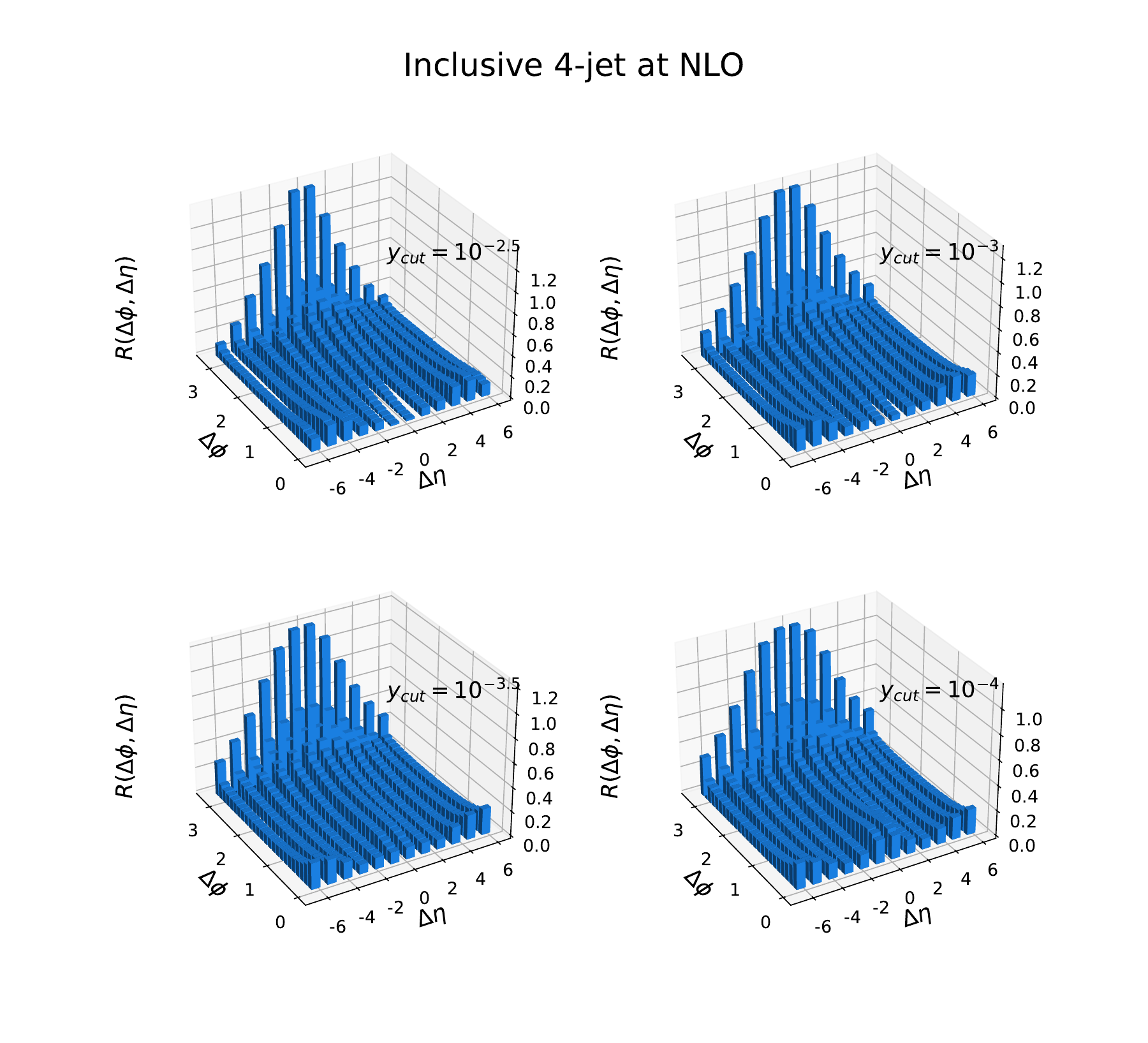}
\caption{The long pseudo-rapidity range $\phi$ correlation (Ridge) distribution in the inclusive 4-jet events at NLO for $y_{\text{cut}}=10^{-2.5}$(upper left), $10^{-3}$(upper right), $10^{-3.5}$(lower left) and $10^{-4}$(lower right).}
\label{fig:ridge-4+5}
\end{figure}

To examine the continuous evolution of the Ridge correlation, we present the distributions with different values of $y_{\text{cut}}$.
Figure~\ref{fig:ridge-3+4+5} and Figure~\ref{fig:ridge-4+5} present the distribution of Ridge in the inclusive 3-jet events and the inclusive 4-jet events, respectively. Specifically, for each figure, the distribution is shown for different values of $y_{\text{cut}}$: $y_{\text{cut}} = 10^{-2.5}$ is presented in the upper-left part of the figure, $y_{\text{cut}} = 10^{-3}$ in the upper-right part, $y_{\text{cut}} = 10^{-3.5}$ in the lower-left part, and $y_{\text{cut}} = 10^{-4}$ in the lower-right part.
In Figure~\ref{fig:ridge-3+4+5}, as the value of $y_{\text{cut}}$ decreases, we observe that the event distributions in the intermediate $\Delta\phi$ region gradually increase, indicating that the proportion of the inclusive 4-jet events is rising. From Table~\ref{table:inclusive-cross-section}, we can see that the proportion of the inclusive 4-jet events becomes predominant when $y_{\text{cut}} = 10^{-4}$. Thus, the distribution with the central scale choice in the lower right panel of Figure~\ref{fig:ridge-3+4+5} is analogous to that in the lower right panel of Figure~\ref{fig:ridge-4+5}.
In the region of $(\Delta\phi,\Delta\eta)= (0,0)$, allowing closer partons to be resolved as separate jets as $y_{\text{cut}}$ decreases, the distribution in this area increases. Similarly, the distribution in the large $\Delta \eta$ region also increases, attributed to the same factor.
We have also calculated the corresponding PYTHIA results. And the ridge phenomenon is observed in all these results (see Appendix~\ref{app:PYTHIAresults}).
Therefore, the physical picture of the fixed order calculation does not change after the parton shower.

\begin{figure}[hbt!]
\centering
\includegraphics[width=1.0\textwidth]{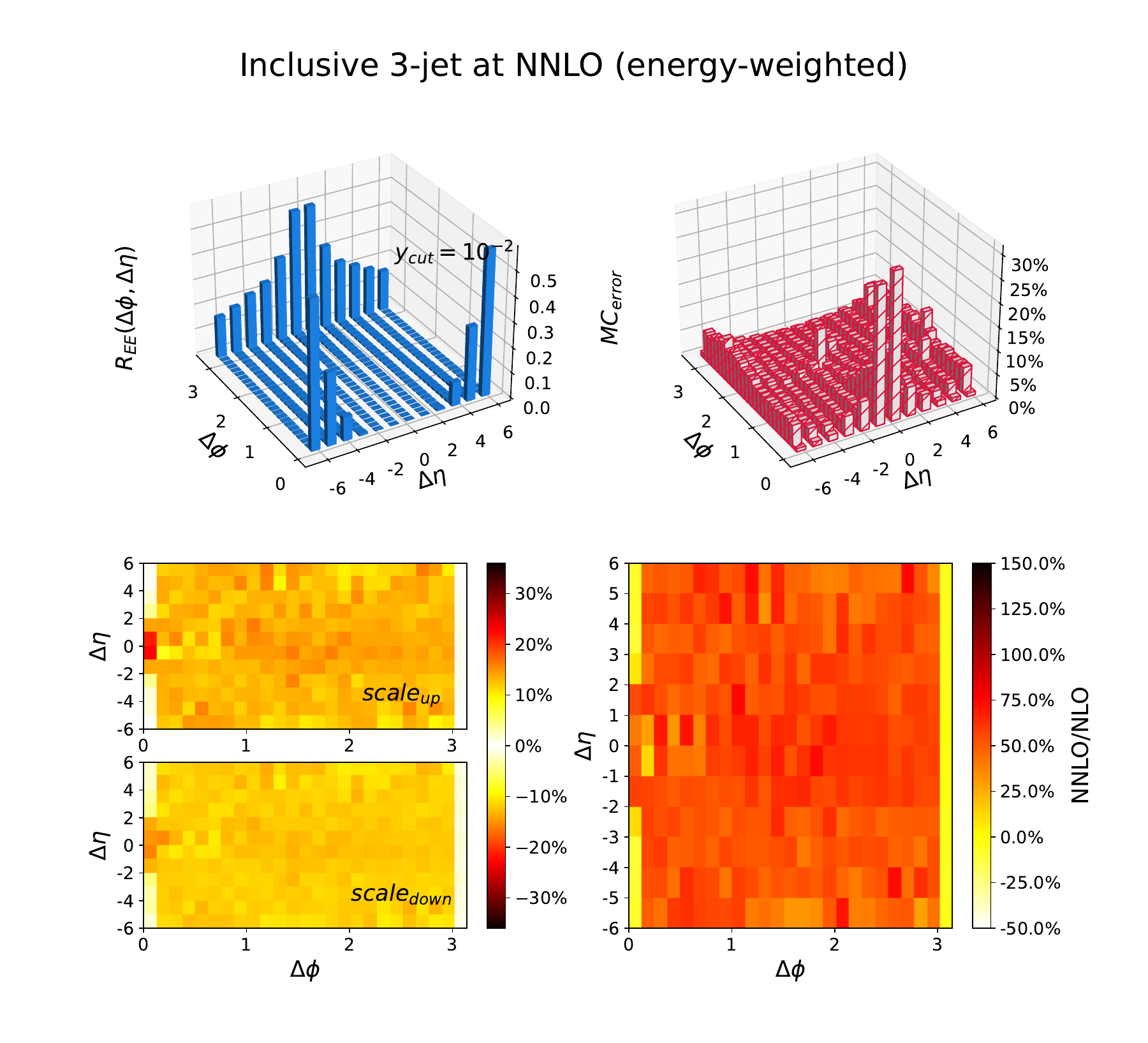}
\caption{The long pseudo-rapidity range $\phi$ correlation weighted by $\frac{4E_{i}E_{j}}{s}$ ($R_{EE}$) in the inclusive 3-jet events at NNLO with \(y_{\text{cut}} = 10^{-2}\). The upper left panel shows the result at the central scale choice; the upper right panel shows the relative statistical error; the lower left panel illustrates the systematic error due to the scale variation, as given by Eq. \eqref{scale}, where $\text{scale}_{\text{up}}$ ($\text{scale}_{\text{down}}$) means the largest positive (negative) correction relative to the central scale; and the lower right panel displays the relative correction of the NNLO results to the NLO results at the central scale.}
\label{fig:ridge-3+4+5-d-3(EEC)}
\end{figure}

\begin{figure}[hbt!]
\centering
\includegraphics[width=1.0\textwidth]{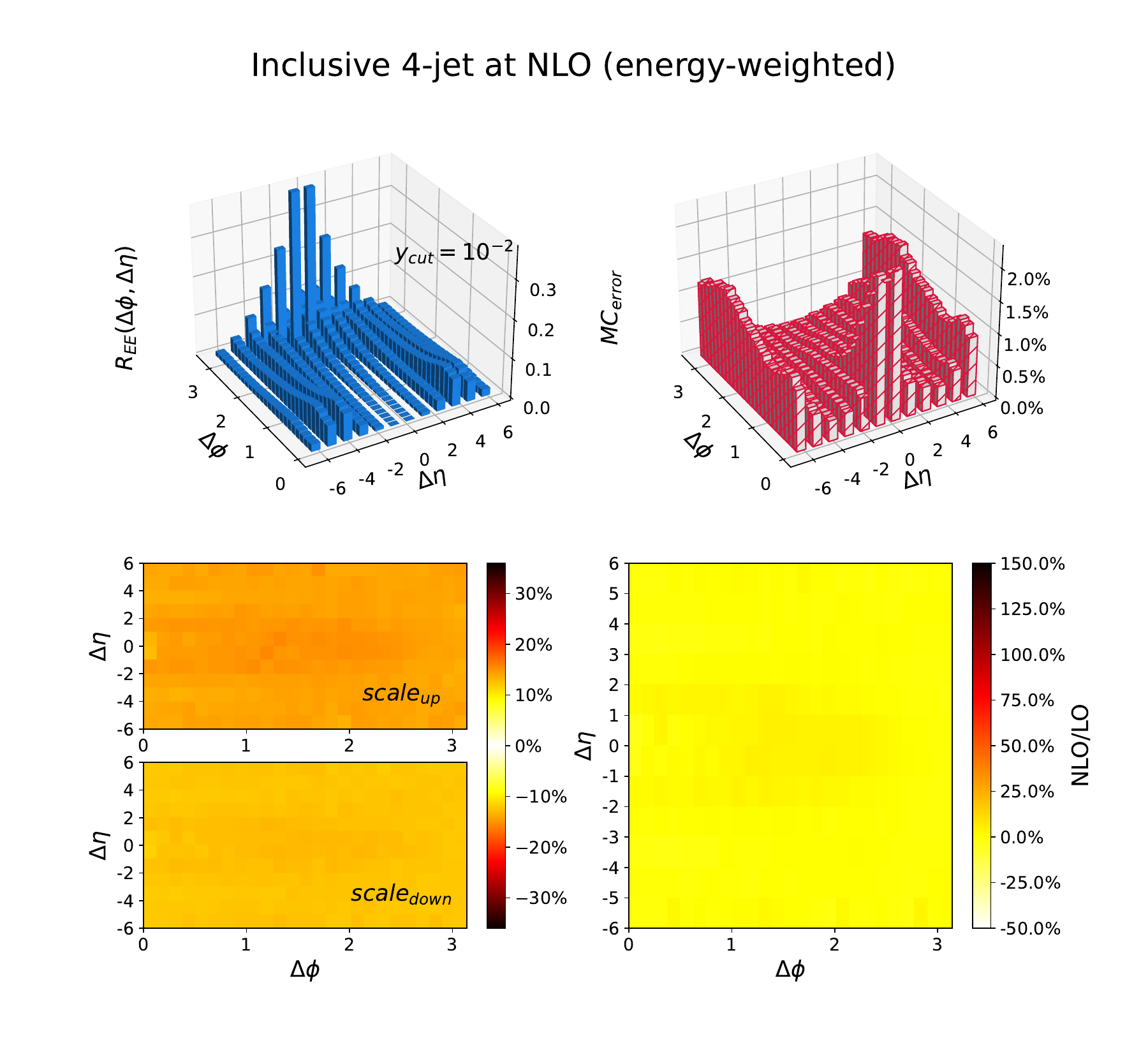}
\caption{The long pseudo-rapidity range $\phi$ correlation weighted by $\frac{4E_{i}E_{j}}{s}$ ($R_{EE}$) in the inclusive 4-jet events at NLO with  \(y_{\text{cut}} = 10^{-2}\). The upper left panel shows the result at the central scale choice; the upper right panel shows the relative statistical error; the lower left panel illustrates the systematic error due to the scale variation, as given by Eq. \eqref{scale}, where $\text{scale}_{\text{up}}$ ($\text{scale}_{\text{down}}$) means the largest positive (negative) correction relative to the central scale; and the lower right panel displays the relative correction of the NLO results to the LO results at the central scale.}
\label{fig:ridge-4+5-d-3(EEC)}
\end{figure}

The joint distribution of $\Delta \phi$ and $\Delta \eta$ between the final state jets exhibits pronounced `Ridge' characteristics, including those with a higher multiplicity of jets.
As mentioned in Section~\ref{sec:event shape}, we can employ the energy-weighted joint distributions to enhance the contribution with hard jets. Figure~\ref{fig:ridge-3+4+5-d-3(EEC)} and Figure~\ref{fig:ridge-4+5-d-3(EEC)} show the distribution of $R_{EE}$ in the inclusive 3-jet events at NNLO and in the inclusive 4-jet events at NLO with $y_{\text{cut}} = 10^{-2}$.
Compared to the relevant plots without the energy weighting, the distribution shows in the large $\Delta\eta$ region for both inclusive 3-jet and inclusive 4-jet events. This leads to a more pronounced `Ridge' shape. This again guarantees that the hard radiations of jets lead the whole phase space of an event to be planar in the $e^+e^-$ annihilation process, since the total momentum conservation.
The systematic error and the high-order correction remain essentially unchanged in the inclusive 3-jet and inclusive 4-jet events, suggesting that the proportion of events with soft jets is small at $y_{\text{cut}} = 10^{-3}$. We have also calculated the corresponding PYTHIA results and more pronounced `Ridge' shape can also be observed.

\FloatBarrier

\section{Conclusions}
\label{sec:conclusions}

The $e^+e^-$ annihilation process provides a clean environment for the study of the anisotropy from the pQCD hard interactions.
In this paper, the precise fixed order QCD predictions of high multiplicity final states are provided.
By employing the event shape observables relevant to the anisotropy of the global shape of the phase space of an event, i.e., the eigenvalues of the sphericity tensor ($\lambda_{1}$, $\lambda_{2}$, $\lambda_{3}$), and the modified planarity ($\tilde{P}$) as the combination of $\lambda_{2}$, $\lambda_{3}$,
we illustrate that the multiple jet (the number of jets
n=3, 4, 5) final states in  $e^+e^-$ annihilation are highly planar. As suggested in \cite{ppt}, we find that this fact leads to a significant Ridge-like correlations among the jet momenta.
These predictions can be directly compared with the LEP data and tested in future $e^+e^-$ colliders.

This study applies NNLO QCD corrections to the inclusive 3-jet production with various values of jet resolution parameter. It is the first time that the NNLO corrections are considered for the Ridge-like correlations among the jet momenta.
The anisotropy in the phase space is determined by the momentum conservation and the intrinsic dynamics of pQCD.
We observe marginal fixed order corrections between the LO and NLO distributions for $\lambda_{1}$, $\lambda_{2}$, $\lambda_{3}$ and $\tilde{P}$ and identify fiducial regions where the resummation of the large algorithmic terms is needed.
For $\lambda_{1}$ and $\lambda_{2}$, we observe that the pQCD corrections converge at NNLO with reduced scale dependence at $\pm 10$\%.
With high precision, we sharpen the fact that the global shapes of most events are oblate rather than isotropic.
The difference is illustrated by the peak regions of the distributions between $\lambda_{2}$ and $\lambda_{3}$.
The peak of the modified planarity $\Tilde{P}$ around 1 (Figure~\ref{fig:P-4jets-lin}) is a strong evidence for the anisotropy in the transverse direction perpendicular to the eigenvector of $\lambda_{1}$.
The Ridge-like correlation  are confirmed in both the inclusive 3-jet final states at NNLO and the inclusive 4-jet final states at NLO.
It is evident that in the region where $\Delta\phi \sim \pi$, there is a significant enhancement attributed to the higher order pQCD corrections.
For $\Delta\phi$ away from $\pi$ and 0, where the inclusive 3-jet results at NNLO are equivalent to the 4-jet case at NLO, we observe the non-flat higher order corrections for both the Ridge and $R_{EE}$ distributions between $(\Delta\phi,\Delta\eta) \sim (0,0)$ and $(\Delta\phi,|\Delta\eta|) > (2,3)$ regions.
In the fixed order region where $y_\text{{cut}}\ge 10^{-3}$, we have not observed the convergence of pQCD for the Ridge and $R_{EE}$ distributions. These distributions may require higher order corrections of $e^+e^-\rightarrow$ 4-jet production at the NNLO QCD accuracy in the future.
At the same time, our study suggests that the resummation effects should be taken into account when $y_\text{{cut}}$ is below $10^{-3}$.
And from the results calculated using PYTHIA, we observe that the parton shower effects do not change the picture that most of events have high planarity for both inclusive 3-jet and 4-jet events.

The Ridge correlations have been employed to study various non-perturbative effects of the strong interaction employing the final state hadron momenta. This means the effects of  soft partons, hadronizations are necessary to be considered. Though these are beyond the approach of pQCD, by clarifying the connections between the
planar property and the Ridge-like correlations for multi-jet systems, this work provides precise fixed order pQCD benchmarks for probing the non-perturbative effects based on hadronic data.
Similar studies of the anisotropy can be extended to hadronic (nuclear) collisions to shed light on various properties of the strong interaction.

\section*{Acknowledgments}
The authors would like to thank Yi Jin, Haitao Li and Qun Wang for enlightening discussions. We are indebted to Elliot Fox, Thomas Gehrmann, Nigel Glover, Alexander Huss and Matteo Marcoli for their feedback and encouragement to pursue this work.
This research was supported by the National Science Foundation of China (NSFC) under contract No.12475085, No.12275157, No.12321005 and No.12235008.
\section*{Appendix}
\appendix
\section{Shengjin Formula}
\label{app:Shengjin Formula}
For a cubic equation with one unknown $X$,
\begin{align}
    a X^3+b X^2+c X+d=0\enspace(a, b, c, d \in \mathbb{R}, a \neq 0),
\end{align}
there are three double root discriminants
\begin{align}
    \left\{\begin{array}{l}
A=b^2-3 a c \\
B=b c-9 a d \\
C=c^2-3 b d
           \end{array}\right.,
\end{align}
and one total discriminant
\begin{align}
    \Delta=B^2-4 \mathrm{AC}.
\end{align}
In the case of $A=B=0$, one could derive
\begin{align}
    X_1=X_2=X_3=\frac{-b}{3 a}=\frac{-c}{b}=\frac{-3 d}{c}.
\end{align}
For $A,B\neq 0$, while $\Delta=0$, we have
\begin{align}
    \begin{aligned}
& X_1=\frac{-b}{a}+K \\
& X_{2,3}=\frac{-K}{2},
\end{aligned}
\end{align}
where
\begin{align}
    K=\frac{B}{A}\enspace(A \neq 0).
\end{align}
For $\Delta>0$, we have
\begin{align}
   \begin{aligned}
& X_1=\frac{-b-\left(\sqrt[3]{Y_+}+\sqrt[3]{Y_-}\right)}{3 a} \\
& X_{2,3}=\frac{-b+\frac{1}{2}\left(\sqrt[3]{Y_+}+\sqrt[3]{Y_-}\right) \pm \mathrm{i}\frac{\sqrt{3}}{2}\left(\sqrt[3]{Y_+}-\sqrt[3]{Y_-}\right) }{3 a},
    \end{aligned}
\end{align}
where
\begin{align}
    Y_{\pm}=A b+3 a\left(\frac{-B \pm \sqrt{B^2-4 A C}}{2}\right).
\end{align}
For $\Delta<0$, we have
\begin{align}
    \begin{aligned}
& X_1=\frac{-b-2 \sqrt{A} \cos \frac{\theta}{3}}{3 a} \\
& X_{2,3}=\frac{-b+\sqrt{A}\left(\cos \frac{\theta}{3} \pm \sqrt{3} \sin \frac{\theta}{3}\right)}{3 a},
    \end{aligned}
\end{align}
where
\begin{align}
    \begin{aligned}
& \theta=\arccos T \\
& T=\frac{2 A b-3 a B}{2 \sqrt{A^3}}\enspace(A>0,\enspace-1<T<1).
    \end{aligned}
\end{align}

\section{$C_{3},C_{4},C_{5}$ comparison with CoLoRFulNNLO and Sherpa}
\label{app:C3,C4,C5}
\subsection{Jet rate}

The production rate of n-jet events in electron-positron annihilation is defined as the ratio of the cross section for the exclusive n-jet events $\sigma_{n-jet}$ divided by the total hadronic cross section $\sigma_{0}$.
The production rates have the following perturbative expansions, as detailed in \cite{Weinzierl:2010cw}:
\begin{align}
\label{jetrate}
    \begin{aligned}
& \frac{\sigma_{3-j e t}(\mu)}{\sigma_0(\mu)}=\frac{\alpha_s(\mu)}{2 \pi} A_3(\mu)+\left(\frac{\alpha_s(\mu)}{2 \pi}\right)^2 B_3(\mu)+\left(\frac{\alpha_s(\mu)}{2 \pi}\right)^3 C_3(\mu)+O\left(\alpha_s^4\right), \\
& \frac{\sigma_{4-j e t}(\mu)}{\sigma_0(\mu)}=\quad\left(\frac{\alpha_s(\mu)}{2 \pi}\right)^2 B_4(\mu)+\left(\frac{\alpha_s(\mu)}{2 \pi}\right)^3 C_4(\mu)+O\left(\alpha_s^4\right), \\
& \frac{\sigma_{5-j e t}(\mu)}{\sigma_0(\mu)}=\quad\left(\frac{\alpha_s(\mu)}{2 \pi}\right)^3 C_5(\mu)+O\left(\alpha_s^4\right),
\end{aligned}
\end{align}
where $\mu$ denotes the renormalization scale.
The corresponding coefficients $C_3$, $C_4$, $C_5$ are derived from the coefficients of the exclusive (NNLO) three-, (NLO) four- and (LO) five-jet rates.
We compare the perturbative coefficients of the jet rate, $C_3$, $C_4$ under NNLOJET and CoLoRFulNNLO, and $C_{5}$ under NNLOJET, CoLoRFulNNLO and Sherpa at Z-pole with different values of $y_\text{cut}$.
In the calculation of the jet rates, we choose the QCD coupling constant $\alpha_{s}=0.118$ and the total hadronic cross section $\sigma_{0} = 4.03538(3)\times 10^{7}$ fb.

\subsection{$C_3$}
Table~\ref{C3} shows the comparison of $C_3$ between NNLOJET and CoLoRFulNNLO \cite{Verbytskyi:2019zhh}. Specifically, the first column lists different values of $y_{\text{cut}}$, while the second and third columns present the calculated values of $C_3$ from NNLOJET and CoLoRFulNNLO, respectively. The last column of the table indicates the deviation between the results of NNLOJET and CoLoRFulNNLO. For $y_{\text{cut}} = 2.49\times 10^{-4}$, there is a large deviation between the results obtained from NNLOJET and CoLoRFulNNLO. However, when $y_{\text{cut}}$ lies in the middle regions ($6.76\times 10^{-4} \leq y_{\text{cut}}\leq 2.243\times10^{-3}$), the deviations between the results of NNLOJET and CoLoRFulNNLO are within $1\sigma$, which indicates that the results of the two methods are in agreement. As the value of $y_{\text{cut}}$ increases, $C_{3}$ approaches zero and then becomes positive. In this case (for $y_{\text{cut}} \geq 6.097\times10^{-3}$), the relative statistical error of $C_{3}$ is significant, which makes it difficult to conduct a reliable comparison. In conclusion, with the current computing power, for $C_{3}$, the accurately-calculated results are consistent in the middle $y_{\text{cut}}$ regions but not at $y_{\text{cut}} = 2.49\times10^{-4}$.

\begin{table}[hbt!]
	\centering
	\begin{threeparttable}
        \renewcommand{\arraystretch}{1.4}
		\begin{tabular}{|c|c|c|c|}\hline
			\makebox[2.8cm][c]{$y_{\text{cut}}$} & \makebox[4cm][c]{$C_{3}$(NNLOJET)} & \makebox[4cm][c]{$C_{3}$(CoLoRFulNNLO)} & \makebox[2cm][c]{Deviation}\\ \hline
			$6.097\cdot10^{-3}$& \makebox[4cm][c]{$-2.92(8)\cdot10^{3}$}& \makebox[4cm][c]{$-3.09\cdot10^{3}$}& \makebox[2cm][c]{$2.13\sigma$}\\ \hline
			$2.243\cdot10^{-3}$& \makebox[4cm][c]{$-1.28(2)\cdot10^{4}$} & \makebox[4cm][c]{$-1.2771\cdot10^{4}$}& \makebox[2cm][c]{$0.15\sigma$}\\ \hline
			$1.836\cdot10^{-3}$& \makebox[4cm][c]{$-1.576(23)\cdot10^{4}$}& \makebox[4cm][c]{$-1.5681\cdot10^{4}$}& \makebox[2cm][c]{$0.34\sigma$}\\ \hline
            $6.76\cdot 10^{-4}$& \makebox[4cm][c]{$-3.415(27)\cdot 10^{4}$}& \makebox[4cm][c]{$-3.3891\cdot 10^{4}$}& \makebox[2cm][c]{$0.96\sigma$} \\ \hline
			$2.49\cdot 10^{-4}$& \makebox[4cm][c]{$-5.364(32)\cdot 10^{4}$}& \makebox[4cm][c]{$-5.2708\cdot 10^{4}$}& \makebox[2cm][c]{$2.91\sigma$} \\ \hline
    \end{tabular}
            \caption{\label{C3}The comparison of $C_3$ between NNLOJET and CoLoRFulNNLO \cite{Verbytskyi:2019zhh}. The first column lists different values of $y_{\text{cut}}$, while the second and third columns present the calculated values of $C_3$ from NNLOJET and CoLoRFulNNLO, respectively. The last column indicates the deviation between the results of NNLOJET and CoLoRFulNNLO.}
	\end{threeparttable}
\end{table}

\FloatBarrier
\subsection{$C_4$}

Table~\ref{C4} shows the comparison of $C_4$ between NNLOJET and CoLoRFulNNLO \cite{Verbytskyi:2019zhh}. Specifically, the first column lists different values of $y_{\text{cut}}$. The second and third columns present the calculated values of $C_4$ obtained from NNLOJET and CoLoRFulNNLO, respectively. The last column of the table indicates the deviation between the results of these two methods. We observe that all deviations between CoLoRFulNNLO and NNLOJET are within $2\sigma$, except for the case corresponding to the smallest value of $y_{\text{cut}}$. For $y_{\text{cut}} = 2.49\times 10^{-4}$, we attribute the large deviation to the insufficient statistical data. Then, we selected a $y_{\text{cut}}$ value slightly larger than \( y_{\text{cut}} = 2.49\times 10^{-4} \) (specifically, $y_{\text{cut}} = 3.04\times 10^{-4}$) and found that the two results are in agreement. Thus, taking into account the influence of statistical errors, the results of $C_{4}$ calculated by NNLOJET and CoLoRFulNNLO are in agreement.

\begin{table}[hbt!]
	\centering
	\begin{threeparttable}
        \renewcommand{\arraystretch}{1.4}
		\begin{tabular}{|c|c|c|c|}\hline
			\makebox[3cm][c]{$y_{\text{cut}}$} & \makebox[4.2cm][c]{$C_{4}$(NNLOJET)} & \makebox[4.2cm][c]{$C_{4}$(CoLoRFulNNLO)} & \makebox[2cm][c]{Deviation}\\ \hline
            $3.6883\cdot10^{-2}$& \makebox[4.2cm][c]{$2.600(7)\cdot10^{2}$}& \makebox[4.2cm][c]{$2.607\cdot10^{2}$} &\makebox[2cm][c]{$0.99\sigma$}\\ \hline
            $1.3569\cdot10^{-2}$& \makebox[4.2cm][c]{$1.876(4)\cdot10^{3}$}& \makebox[4.2cm][c]{$1.8684\cdot10^{3}$} &\makebox[2cm][c]{$1.63\sigma$}\\ \hline
            $6.097\cdot10^{-3}$& \makebox[4.2cm][c]{$4.90(1)\cdot10^{3}$}& \makebox[4.2cm][c]{$4.8976\cdot10^{4}$} &\makebox[2cm][c]{$0.63\sigma$}\\ \hline
            $2.243\cdot10^{-3}$& \makebox[4.2cm][c]{$1.004(5)\cdot10^{4}$}& \makebox[4.2cm][c]{$1.0084\cdot10^{4}$} &\makebox[2cm][c]{$1.00\sigma$}\\ \hline
            $1.231\cdot10^{-3}$& \makebox[4.2cm][c]{$1.180(6)\cdot10^{4}$}& \makebox[4.2cm][c]{$1.1905\cdot10^{4}$} &\makebox[2cm][c]{$1.65\sigma$}\\ \hline
            $6.76\cdot10^{-4}$& \makebox[4.2cm][c]{$9.40(8)\cdot 10^{3}$}& \makebox[4.2cm][c]{$9.2938\cdot10^{3}$} &\makebox[2cm][c]{$1.33\sigma$}\\ \hline
            $3.04\cdot10^{-4}$& \makebox[4.2cm][c]{$-9.63(19)\cdot10^{3}$}& \makebox[4.2cm][c]{$-9.434\cdot10^{3}$} &\makebox[2cm][c]{$0.94\sigma$}\\ \hline
            $2.49\cdot10^{-4}$& \makebox[4.2cm][c]{$-1.86(2)\cdot 10^{4}$}& \makebox[4.2cm][c]{$-1.8059\cdot 10^{4}$} &\makebox[2cm][c]{$2.39\sigma$} \\ \hline
    \end{tabular}
            \caption{\label{C4}The comparison of $C_4$ between NNLOJET and CoLoRFulNNLO \cite{Verbytskyi:2019zhh}. The first column lists different values of $y_{\text{cut}}$. The second and third columns present the calculated values of $C_4$ obtained from NNLOJET and CoLoRFulNNLO, respectively. The last column indicates the deviation between the results of these two methods.}
	\end{threeparttable}
\end{table}

\FloatBarrier
\subsection{$C_5$}

Table~\ref{N-S-C5} shows the comparison of the total cross sections for the exclusive 5-jet events at LO between NNLOJET and Sherpa. Specifically, the first column lists the different Durham $y_{\text{cut}}$ values. The second and third columns present the total cross sections obtained from NNLOJET and Sherpa, respectively. The last column indicates the deviation between the results of NNLOJET and Sherpa.
We observe that the deviation is within $2\sigma$ for all values of $y_{\text{cut}}$. Thus, the results of the total cross sections for the exclusive 5-jet events at LO between NNLOJET and Sherpa are consistent.

\begin{table}[hbt!]
	\centering
	\begin{threeparttable}
        \renewcommand{\arraystretch}{1.4}
		\begin{tabular}{|c|c|c|c|}\hline
			\makebox[2.8cm][c]{$y_{\text{cut}}$} & \makebox[4.2cm][c]{$\sigma_{\text{total}}$ (NNLOJET)} & \makebox[4.2cm][c]{$\sigma_{\text{total}}$ (Sherpa)} & \makebox[2cm][c]{Deviation}\\ \hline
			$1\cdot10^{-2.0}$& \makebox[4.2cm][c]{$4.789(1)\cdot 10^{4}$}& \makebox[4.2cm][c]{$4.794(4)\cdot 10^{4}$} &\makebox[2cm][c]{$1.21\sigma$} \\ \hline
			$1\cdot10^{-2.5}$& \makebox[4.2cm][c]{$5.523(4)\cdot10^{5}$}& \makebox[4.2cm][c]{$5.526(6)\cdot10^{5}$} &\makebox[2cm][c]{$0.41\sigma$}\\ \hline
			$1\cdot10^{-3.0}$& \makebox[4.2cm][c]{$3.017(2)\cdot 10^{6}$}& \makebox[4.2cm][c]{$3.018(4)\cdot10^{6}$} &\makebox[2cm][c]{$0.23\sigma$}\\ \hline
			$1\cdot10^{-3.5}$& \makebox[4.2cm][c]{$1.1181(5)\cdot10^{7}$}& \makebox[4.2cm][c]{$1.118(1)\cdot10^{7}$} &\makebox[2cm][c]{$0.09\sigma$}\\ \hline
			$1\cdot10^{-4.0}$& \makebox[4.2cm][c]{$3.261(5)\cdot10^{7}$}& \makebox[4.2cm][c]{$3.259(5)\cdot10^{7}$} &\makebox[2cm][c]{$0.20\sigma$}\\ \hline
    \end{tabular}
            \caption{\label{N-S-C5}The comparison of the total cross sections for the exclusive 5-jet events at LO between NNLOJET and Sherpa. The first column lists the different Durham $y_{\text{cut}}$ values. The second and third columns present the total cross sections obtained from NNLOJET and Sherpa, respectively. The last column indicates the deviation between the results of NNLOJET and Sherpa.}
	\end{threeparttable}
\end{table}

Table~\ref{C5} presents the comparison of the $C_5$ values among CoLoRFulNNLO \cite{Verbytskyi:2019zhh}, Sherpa, and NNLOJET at different Durham $y_{\text{cut}}$ values (listed in the first column). Specifically, the second, third, and fourth columns show the $C_5$ values calculated by NNLOJET, CoLoRFulNNLO, and Sherpa, respectively. The fifth, sixth, and seventh columns display the deviations between (Sherpa and CoLoRFulNNLO), (NNLOJET and CoLoRFulNNLO), and (NNLOJET and Sherpa), respectively.
We observe that the deviation between NNLOJET and Sherpa is within $2\sigma$ across all the values of $y_{\text{cut}}$ except for $y_{\text{cut}} = 6.097\times10^{-4}$.
Similar to the results of $C_4$, when comparing CoLoRFulNNLO with Sherpa and NNLOJET separately, the deviation is larger for $y_{\text{cut}} = 2.49\times 10^{-4}$ than when it takes other values.
We also selected $y_{\text{cut}} = 3.04\times 10^{-4}$, which is slightly larger than \( y_{\text{cut}} = 2.49\times 10^{-4} \), but found that the results of NNLOJET and CoLoRFulNNLO are not in agreement.
In summary, taking into account the influence of statistical errors, the results of $C_{5}$ calculated by NNLOJET, CoLoRFulNNLO, and Sherpa are consistent at relatively large values of $y_{\text{cut}}$. Specifically, for $y_{\text{cut}}\sim 3\times 10^{-4}$, the results from NNLOJET and Sherpa are in agreement within $2\sigma$. However, the results from NNLOJET do not agree with those from CoLoRFulNNLO.

\begin{table}[hbt!]
	\centering
	\begin{threeparttable}
        \renewcommand{\arraystretch}{1.4}
		\begin{tabular}{|c|c|c|c|c|c|c|}\hline
			$y_{\text{cut}}$ & $C_{5}(N)$& $C_{5}(C)$ & $C_{5}(S)$&$\text{Dev}_{S-C}$&$\text{Dev}_{N-C}$&$\text{Dev}_{S-N}$\\ \hline		
			$1.3569\cdot10^{-2}$& $7.406(6)\cdot10^{1}$& $7.407\cdot10^{1}$&$7.4068(6)\cdot10^{1}$& $0.33\sigma$&$0.17\sigma$&$0.13\sigma$\\ \hline
			$6.097\cdot10^{-3}$& $5.8639(24)\cdot 10^{2}$& $5.869\cdot10^{2}$ &$5.855(5)\cdot10^{2}$&$2.8\sigma$&$2.13\sigma$&$1.60\sigma$\\ \hline
			$1.836\cdot10^{-3}$& $4.906(2)\cdot10^{3}$& $4.91\cdot10^{3}$&$4.904(5)\cdot10^{3}$&$1.2\sigma$&$2.0\sigma$&$0.37\sigma$\\ \hline
	        $6.76\cdot 10^{-4}$& $1.8231(5)\cdot 10^{4}$& $1.8248\cdot 10^{4}$&$1.8258(8)\cdot 10^{4}$&$1.25\sigma$ &$3.4\sigma$&$2.86\sigma$ \\ \hline
		    $3.04\cdot10^{-4}$& $4.357(1)\cdot10^{4}$& $4.3613\cdot10^{4}$&$4.3607(19)\cdot10^{4}$&$0.32\sigma$&$4.3\sigma$&$1.79\sigma$\\ \hline
        	$2.49\cdot 10^{-4}$& $5.314(1)\cdot 10^{4}$& $5.2659\cdot 10^{4}$&$5.316(2)\cdot 10^{4}$&$25\sigma$ &$48\sigma$&$0.89\sigma$ \\ \hline
	    \end{tabular}
            \caption{\label{C5}The comparison of $C_5$ values among CoLoRFulNNLO \cite{Verbytskyi:2019zhh}, Sherpa, and NNLOJET at different Durham $y_{\text{cut}}$ values (listed in the first column). The second, third, and fourth columns show the $C_5$ values calculated by NNLOJET, CoLoRFulNNLO, and Sherpa, respectively. The fifth, sixth, and seventh columns display the deviations between (Sherpa and CoLoRFulNNLO), (NNLOJET and CoLoRFulNNLO), and (NNLOJET and Sherpa), respectively.}
	\end{threeparttable}
\end{table}

\section{PYTHIA results}

\begin{figure}[hbt!]
\includegraphics[width=0.49\textwidth]{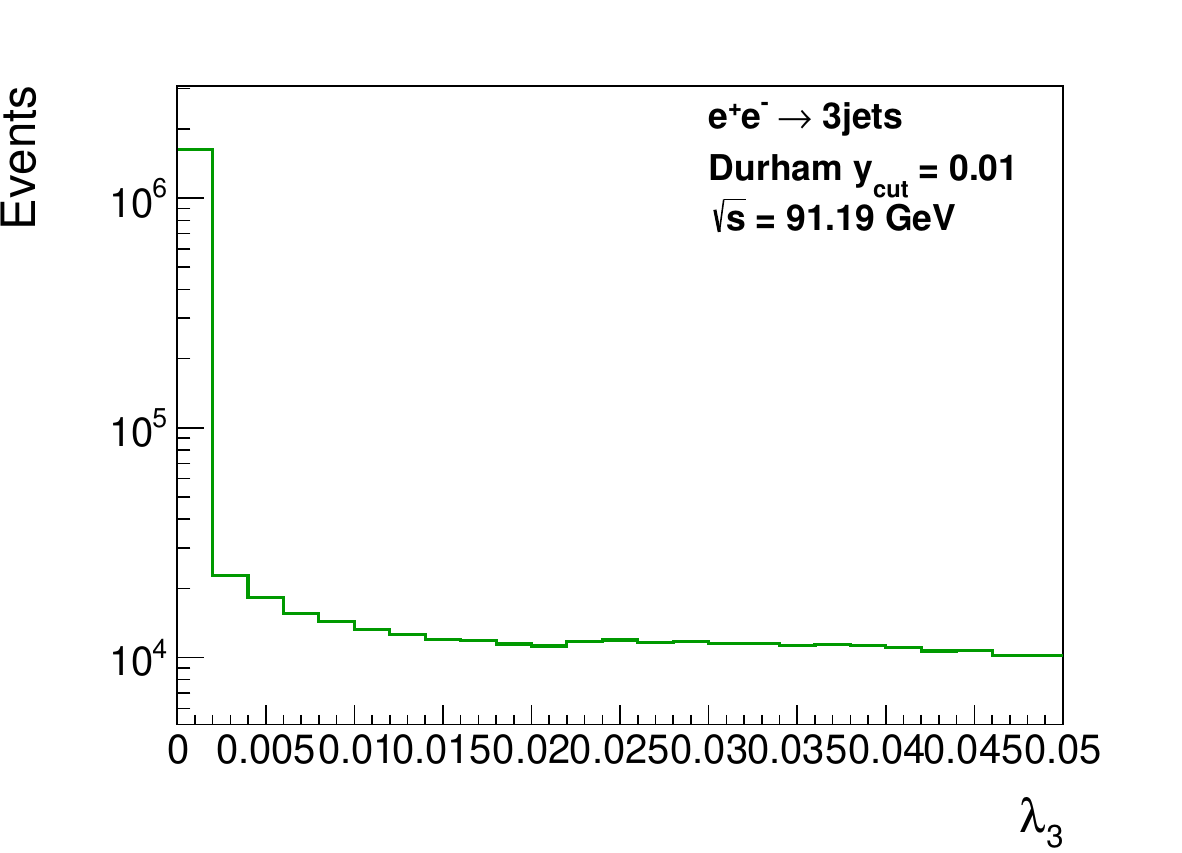}
\includegraphics[width=0.49\textwidth]{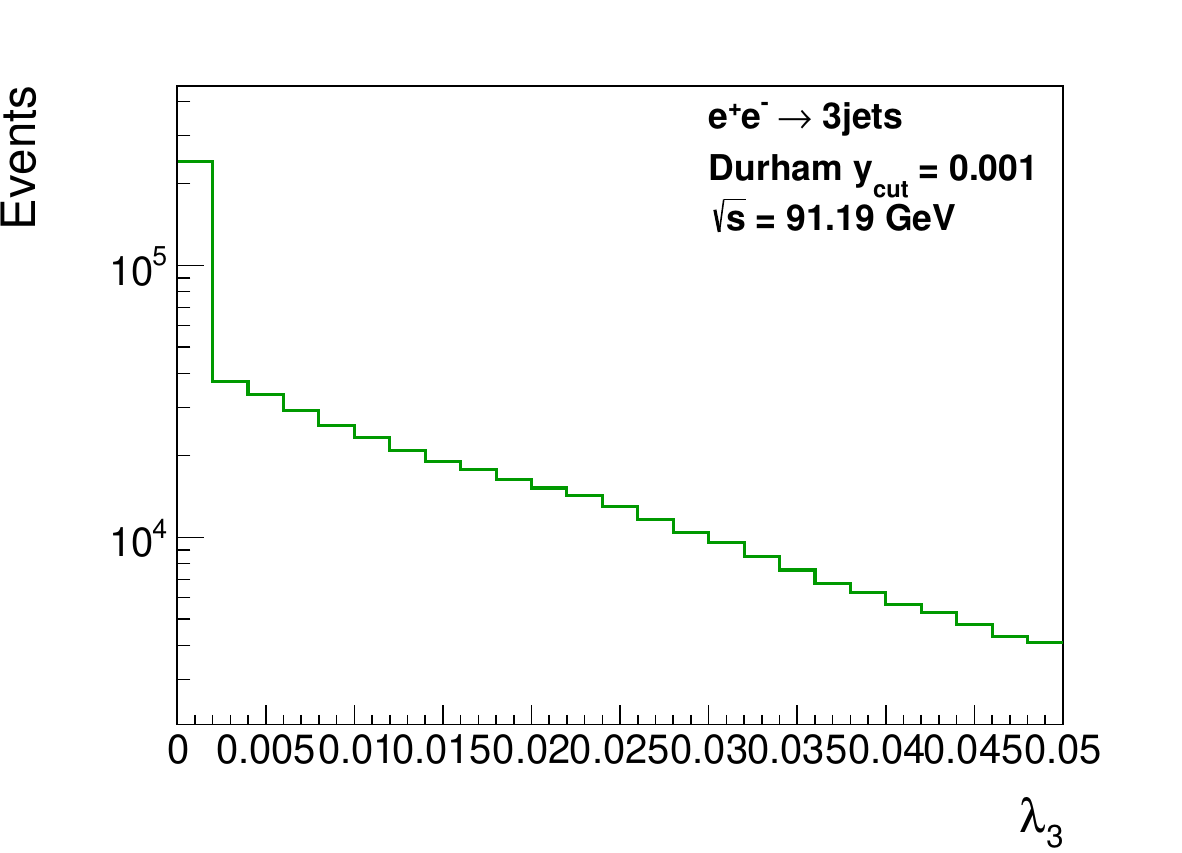}
\caption{The distribution of the event shape observable \(\lambda_3\) in the inclusive 3-jet events at the center-of-mass energy $\sqrt{s}=m_Z$ for \(y_{\text{cut}} = 10^{-2}\) (left panel) and \(10^{-3}\) (right panel), based on PYTHIA 8.3~\cite{Bierlich:2022pfr}.}
\label{fig:lambda3-3jets-lin-pythia}
\end{figure}

\begin{figure}[hbt!]
\includegraphics[width=0.49\textwidth]{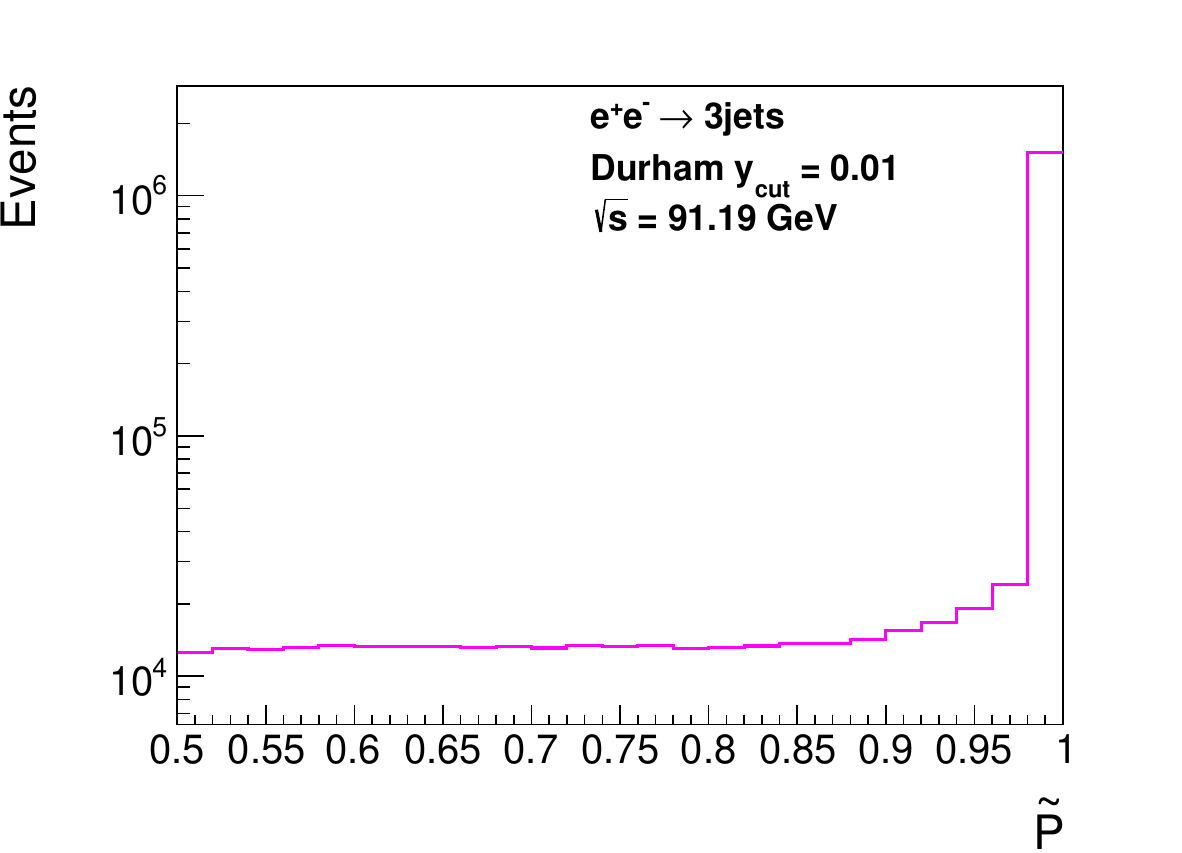}
\includegraphics[width=0.49\textwidth]{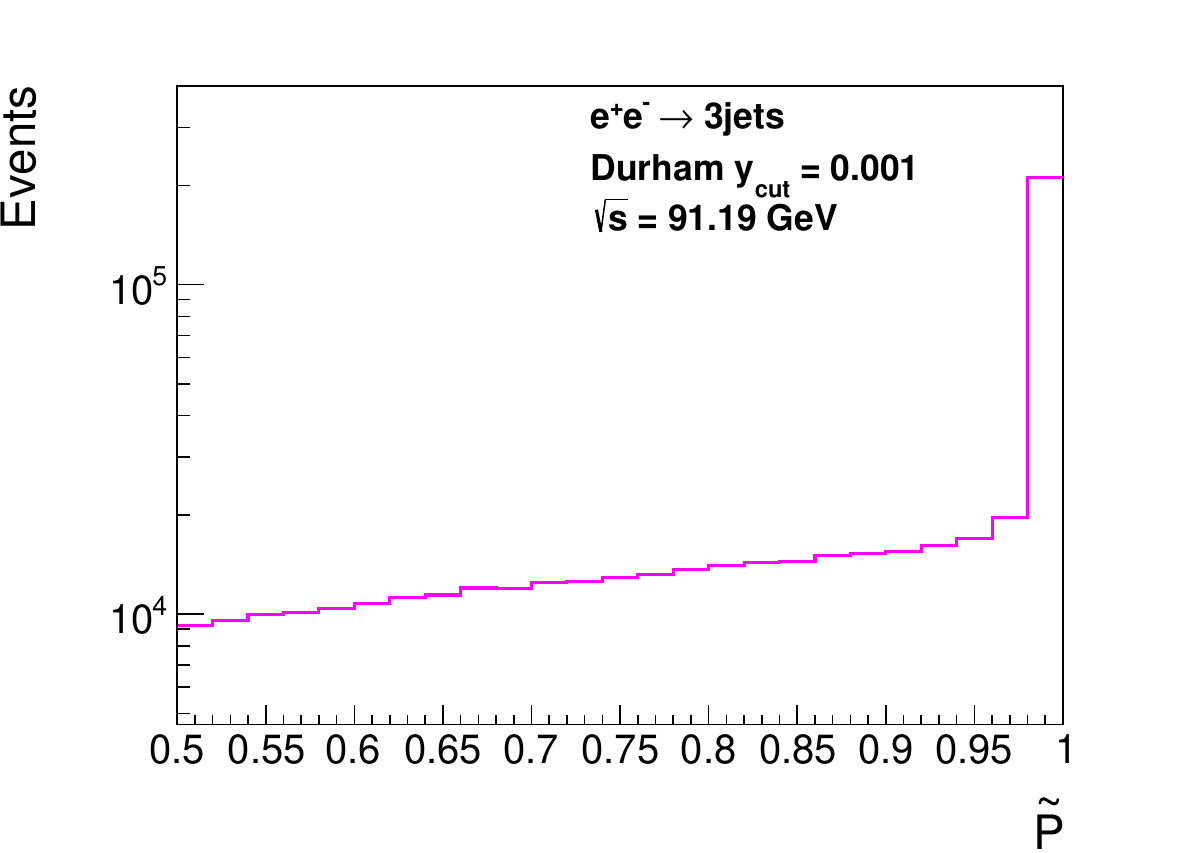}
\caption{The distribution of the event shape observable \(\tilde{P}\) in the inclusive 3-jet events at the center-of-mass energy $\sqrt{s}=m_Z$ for \(y_{\text{cut}} = 10^{-2}\) (left panel) and \(10^{-3}\) (right panel), based on PYTHIA 8.3~\cite{Bierlich:2022pfr}.}
\label{fig:P-3jets-lin-pythia}
\end{figure}

\begin{figure}[hbt!]
\includegraphics[width=0.49\textwidth]{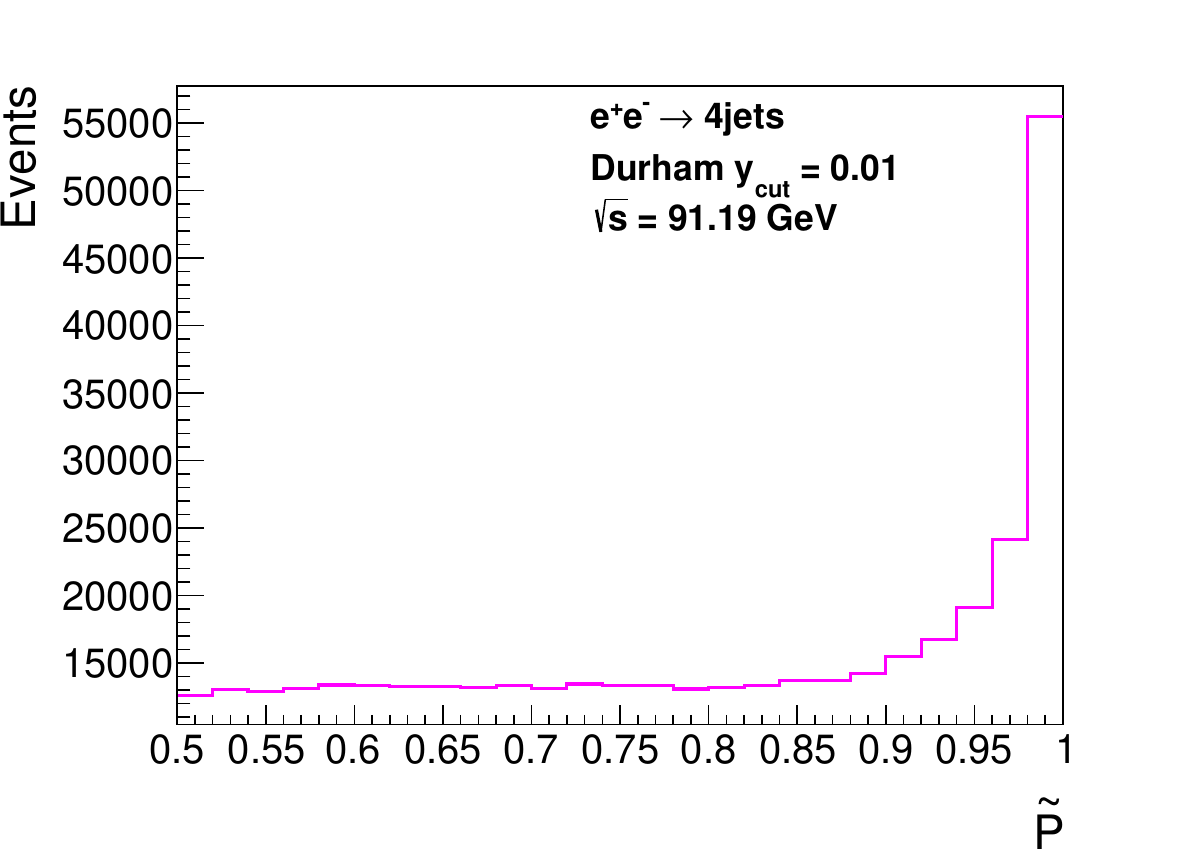}
\includegraphics[width=0.49\textwidth]{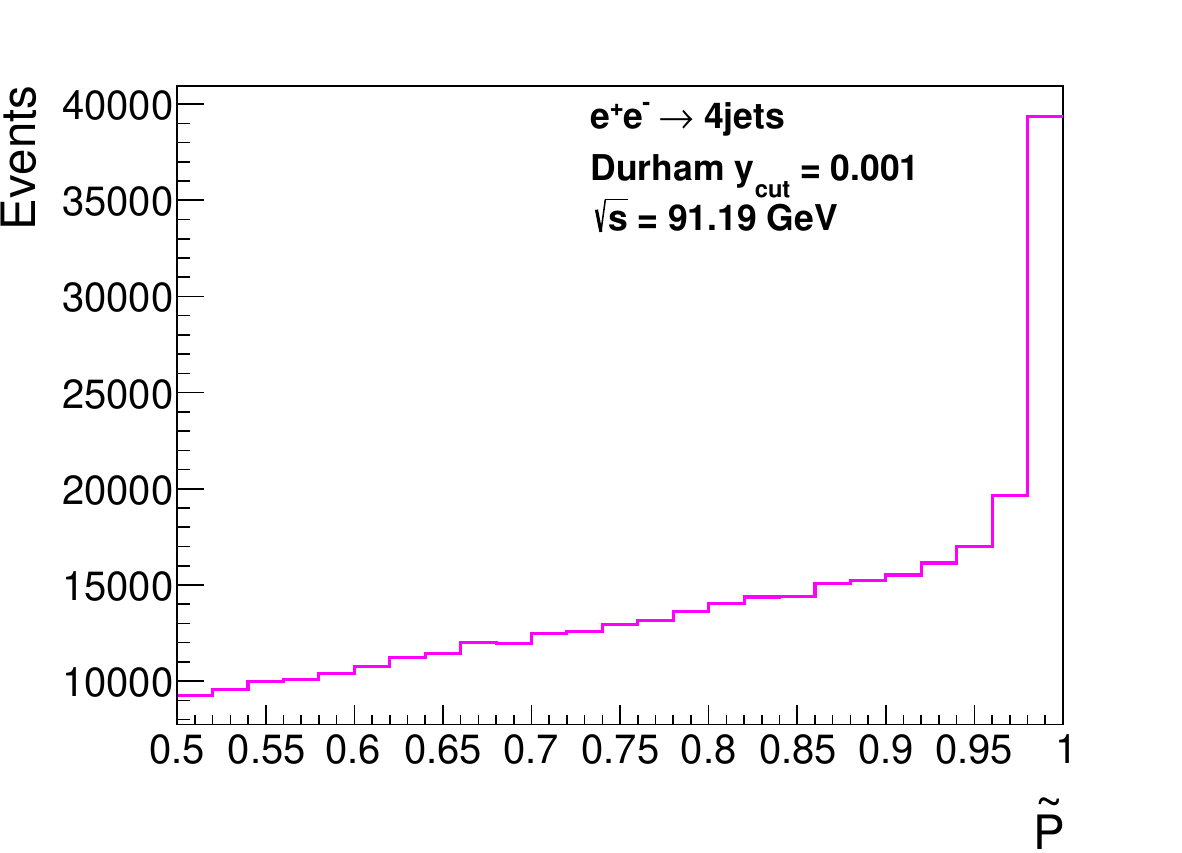}
\caption{The distribution of the event shape observable \(\tilde{P}\) in the inclusive 4-jet events at the center-of-mass energy $\sqrt{s}=m_Z$ for \(y_{\text{cut}} = 10^{-2}\) (left panel) and \(10^{-3}\) (right panel), based on PYTHIA 8.3~\cite{Bierlich:2022pfr}.}
\label{fig:P-4jets-lin-pythia}
\end{figure}

Large logarithmic divergences in the back-to-back fiducial regions limit the prediction power of fixed order QCD calculations. We use PYTHIA 8.3~\cite{Bierlich:2022pfr} to resum the leading log divergences for inclusive 3, 4-jet events to extend and exam our study. Figure~\ref{fig:lambda1-3jets-lin-pythia} and Figure~\ref{fig:lambda2-3jets-lin-pythia} show the distributions of \(\lambda_1\) and \(\lambda_2\) in inclusive 3-jet events for \(y_{\text{cut}} = 10^{-2}\) (left panel) and \(10^{-3}\) (right panel). Figure~\ref{fig:lambda3-3jets-lin-pythia} shows the distributions of \(\lambda_3\) in inclusive 3-jet events for \(y_{\text{cut}} = 10^{-2}\) (left panel) and \(10^{-3}\) (right panel). Figure~\ref{fig:P-3jets-lin-pythia} and Figure~\ref{fig:P-4jets-lin-pythia} show the distributions of \(\tilde{P}\) in inclusive 3, 4-jet events for \(y_{\text{cut}} = 10^{-2}\) (left panel) and \(10^{-3}\) (right panel). Figure~\ref{fig:lambda3-3jets-lin-pythia} shows the distributions of the Ridge correlation in inclusive 3-jet events (left panel) and inclusive 4-jet (right panel) for \(y_{\text{cut}} = 10^{-2}\). For the distributions of \(\lambda_1\), \(\lambda_2\), \(\lambda_3\) and $\tilde{P}$, we observe the same shape and peak region as in the above fixed order calculations. For the distributions of the Ridge correlation, we observe the ridge phenomenon both in inclusive 3-jet and 4-jet events. Thus, the parton shower does not change the picture that most of events have high planarity in both inclusive 3-jet and 4-jet events for $y_{\rm cut}=10^{-2} \sim 10^{-3} $.

\label{app:PYTHIAresults}

\begin{figure}[hbt!]
\includegraphics[width=0.49\textwidth]{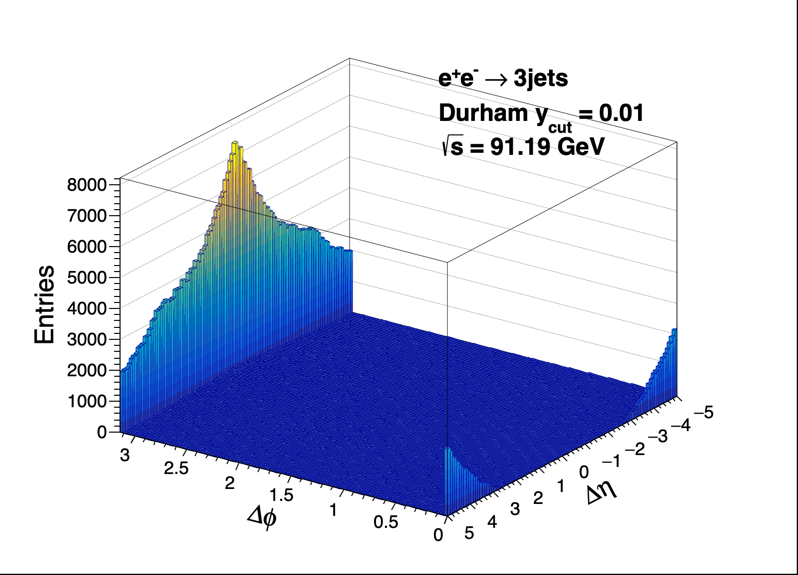}
\includegraphics[width=0.49\textwidth]{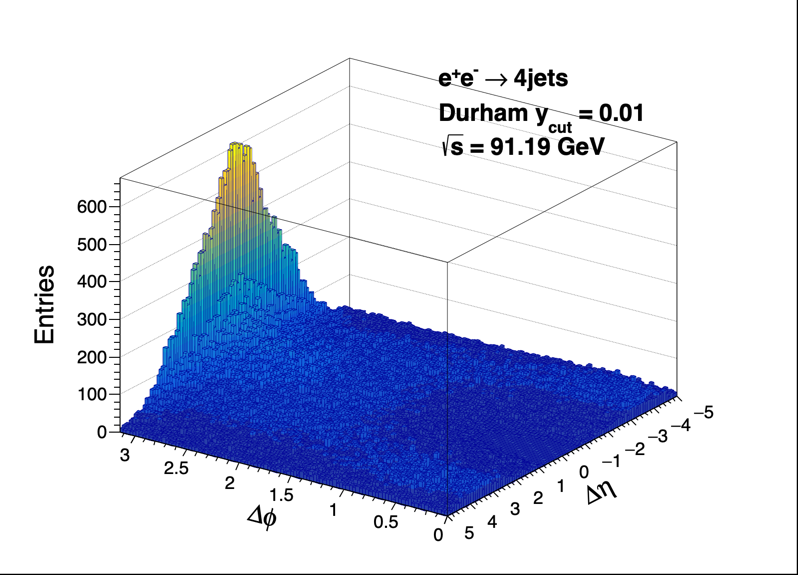}
\caption{The distributions of the Ridge correlation in inclusive 3-jet events (left panel) and inclusive 4-jet (right panel) at the center-of-mass energy $\sqrt{s}=m_Z$ for \(y_{\text{cut}} = 10^{-2}\) based on PYTHIA 8.3~\cite{Bierlich:2022pfr}.}
\label{fig:Ridge-jet-PYTHIA}
\end{figure}

\clearpage

\bibliographystyle{elsarticle-num}

\bibliography{ridge.bib}

\begin{thebibliography}{10}
\expandafter\ifx\csname url\endcsname\relax
  \def\url#1{\texttt{#1}}\fi
\expandafter\ifx\csname urlprefix\endcsname\relax\def\urlprefix{URL }\fi
\expandafter\ifx\csname href\endcsname\relax
  \def\href#1#2{#2} \def\path#1{#1}\fi

\bibitem{Adams:2005ph}
J.~Adams, et~al., {Distributions of charged hadrons associated with high transverse momentum particles in pp and AuAu collisions at $\sqrt{s} = 200~\rm{GeV}$}, Phys. Rev. Lett. 95 (2005) 152301.
\newblock \href {http://arxiv.org/abs/nucl-ex/0501016} {\path{arXiv:nucl-ex/0501016}}, \href {https://doi.org/10.1103/PhysRevLett.95.152301} {\path{doi:10.1103/PhysRevLett.95.152301}}.

\bibitem{Alver:2008gk}
B.~Alver, et~al., {System size dependence of cluster properties from two-particle angular correlations in CuCu and AuAu collisions at $\sqrt{s} = 200~\rm{GeV}$}, Phys. Rev. C 81 (2010) 024904.
\newblock \href {http://arxiv.org/abs/0812.1172} {\path{arXiv:0812.1172}}, \href {https://doi.org/10.1103/PhysRevC.81.024904} {\path{doi:10.1103/PhysRevC.81.024904}}.

\bibitem{Abelev:2009af}
B.~Abelev, et~al., {Long range rapidity correlations and jet production in high energy nuclear collisions}, Phys. Rev. C 80 (2009) 064912.
\newblock \href {http://arxiv.org/abs/0909.0191} {\path{arXiv:0909.0191}}, \href {https://doi.org/10.1103/PhysRevC.80.064912} {\path{doi:10.1103/PhysRevC.80.064912}}.

\bibitem{Alver:2009id}
B.~Alver, et~al., {High transverse momentum triggered correlations over a large pseudorapidity acceptance in AuAu collisions at $\sqrt{s}= 200~\rm{GeV}$}, Phys. Rev. Lett. 104 (2010) 062301.
\newblock \href {http://arxiv.org/abs/0903.2811} {\path{arXiv:0903.2811}}, \href {https://doi.org/10.1103/PhysRevLett.104.062301} {\path{doi:10.1103/PhysRevLett.104.062301}}.

\bibitem{Chatrchyan:2011eka}
{CMS Collaboration}, {Long-range and short-range dihadron angular correlations in central PbPb collisions at a nucleon-nucleon center of mass energy of $2.76~\rm{TeV}$}, JHEP 07 (2011) 076.
\newblock \href {http://arxiv.org/abs/1105.2438} {\path{arXiv:1105.2438}}, \href {https://doi.org/10.1007/JHEP07(2011)076} {\path{doi:10.1007/JHEP07(2011)076}}.

\bibitem{Chatrchyan:2012wg}
{CMS Collaboration}, {Centrality dependence of dihadron correlations and azimuthal anisotropy harmonics in PbPb collisions at $\sqrt{s_{NN}} = 2.76 ~\rm{TeV}$}, Eur. Phys. J. C 72 (2012) 2012.
\newblock \href {http://arxiv.org/abs/1201.3158} {\path{arXiv:1201.3158}}, \href {https://doi.org/10.1140/epjc/s10052-012-2012-3} {\path{doi:10.1140/epjc/s10052-012-2012-3}}.

\bibitem{Aamodt:2010pa}
{ALICE Collaboration}, {Elliptic flow of charged particles in PbPb collisions at $2.76 ~\rm{TeV}$}, Phys. Rev. Lett. 105 (2010) 252302.
\newblock \href {http://arxiv.org/abs/1011.3914} {\path{arXiv:1011.3914}}, \href {https://doi.org/10.1103/PhysRevLett.105.252302} {\path{doi:10.1103/PhysRevLett.105.252302}}.

\bibitem{ATLAS:2012at}
{ATLAS Collaboration}, {Measurement of the azimuthal anisotropy for charged particle production in $\sqrt{s} = 2.76~\rm{TeV}$ lead-lead collisions with the ATLAS detector}, Phys. Rev. C 86 (2012) 014907.
\newblock \href {http://arxiv.org/abs/1203.3087} {\path{arXiv:1203.3087}}, \href {https://doi.org/10.1103/PhysRevC.86.014907} {\path{doi:10.1103/PhysRevC.86.014907}}.

\bibitem{CMS:2013bza}
{CMS Collaboration}, {Studies of azimuthal dihadron correlations in ultra-central PbPb collisions at $\sqrt{s_{NN}} = 2.76~\rm{TeV}$}, JHEP 02 (2014) 088.
\newblock \href {http://arxiv.org/abs/1312.1845} {\path{arXiv:1312.1845}}, \href {https://doi.org/10.1007/JHEP02(2014)088} {\path{doi:10.1007/JHEP02(2014)088}}.

\bibitem{Khachatryan:2010gv}
{CMS Collaboration}, {Observation of long-range near-side angular correlations in proton-proton collisions at the LHC}, JHEP 09 (2010) 091.
\newblock \href {http://arxiv.org/abs/1009.4122} {\path{arXiv:1009.4122}}, \href {https://doi.org/10.1007/JHEP09(2010)091} {\path{doi:10.1007/JHEP09(2010)091}}.

\bibitem{Aad:2015gqa}
G.~Aad, et~al., {Observation of Long-Range Elliptic Azimuthal Anisotropies in $\sqrt{s}=$13 and 2.76 TeV $pp$ Collisions with the ATLAS Detector}, Phys. Rev. Lett. 116~(17) (2016) 172301.
\newblock \href {http://arxiv.org/abs/1509.04776} {\path{arXiv:1509.04776}}, \href {https://doi.org/10.1103/PhysRevLett.116.172301} {\path{doi:10.1103/PhysRevLett.116.172301}}.

\bibitem{Khachatryan:2015lva}
{CMS Collaboration}, {Measurement of long-range near-side two-particle angular correlations in pp collisions at $\sqrt{s} =13 ~\rm{TeV}$}, Phys. Rev. Lett. 116 (2016) 172302.
\newblock \href {http://arxiv.org/abs/1510.03068} {\path{arXiv:1510.03068}}, \href {https://doi.org/10.1103/PhysRevLett.116.172302} {\path{doi:10.1103/PhysRevLett.116.172302}}.

\bibitem{Khachatryan:2016txc}
{CMS Collaboration}, {Evidence for collectivity in pp collisions at the LHC}, Phys. Lett. B 765 (2017) 193.
\newblock \href {http://arxiv.org/abs/1606.06198} {\path{arXiv:1606.06198}}, \href {https://doi.org/10.1016/j.physletb.2016.12.009} {\path{doi:10.1016/j.physletb.2016.12.009}}.

\bibitem{CMS:2012qk}
{CMS Collaboration}, {Observation of long-range near-side angular correlations in proton-lead collisions at the LHC}, Phys. Lett. B 718 (2013) 795.
\newblock \href {http://arxiv.org/abs/1210.5482} {\path{arXiv:1210.5482}}, \href {https://doi.org/10.1016/j.physletb.2012.11.025} {\path{doi:10.1016/j.physletb.2012.11.025}}.

\bibitem{alice:2012qe}
{ALICE Collaboration}, {Long-range angular correlations on the near and away side in pPb collisions at $\sqrt{s}=5.02~\rm{TeV}$}, Phys. Lett. B 719 (2013) 29.
\newblock \href {http://arxiv.org/abs/1212.2001} {\path{arXiv:1212.2001}}, \href {https://doi.org/10.1016/j.physletb.2013.01.012} {\path{doi:10.1016/j.physletb.2013.01.012}}.

\bibitem{Aad:2012gla}
{ATLAS Collaboration}, {Observation of associated near-side and away-side long-range correlations in $\sqrt{s_{NN}}=5.02~\rm{TeV}$ proton-lead collisions with the ATLAS detector}, Phys. Rev. Lett. 110 (2013) 182302.
\newblock \href {http://arxiv.org/abs/1212.5198} {\path{arXiv:1212.5198}}, \href {https://doi.org/10.1103/PhysRevLett.110.182302} {\path{doi:10.1103/PhysRevLett.110.182302}}.

\bibitem{Aaij:2015qcq}
{LHCb Collaboration}, {Measurements of long-range near-side angular correlations in $\sqrt{s_{NN}}=5~\rm{TeV}$ proton-lead collisions in the forward region}, Phys. Lett. B 762 (2016) 473.
\newblock \href {http://arxiv.org/abs/1512.00439} {\path{arXiv:1512.00439}}, \href {https://doi.org/10.1016/j.physletb.2016.09.064} {\path{doi:10.1016/j.physletb.2016.09.064}}.

\bibitem{ABELEV:2013wsa}
{ALICE Collaboration}, {Long-range angular correlations of $\pi$, K and p in p-Pb collisions at $\sqrt{s_{NN}}=5.02~\rm{TeV}$}, Phys. Lett. B 726 (2013) 164.
\newblock \href {http://arxiv.org/abs/1307.3237} {\path{arXiv:1307.3237}}, \href {https://doi.org/10.1016/j.physletb.2013.08.024} {\path{doi:10.1016/j.physletb.2013.08.024}}.

\bibitem{Khachatryan:2014jra}
{CMS Collaboration}, {Long-range two-particle correlations of strange hadrons with charged particles in pPb and PbPb collisions at LHC energies}, Phys. Lett. B 742 (2015) 200.
\newblock \href {http://arxiv.org/abs/1409.3392} {\path{arXiv:1409.3392}}, \href {https://doi.org/10.1016/j.physletb.2015.01.034} {\path{doi:10.1016/j.physletb.2015.01.034}}.

\bibitem{Khachatryan:2015waa}
{CMS Collaboration}, {Evidence for collective multi-particle correlations in pPb collisions}, Phys. Rev. Lett. 115 (2015) 012301.
\newblock \href {http://arxiv.org/abs/1502.05382} {\path{arXiv:1502.05382}}, \href {https://doi.org/10.1103/PhysRevLett.115.012301} {\path{doi:10.1103/PhysRevLett.115.012301}}.

\bibitem{ATLAS:2017hap}
M.~Aaboud, et~al., {Measurement of multi-particle azimuthal correlations in $pp$, $p+$Pb and low-multiplicity Pb$+$Pb collisions with the ATLAS detector}, Eur. Phys. J. C 77~(6) (2017) 428.
\newblock \href {http://arxiv.org/abs/1705.04176} {\path{arXiv:1705.04176}}, \href {https://doi.org/10.1140/epjc/s10052-017-4988-1} {\path{doi:10.1140/epjc/s10052-017-4988-1}}.

\bibitem{ATLAS:2017rtr}
M.~Aaboud, et~al., {Measurement of long-range multiparticle azimuthal correlations with the subevent cumulant method in $pp$ and $p + Pb$ collisions with the ATLAS detector at the CERN Large Hadron Collider}, Phys. Rev. C 97~(2) (2018) 024904.
\newblock \href {http://arxiv.org/abs/1708.03559} {\path{arXiv:1708.03559}}, \href {https://doi.org/10.1103/PhysRevC.97.024904} {\path{doi:10.1103/PhysRevC.97.024904}}.

\bibitem{PHENIX:2018lia}
C.~Aidala, et~al., {Creation of quark\textendash{}gluon plasma droplets with three distinct geometries}, Nature Phys. 15~(3) (2019) 214--220.
\newblock \href {http://arxiv.org/abs/1805.02973} {\path{arXiv:1805.02973}}, \href {https://doi.org/10.1038/s41567-018-0360-0} {\path{doi:10.1038/s41567-018-0360-0}}.

\bibitem{Adamczyk:2015xjc}
L.~Adamczyk, et~al., {Long-range pseudorapidity dihadron correlations in $d$Au collisions at $\sqrt{s}=200~\rm{GeV}$}, Phys. Lett. B 747 (2015) 265.
\newblock \href {http://arxiv.org/abs/1502.07652} {\path{arXiv:1502.07652}}, \href {https://doi.org/10.1016/j.physletb.2015.05.075} {\path{doi:10.1016/j.physletb.2015.05.075}}.

\bibitem{Adare:2015ctn}
A.~Adare, et~al., {Measurements of elliptic and triangular flow in high-multiplicity $^{3}$HeAu collisions at $\sqrt{s}=200~\rm{GeV}$}, Phys. Rev. Lett. 115 (2015) 142301.
\newblock \href {http://arxiv.org/abs/1507.06273} {\path{arXiv:1507.06273}}, \href {https://doi.org/10.1103/PhysRevLett.115.142301} {\path{doi:10.1103/PhysRevLett.115.142301}}.

\bibitem{Aidala:2017ajz}
C.~Aidala, et~al., {Measurements of Multiparticle Correlations in $d+\mathrm{Au}$ Collisions at 200, 62.4, 39, and 19.6 GeV and $p+\mathrm{Au}$ Collisions at 200 GeV and Implications for Collective Behavior}, Phys. Rev. Lett. 120~(6) (2018) 062302.
\newblock \href {http://arxiv.org/abs/1707.06108} {\path{arXiv:1707.06108}}, \href {https://doi.org/10.1103/PhysRevLett.120.062302} {\path{doi:10.1103/PhysRevLett.120.062302}}.

\bibitem{Badea:2019vey}
A.~Badea, A.~Baty, P.~Chang, G.~M. Innocenti, M.~Maggi, C.~Mcginn, M.~Peters, T.-A. Sheng, J.~Thaler, Y.-J. Lee, {Measurements of two-particle correlations in $e^+e^-$ collisions at 91 GeV with ALEPH archived data}, Phys. Rev. Lett. 123~(21) (2019) 212002.
\newblock \href {http://arxiv.org/abs/1906.00489} {\path{arXiv:1906.00489}}, \href {https://doi.org/10.1103/PhysRevLett.123.212002} {\path{doi:10.1103/PhysRevLett.123.212002}}.

\bibitem{Belle:2022fvl}
Y.~C. Chen, et~al., {Measurement of Two-Particle Correlations of Hadrons in $e^{+}e^{-}$ Collisions at Belle}, Phys. Rev. Lett. 128~(14) (2022) 142005.
\newblock \href {http://arxiv.org/abs/2201.01694} {\path{arXiv:2201.01694}}, \href {https://doi.org/10.1103/PhysRevLett.128.142005} {\path{doi:10.1103/PhysRevLett.128.142005}}.

\bibitem{ZEUS:2019jya}
I.~Abt, et~al., {Two-particle azimuthal correlations as a probe of collective behaviour in deep inelastic $ep$ scattering at HERA}, JHEP 04 (2020) 070.
\newblock \href {http://arxiv.org/abs/1912.07431} {\path{arXiv:1912.07431}}, \href {https://doi.org/10.1007/JHEP04(2020)070} {\path{doi:10.1007/JHEP04(2020)070}}.

\bibitem{ZEUS:2021qzg}
I.~Abt, et~al., {Azimuthal correlations in photoproduction and deep inelastic $ep$ scattering at HERA}, JHEP 12 (2021) 102.
\newblock \href {http://arxiv.org/abs/2106.12377} {\path{arXiv:2106.12377}}, \href {https://doi.org/10.1007/JHEP12(2021)102} {\path{doi:10.1007/JHEP12(2021)102}}.

\bibitem{CMS:2022doq}
A.~Tumasyan, et~al., {Two-particle azimuthal correlations in \ensuremath{\gamma}p interactions using pPb collisions at $\sqrt{s_{NN}}$=8.16 TeV}, Phys. Lett. B 844 (2023) 137905.
\newblock \href {http://arxiv.org/abs/2204.13486} {\path{arXiv:2204.13486}}, \href {https://doi.org/10.1016/j.physletb.2023.137905} {\path{doi:10.1016/j.physletb.2023.137905}}.

\bibitem{ATLAS:2021jhn}
{ATLAS Collaboration}, {Two-particle azimuthal correlations in photonuclear ultraperipheral PbPb collisions at $5.02 ~\rm{TeV}$ with ATLAS}, Phys. Rev. C 104 (2021) 014903.
\newblock \href {http://arxiv.org/abs/2101.10771} {\path{arXiv:2101.10771}}, \href {https://doi.org/10.1103/PhysRevC.104.014903} {\path{doi:10.1103/PhysRevC.104.014903}}.

\bibitem{Baty:2021ugw}
A.~Baty, P.~Gardner, W.~Li, {Novel observables for exploring QCD collective evolution and quantum entanglement within individual jets}, Phys. Rev. C 107~(6) (2023) 064908.
\newblock \href {http://arxiv.org/abs/2104.11735} {\path{arXiv:2104.11735}}, \href {https://doi.org/10.1103/PhysRevC.107.064908} {\path{doi:10.1103/PhysRevC.107.064908}}.

\bibitem{CMS:2023iam}
A.~Hayrapetyan, et~al., {Observation of enhanced long-range elliptic anisotropies inside high-multiplicity jets in pp collisions at $\sqrt{s} = 13$ TeV} (12 2023).
\newblock \href {http://arxiv.org/abs/2312.17103} {\path{arXiv:2312.17103}}.

\bibitem{Chen:2023njr}
Y.-C. Chen, et~al., {Long-range near-side correlation in e+e\ensuremath{-} collisions at 183-209 GeV with ALEPH archived data}, Phys. Lett. B 856 (2024) 138957.
\newblock \href {http://arxiv.org/abs/2312.05084} {\path{arXiv:2312.05084}}, \href {https://doi.org/10.1016/j.physletb.2024.138957} {\path{doi:10.1016/j.physletb.2024.138957}}.

\bibitem{Vertesi:2024fwl}
R.~Vertesi, A.~Ortiz, {Thermodynamical string fragmentation and QGP-like effects in jets} (8 2024).
\newblock \href {http://arxiv.org/abs/2408.06340} {\path{arXiv:2408.06340}}.

\bibitem{Zhao:2024wqs}
W.~Zhao, Z.-W. Lin, X.-N. Wang, {Collectivity inside high-multiplicity jets in high-energy proton-proton collisions} (1 2024).
\newblock \href {http://arxiv.org/abs/2401.13137} {\path{arXiv:2401.13137}}.

\bibitem{Zheng:2024xyv}
L.~Zheng, L.~Liu, Z.-W. Lin, Q.-Y. Shou, Z.-B. Yin, {Disentangling the development of collective flow in high energy proton proton collisions with a multiphase transport model}, Eur. Phys. J. C 84~(10) (2024) 1029.
\newblock \href {http://arxiv.org/abs/2404.18829} {\path{arXiv:2404.18829}}, \href {https://doi.org/10.1140/epjc/s10052-024-13378-1} {\path{doi:10.1140/epjc/s10052-024-13378-1}}.

\bibitem{Nagle:2013lja}
J.~L. Nagle, A.~Adare, S.~Beckman, T.~Koblesky, J.~Orjuela~Koop, D.~McGlinchey, P.~Romatschke, J.~Carlson, J.~E. Lynn, M.~McCumber, {Exploiting Intrinsic Triangular Geometry in Relativistic He3+Au Collisions to Disentangle Medium Properties}, Phys. Rev. Lett. 113~(11) (2014) 112301.
\newblock \href {http://arxiv.org/abs/1312.4565} {\path{arXiv:1312.4565}}, \href {https://doi.org/10.1103/PhysRevLett.113.112301} {\path{doi:10.1103/PhysRevLett.113.112301}}.

\bibitem{Schenke:2014zha}
B.~Schenke, R.~Venugopalan, {Eccentric protons? Sensitivity of flow to system size and shape in p+p, p+Pb and Pb+Pb collisions}, Phys. Rev. Lett. 113 (2014) 102301.
\newblock \href {http://arxiv.org/abs/1405.3605} {\path{arXiv:1405.3605}}, \href {https://doi.org/10.1103/PhysRevLett.113.102301} {\path{doi:10.1103/PhysRevLett.113.102301}}.

\bibitem{Shen:2016zpp}
C.~Shen, J.-F. Paquet, G.~S. Denicol, S.~Jeon, C.~Gale, {Collectivity and electromagnetic radiation in small systems}, Phys. Rev. C 95~(1) (2017) 014906.
\newblock \href {http://arxiv.org/abs/1609.02590} {\path{arXiv:1609.02590}}, \href {https://doi.org/10.1103/PhysRevC.95.014906} {\path{doi:10.1103/PhysRevC.95.014906}}.

\bibitem{Weller:2017tsr}
R.~D. Weller, P.~Romatschke, {One fluid to rule them all: viscous hydrodynamic description of event-by-event central p+p, p+Pb and Pb+Pb collisions at $\sqrt{s}=5.02$ TeV}, Phys. Lett. B 774 (2017) 351--356.
\newblock \href {http://arxiv.org/abs/1701.07145} {\path{arXiv:1701.07145}}, \href {https://doi.org/10.1016/j.physletb.2017.09.077} {\path{doi:10.1016/j.physletb.2017.09.077}}.

\bibitem{Mantysaari:2017cni}
H.~M\"antysaari, B.~Schenke, C.~Shen, P.~Tribedy, {Imprints of fluctuating proton shapes on flow in proton-lead collisions at the LHC}, Phys. Lett. B 772 (2017) 681--686.
\newblock \href {http://arxiv.org/abs/1705.03177} {\path{arXiv:1705.03177}}, \href {https://doi.org/10.1016/j.physletb.2017.07.038} {\path{doi:10.1016/j.physletb.2017.07.038}}.

\bibitem{Zhao:2017rgg}
W.~Zhao, Y.~Zhou, H.~Xu, W.~Deng, H.~Song, {Hydrodynamic collectivity in proton\textendash{}proton collisions at $13$ TeV}, Phys. Lett. B 780 (2018) 495--500.
\newblock \href {http://arxiv.org/abs/1801.00271} {\path{arXiv:1801.00271}}, \href {https://doi.org/10.1016/j.physletb.2018.03.022} {\path{doi:10.1016/j.physletb.2018.03.022}}.

\bibitem{Bierlich:2019wld}
C.~Bierlich, C.~O. Rasmussen, {Dipole evolution: perspectives for collectivity and $\gamma^*$A collisions}, JHEP 10 (2019) 026.
\newblock \href {http://arxiv.org/abs/1907.12871} {\path{arXiv:1907.12871}}, \href {https://doi.org/10.1007/JHEP10(2019)026} {\path{doi:10.1007/JHEP10(2019)026}}.

\bibitem{Katz:2019qwv}
R.~Katz, C.~A.~G. Prado, J.~Noronha-Hostler, A.~A.~P. Suaide, {System-size scan of $D$ meson $R_{AA}$ and $v_n$ using PbPb, XeXe, ArAr, and OO collisions at energies available at the CERN Large Hadron Collider}, Phys. Rev. C 102~(4) (2020) 041901.
\newblock \href {http://arxiv.org/abs/1907.03308} {\path{arXiv:1907.03308}}, \href {https://doi.org/10.1103/PhysRevC.102.041901} {\path{doi:10.1103/PhysRevC.102.041901}}.

\bibitem{Zhao:2022ayk}
W.~Zhao, C.~Shen, B.~Schenke, {Collectivity in Ultraperipheral Pb+Pb Collisions at the Large Hadron Collider}, Phys. Rev. Lett. 129~(25) (2022) 252302.
\newblock \href {http://arxiv.org/abs/2203.06094} {\path{arXiv:2203.06094}}, \href {https://doi.org/10.1103/PhysRevLett.129.252302} {\path{doi:10.1103/PhysRevLett.129.252302}}.

\bibitem{Zhao:2022ugy}
W.~Zhao, S.~Ryu, C.~Shen, B.~Schenke, {3D structure of anisotropic flow in small collision systems at energies available at the BNL Relativistic Heavy Ion Collider}, Phys. Rev. C 107~(1) (2023) 014904.
\newblock \href {http://arxiv.org/abs/2211.16376} {\path{arXiv:2211.16376}}, \href {https://doi.org/10.1103/PhysRevC.107.014904} {\path{doi:10.1103/PhysRevC.107.014904}}.

\bibitem{Bzdak:2014dia}
A.~Bzdak, G.-L. Ma, {Elliptic and triangular flow in $p$+Pb and peripheral Pb+Pb collisions from parton scatterings}, Phys. Rev. Lett. 113~(25) (2014) 252301.
\newblock \href {http://arxiv.org/abs/1406.2804} {\path{arXiv:1406.2804}}, \href {https://doi.org/10.1103/PhysRevLett.113.252301} {\path{doi:10.1103/PhysRevLett.113.252301}}.

\bibitem{He:2015hfa}
L.~He, T.~Edmonds, Z.-W. Lin, F.~Liu, D.~Molnar, F.~Wang, {Anisotropic parton escape is the dominant source of azimuthal anisotropy in transport models}, Phys. Lett. B 753 (2016) 506--510.
\newblock \href {http://arxiv.org/abs/1502.05572} {\path{arXiv:1502.05572}}, \href {https://doi.org/10.1016/j.physletb.2015.12.051} {\path{doi:10.1016/j.physletb.2015.12.051}}.

\bibitem{Lin:2015ucn}
Z.-W. Lin, L.~He, T.~Edmonds, F.~Liu, D.~Molnar, F.~Wang, {Elliptic Anisotropy $v_2$ May Be Dominated by Particle Escape instead of Hydrodynamic Flow}, Nucl. Phys. A 956 (2016) 316--319.
\newblock \href {http://arxiv.org/abs/1512.06465} {\path{arXiv:1512.06465}}, \href {https://doi.org/10.1016/j.nuclphysa.2016.01.017} {\path{doi:10.1016/j.nuclphysa.2016.01.017}}.

\bibitem{Nagle:2017sjv}
J.~L. Nagle, R.~Belmont, K.~Hill, J.~Orjuela~Koop, D.~V. Perepelitsa, P.~Yin, Z.-W. Lin, D.~McGlinchey, {Minimal conditions for collectivity in $e^+e^-$ and $p+p$ collisions}, Phys. Rev. C 97~(2) (2018) 024909.
\newblock \href {http://arxiv.org/abs/1707.02307} {\path{arXiv:1707.02307}}, \href {https://doi.org/10.1103/PhysRevC.97.024909} {\path{doi:10.1103/PhysRevC.97.024909}}.

\bibitem{Kurkela:2018qeb}
A.~Kurkela, U.~A. Wiedemann, B.~Wu, {Opacity dependence of elliptic flow in kinetic theory}, Eur. Phys. J. C 79~(9) (2019) 759.
\newblock \href {http://arxiv.org/abs/1805.04081} {\path{arXiv:1805.04081}}, \href {https://doi.org/10.1140/epjc/s10052-019-7262-x} {\path{doi:10.1140/epjc/s10052-019-7262-x}}.

\bibitem{Kurkela:2019kip}
A.~Kurkela, U.~A. Wiedemann, B.~Wu, {Flow in AA and pA as an interplay of fluid-like and non-fluid like excitations}, Eur. Phys. J. C 79~(11) (2019) 965.
\newblock \href {http://arxiv.org/abs/1905.05139} {\path{arXiv:1905.05139}}, \href {https://doi.org/10.1140/epjc/s10052-019-7428-6} {\path{doi:10.1140/epjc/s10052-019-7428-6}}.

\bibitem{Zhao:2021bef}
X.-L. Zhao, Z.-W. Lin, L.~Zheng, G.-L. Ma, {A transport model study of multiparticle cumulants in p+p collisions at $13$ TeV}, Phys. Lett. B 839 (2023) 137799.
\newblock \href {http://arxiv.org/abs/2112.01232} {\path{arXiv:2112.01232}}, \href {https://doi.org/10.1016/j.physletb.2023.137799} {\path{doi:10.1016/j.physletb.2023.137799}}.

\bibitem{Schenke:2019pmk}
B.~Schenke, C.~Shen, P.~Tribedy, {Hybrid Color Glass Condensate and hydrodynamic description of the Relativistic Heavy Ion Collider small system scan}, Phys. Lett. B 803 (2020) 135322.
\newblock \href {http://arxiv.org/abs/1908.06212} {\path{arXiv:1908.06212}}, \href {https://doi.org/10.1016/j.physletb.2020.135322} {\path{doi:10.1016/j.physletb.2020.135322}}.

\bibitem{Giacalone:2020byk}
G.~Giacalone, B.~Schenke, C.~Shen, {Observable signatures of initial state momentum anisotropies in nuclear collisions}, Phys. Rev. Lett. 125~(19) (2020) 192301.
\newblock \href {http://arxiv.org/abs/2006.15721} {\path{arXiv:2006.15721}}, \href {https://doi.org/10.1103/PhysRevLett.125.192301} {\path{doi:10.1103/PhysRevLett.125.192301}}.

\bibitem{Schenke:2021mxx}
B.~Schenke, {The smallest fluid on Earth}, Rept. Prog. Phys. 84~(8) (2021) 082301.
\newblock \href {http://arxiv.org/abs/2102.11189} {\path{arXiv:2102.11189}}, \href {https://doi.org/10.1088/1361-6633/ac14c9} {\path{doi:10.1088/1361-6633/ac14c9}}.

\bibitem{Grosse-Oetringhaus:2024bwr}
J.~F. Grosse-Oetringhaus, U.~A. Wiedemann, {A Decade of Collectivity in Small Systems} (7 2024).
\newblock \href {http://arxiv.org/abs/2407.07484} {\path{arXiv:2407.07484}}.

\bibitem{ppt}
S.-Y. Li, \href{https://moriond.in2p3.fr/QCD/2024/SaturdayAfternoon/Li.pdf}{Near-sided long-pseudorapidity-range $\phi$ correlation (ridge) of planar events via hard radiations in $e^+e^-$ annihilation, moriond qcd2024 - proceedings} (2024).
\newline\urlprefix\url{https://moriond.in2p3.fr/QCD/2024/SaturdayAfternoon/Li.pdf}

\bibitem{Ellis:1976uc}
J.~R. Ellis, M.~K. Gaillard, G.~G. Ross, {Search for Gluons in $e^+e^-$ Annihilation}, Nucl. Phys. B 111 (1976) 253, [Erratum: Nucl.Phys.B 130, 516 (1977)].
\newblock \href {https://doi.org/10.1016/0550-3213(77)90253-X} {\path{doi:10.1016/0550-3213(77)90253-X}}.

\bibitem{Wu:2015yoa}
S.~L. Wu, {Discovery of the Gluon}, World Scientific Publising Company, 2015, pp. 199--226.
\newblock \href {https://doi.org/10.1142/9789814618113_0013} {\path{doi:10.1142/9789814618113_0013}}.

\bibitem{SJOSTRAND2015159}
T.~Sjöstrand, S.~Ask, J.~R. Christiansen, R.~Corke, N.~Desai, P.~Ilten, S.~Mrenna, S.~Prestel, C.~O. Rasmussen, P.~Z. Skands, \href{https://www.sciencedirect.com/science/article/pii/S0010465515000442}{An introduction to pythia 8.2}, Computer Physics Communications 191 (2015) 159--177.
\newblock \href {http://arxiv.org/abs/1410.3012} {\path{arXiv:1410.3012}}, \href {https://doi.org/https://doi.org/10.1016/j.cpc.2015.01.024} {\path{doi:https://doi.org/10.1016/j.cpc.2015.01.024}}.
\newline\urlprefix\url{https://www.sciencedirect.com/science/article/pii/S0010465515000442}

\bibitem{ATLAS:2012tch}
G.~Aad, et~al., {Measurement of event shapes at large momentum transfer with the ATLAS detector in $pp$ collisions at $\sqrt{s}=7$ TeV}, Eur. Phys. J. C 72 (2012) 2211.
\newblock \href {http://arxiv.org/abs/1206.2135} {\path{arXiv:1206.2135}}, \href {https://doi.org/10.1140/epjc/s10052-012-2211-y} {\path{doi:10.1140/epjc/s10052-012-2211-y}}.

\bibitem{Czakon:2021mjy}
M.~Czakon, A.~Mitov, R.~Poncelet, {Next-to-Next-to-Leading Order Study of Three-Jet Production at the LHC}, Phys. Rev. Lett. 127~(15) (2021) 152001, [Erratum: Phys.Rev.Lett. 129, 119901 (2022)].
\newblock \href {http://arxiv.org/abs/2106.05331} {\path{arXiv:2106.05331}}, \href {https://doi.org/10.1103/PhysRevLett.127.152001} {\path{doi:10.1103/PhysRevLett.127.152001}}.

\bibitem{Chen:2022ktf}
X.~Chen, T.~Gehrmann, E.~W.~N. Glover, A.~Huss, M.~Marcoli, {Automation of antenna subtraction in colour space: gluonic processes}, JHEP 10 (2022) 099.
\newblock \href {http://arxiv.org/abs/2203.13531} {\path{arXiv:2203.13531}}, \href {https://doi.org/10.1007/JHEP10(2022)099} {\path{doi:10.1007/JHEP10(2022)099}}.

\bibitem{Gehrmann:2023dxm}
T.~Gehrmann, E.~W.~N. Glover, M.~Marcoli, {The colourful antenna subtraction method}, JHEP 03 (2024) 114.
\newblock \href {http://arxiv.org/abs/2310.19757} {\path{arXiv:2310.19757}}, \href {https://doi.org/10.1007/JHEP03(2024)114} {\path{doi:10.1007/JHEP03(2024)114}}.

\bibitem{Gehrmann-DeRidder:2007nzq}
A.~Gehrmann-De~Ridder, T.~Gehrmann, E.~W.~N. Glover, G.~Heinrich, {Second-order QCD corrections to the thrust distribution}, Phys. Rev. Lett. 99 (2007) 132002.
\newblock \href {http://arxiv.org/abs/0707.1285} {\path{arXiv:0707.1285}}, \href {https://doi.org/10.1103/PhysRevLett.99.132002} {\path{doi:10.1103/PhysRevLett.99.132002}}.

\bibitem{Gehrmann-DeRidder:2007vsv}
A.~Gehrmann-De~Ridder, T.~Gehrmann, E.~W.~N. Glover, G.~Heinrich, {NNLO corrections to event shapes in $e^+ e^-$ annihilation}, JHEP 12 (2007) 094.
\newblock \href {http://arxiv.org/abs/0711.4711} {\path{arXiv:0711.4711}}, \href {https://doi.org/10.1088/1126-6708/2007/12/094} {\path{doi:10.1088/1126-6708/2007/12/094}}.

\bibitem{Gehrmann-DeRidder:2008qsl}
A.~Gehrmann-De~Ridder, T.~Gehrmann, E.~W.~N. Glover, G.~Heinrich, {Jet rates in electron-positron annihilation at O($\alpha_{s}^{3}$) in QCD}, Phys. Rev. Lett. 100 (2008) 172001.
\newblock \href {http://arxiv.org/abs/0802.0813} {\path{arXiv:0802.0813}}, \href {https://doi.org/10.1103/PhysRevLett.100.172001} {\path{doi:10.1103/PhysRevLett.100.172001}}.

\bibitem{Weinzierl_2008}
S.~Weinzierl, \href{http://dx.doi.org/10.1103/PhysRevLett.101.162001}{Next-to-next-to-leading order corrections to three-jet observables in electron-positron annihilation}, Physical Review Letters 101~(16) (Oct. 2008).
\newblock \href {http://arxiv.org/abs/0807.3241} {\path{arXiv:0807.3241}}, \href {https://doi.org/10.1103/physrevlett.101.162001} {\path{doi:10.1103/physrevlett.101.162001}}.
\newline\urlprefix\url{http://dx.doi.org/10.1103/PhysRevLett.101.162001}

\bibitem{Gehrmann-DeRidder:2009fgd}
A.~Gehrmann-De~Ridder, T.~Gehrmann, E.~W.~N. Glover, G.~Heinrich, {NNLO moments of event shapes in $e^{+}e^{-}$ annihilation}, JHEP 05 (2009) 106.
\newblock \href {http://arxiv.org/abs/0903.4658} {\path{arXiv:0903.4658}}, \href {https://doi.org/10.1088/1126-6708/2009/05/106} {\path{doi:10.1088/1126-6708/2009/05/106}}.

\bibitem{Weinzierl:2009ms}
S.~Weinzierl, {Event shapes and jet rates in electron-positron annihilation at NNLO}, JHEP 06 (2009) 041.
\newblock \href {http://arxiv.org/abs/0904.1077} {\path{arXiv:0904.1077}}, \href {https://doi.org/10.1088/1126-6708/2009/06/041} {\path{doi:10.1088/1126-6708/2009/06/041}}.

\bibitem{Gehrmann-DeRidder:2014hxk}
A.~Gehrmann-De~Ridder, T.~Gehrmann, E.~W.~N. Glover, G.~Heinrich, {EERAD3: Event shapes and jet rates in electron-positron annihilation at order $\alpha_s^3$}, Comput. Phys. Commun. 185 (2014) 3331.
\newblock \href {http://arxiv.org/abs/1402.4140} {\path{arXiv:1402.4140}}, \href {https://doi.org/10.1016/j.cpc.2014.07.024} {\path{doi:10.1016/j.cpc.2014.07.024}}.

\bibitem{DelDuca2016ThreeJetPI}
V.~D. Duca, C.~Duhr, A.~Kardos, G.~Somogyi, Z.~L. Trocsanyi, \href{https://api.semanticscholar.org/CorpusID:33336336}{Three-jet production in electron-positron collisions at next-to-next-to-leading order accuracy}, Physical review letters 117 15 (2016) 152004.
\newblock \href {http://arxiv.org/abs/1603.08927} {\path{arXiv:1603.08927}}, \href {https://doi.org/10.1103/PhysRevLett.117.152004} {\path{doi:10.1103/PhysRevLett.117.152004}}.
\newline\urlprefix\url{https://api.semanticscholar.org/CorpusID:33336336}

\bibitem{Gehrmann:2017xfb}
T.~Gehrmann, E.~W.~N. Glover, A.~Huss, J.~Niehues, H.~Zhang, {NNLO QCD corrections to event orientation in $e^+ e^-$ annihilation}, Phys. Lett. B 775 (2017) 185--189.
\newblock \href {http://arxiv.org/abs/1709.01097} {\path{arXiv:1709.01097}}, \href {https://doi.org/10.1016/j.physletb.2017.10.069} {\path{doi:10.1016/j.physletb.2017.10.069}}.

\bibitem{Huss:2025iov}
A.~Huss, et~al., {NNLOJET: a parton-level event generator for jet cross sections at NNLO QCD accuracy} (3 2025).
\newblock \href {http://arxiv.org/abs/2503.22804} {\path{arXiv:2503.22804}}.

\bibitem{GehrmannDeRidder:2005cm}
A.~Gehrmann-De~Ridder, T.~Gehrmann, E.~W.~N. Glover, {Antenna subtraction at NNLO}, JHEP 09 (2005) 056.
\newblock \href {http://arxiv.org/abs/hep-ph/0505111} {\path{arXiv:hep-ph/0505111}}.

\bibitem{Gehrmann-DeRidder:2005svg}
A.~Gehrmann-De~Ridder, T.~Gehrmann, E.~W.~N. Glover, {Quark-gluon antenna functions from neutralino decay}, Phys. Lett. B 612 (2005) 36--48.
\newblock \href {http://arxiv.org/abs/hep-ph/0501291} {\path{arXiv:hep-ph/0501291}}, \href {https://doi.org/10.1016/j.physletb.2005.02.039} {\path{doi:10.1016/j.physletb.2005.02.039}}.

\bibitem{Daleo:2006xa}
A.~Daleo, T.~Gehrmann, D.~Maitre, {Antenna subtraction with hadronic initial states}, JHEP 04 (2007) 016.
\newblock \href {http://arxiv.org/abs/hep-ph/0612257} {\path{arXiv:hep-ph/0612257}}, \href {https://doi.org/10.1088/1126-6708/2007/04/016} {\path{doi:10.1088/1126-6708/2007/04/016}}.

\bibitem{Currie:2013vh}
J.~Currie, E.~W.~N. Glover, S.~Wells, {Infrared Structure at NNLO Using Antenna Subtraction}, JHEP 04 (2013) 066.
\newblock \href {http://arxiv.org/abs/1301.4693} {\path{arXiv:1301.4693}}, \href {https://doi.org/10.1007/JHEP04(2013)066} {\path{doi:10.1007/JHEP04(2013)066}}.

\bibitem{Lepage:1977sw}
G.~P. Lepage, {A New Algorithm for Adaptive Multidimensional Integration}, J. Comput. Phys. 27 (1978) 192.
\newblock \href {https://doi.org/10.1016/0021-9991(78)90004-9} {\path{doi:10.1016/0021-9991(78)90004-9}}.

\bibitem{Fox:2024bfp}
E.~Fox, N.~Glover, M.~Marcoli, {Generalised antenna functions for higher-order calculations}, JHEP 12 (2025) 225.
\newblock \href {http://arxiv.org/abs/2410.12904} {\path{arXiv:2410.12904}}, \href {https://doi.org/10.1007/JHEP12(2024)225} {\path{doi:10.1007/JHEP12(2024)225}}.

\bibitem{Bjorken:1969wi}
J.~D. Bjorken, S.~J. Brodsky, {Statistical Model for electron-Positron Annihilation Into Hadrons}, Phys. Rev. D 1 (1970) 1416--1420.
\newblock \href {https://doi.org/10.1103/PhysRevD.1.1416} {\path{doi:10.1103/PhysRevD.1.1416}}.

\bibitem{Andersson1983PartonFA}
B.~Andersson, G.~Gustafson, G.~Ingelman, T.~Sj{\"o}strand, \href{https://api.semanticscholar.org/CorpusID:121092649}{Parton fragmentation and string dynamics}, Physics Reports 97 (1983) 31--145.
\newblock \href {http://arxiv.org/abs/1410.3012} {\path{arXiv:1410.3012}}.
\newline\urlprefix\url{https://api.semanticscholar.org/CorpusID:121092649}

\bibitem{fan1989new}
S.~Fan, A new extracting formula and a new distinguishing means on the one variable cubic equation, Natural Science Journal of Hainan Teachers College 2~(2) (1989) 91--98.

\bibitem{golub2013matrix}
G.~H. Golub, C.~F. Van~Loan, Matrix Computations, 4th Edition, Johns Hopkins University Press, Baltimore, MD, 2013.

\bibitem{Basham:1978bw}
C.~L. Basham, L.~S. Brown, S.~D. Ellis, S.~T. Love, {Energy Correlations in electron - Positron Annihilation: Testing QCD}, Phys. Rev. Lett. 41 (1978) 1585.
\newblock \href {https://doi.org/10.1103/PhysRevLett.41.1585} {\path{doi:10.1103/PhysRevLett.41.1585}}.

\bibitem{Basham:1978zq}
C.~L. Basham, L.~S. Brown, S.~D. Ellis, S.~T. Love, {Energy Correlations in electron-Positron Annihilation in Quantum Chromodynamics: Asymptotically Free Perturbation Theory}, Phys. Rev. D 19 (1979) 2018.
\newblock \href {https://doi.org/10.1103/PhysRevD.19.2018} {\path{doi:10.1103/PhysRevD.19.2018}}.

\bibitem{Buccioni:2019sur}
F.~Buccioni, J.-N. Lang, J.~M. Lindert, P.~Maierh\"ofer, S.~Pozzorini, H.~Zhang, M.~F. Zoller, {OpenLoops 2}, Eur. Phys. J. C 79~(10) (2019) 866.
\newblock \href {http://arxiv.org/abs/1907.13071} {\path{arXiv:1907.13071}}, \href {https://doi.org/10.1140/epjc/s10052-019-7306-2} {\path{doi:10.1140/epjc/s10052-019-7306-2}}.

\bibitem{WJStirling_1991}
W.~J. Stirling, \href{https://dx.doi.org/10.1088/0954-3899/17/10/014}{Hard qcd working group-theory summary}, Journal of Physics G: Nuclear and Particle Physics 17~(10) (1991) 1567.
\newblock \href {https://doi.org/10.1088/0954-3899/17/10/014} {\path{doi:10.1088/0954-3899/17/10/014}}.
\newline\urlprefix\url{https://dx.doi.org/10.1088/0954-3899/17/10/014}

\bibitem{ALEPH:2003obs}
A.~Heister, et~al., {Studies of QCD at $e^{+}e^{-}$ centre-of-mass energies between 91-GeV and 209-GeV}, Eur. Phys. J. C 35 (2004) 457--486.
\newblock \href {https://doi.org/10.1140/epjc/s2004-01891-4} {\path{doi:10.1140/epjc/s2004-01891-4}}.

\bibitem{Weinzierl:2010cw}
S.~Weinzierl, {Jet algorithms in electron-positron annihilation: Perturbative higher order predictions}, Eur. Phys. J. C 71 (2011) 1565, [Erratum: Eur.Phys.J.C 71, 1717 (2011)].
\newblock \href {http://arxiv.org/abs/1011.6247} {\path{arXiv:1011.6247}}, \href {https://doi.org/10.1140/epjc/s10052-011-1717-z} {\path{doi:10.1140/epjc/s10052-011-1717-z}}.

\bibitem{Verbytskyi:2019zhh}
A.~Verbytskyi, A.~Banfi, A.~Kardos, P.~F. Monni, S.~Kluth, G.~Somogyi, Z.~Sz\H{o}r, Z.~Tr\'ocs\'anyi, Z.~Tulip\'ant, G.~Zanderighi, {High precision determination of $\alpha_s$ from a global fit of jet rates}, JHEP 08 (2019) 129.
\newblock \href {http://arxiv.org/abs/1902.08158} {\path{arXiv:1902.08158}}, \href {https://doi.org/10.1007/JHEP08(2019)129} {\path{doi:10.1007/JHEP08(2019)129}}.

\bibitem{Sherpa:2019gpd}
E.~Bothmann, et~al., {Event Generation with Sherpa 2.2}, SciPost Phys. 7~(3) (2019) 034.
\newblock \href {http://arxiv.org/abs/1905.09127} {\path{arXiv:1905.09127}}, \href {https://doi.org/10.21468/SciPostPhys.7.3.034} {\path{doi:10.21468/SciPostPhys.7.3.034}}.

\bibitem{Bierlich:2022pfr}
C.~Bierlich, et~al., {A comprehensive guide to the physics and usage of PYTHIA 8.3}, SciPost Phys. Codeb. 2022 (2022) 8.
\newblock \href {http://arxiv.org/abs/2203.11601} {\path{arXiv:2203.11601}}, \href {https://doi.org/10.21468/SciPostPhysCodeb.8} {\path{doi:10.21468/SciPostPhysCodeb.8}}.

\end{thebibliography}


\end{document}